\pdfoutput=1
\documentclass[12pt,preprint]{aastex}

\usepackage{graphicx}

\usepackage{epstopdf}

\usepackage{amsmath}

\newcommand{\numsne}{134}
\newcommand{\numsneir}{120}
\newcommand{\dm}{$\Delta m_{15}$}
\newcommand{\dmb}{$\Delta m_{15}(B)$}
\newcommand{\csp}{\mbox{CSP-I}}

\shorttitle{Photometry of \csp\ Type Ia SNe}
\shortauthors{Krisciunas et al.}
\begin{document}

\title{The Carnegie Supernova Project~I:\\ Third Photometry Data Release of
 Low-Redshift Type~Ia Supernovae and Other White Dwarf Explosions}

\author{Kevin Krisciunas\altaffilmark{1},
Carlos~Contreras\altaffilmark{2,3},
Christopher~R.~Burns\altaffilmark{4},
M.~M.~Phillips\altaffilmark{2},
Maximilian~D.~Stritzinger\altaffilmark{2,3},
Nidia~Morrell\altaffilmark{2},
Mario~Hamuy\altaffilmark{5},
Jorge~Anais\altaffilmark{2},
Luis~Boldt\altaffilmark{2},
Luis~Busta\altaffilmark{2},
Abdo~Campillay\altaffilmark{2},
Sergio~Castell\'on\altaffilmark{2},
Gast\'on~Folatelli\altaffilmark{2,6},
Wendy~L.~Freedman\altaffilmark{4,7},
Consuelo~Gonz\'{a}lez\altaffilmark{2},
Eric~Y.~Hsiao\altaffilmark{2,3,8},
Wojtek~Krzeminski\altaffilmark{2,9},
Sven~Eric~Persson\altaffilmark{4},
Miguel~Roth\altaffilmark{2,10},
Francisco~Salgado\altaffilmark{2,11},
Jacqueline~Ser\'{o}n,\altaffilmark{2,12},
Nicholas~B.~Suntzeff\altaffilmark{1},
Sim\'{o}n~Torres,\altaffilmark{2,13},
Alexei~V.~Filippenko\altaffilmark{14,15}, 
Weidong~Li\altaffilmark{14,9},  
Barry~F.~Madore\altaffilmark{4,16},
D.~L.~DePoy\altaffilmark{1},
Jennifer~L.~Marshall\altaffilmark{1},
Jean-Philippe~Rheault\altaffilmark{1}, and
Steven~Villanueva\altaffilmark{1,17} }

\altaffiltext{1}{
George P. and Cynthia Woods Mitchell Institute for Fundamental Physics and Astronomy, 
Department of Physics and Astronomy, Texas A\&M University, College Station, TX 77843, USA;
{krisciunas@physics.tamu.edu}}
\altaffiltext{2}{Carnegie Observatories, Las Campanas Observatory, Casilla 601, 
La Serena, Chile}
\altaffiltext{3}{Department of Physics and Astronomy, Aarhus University, Ny Munkegade 120,
DK-8000 Aarhus C, Denmark}
\altaffiltext{4}{Observatories of the Carnegie Institution for
 Science, 813 Santa Barbara St., Pasadena, CA 91101, USA}
 \altaffiltext{5}{Departamento de Astronom\'{\i}a, Universidad de Chile,
  Casilla 36-D, Santiago, Chile}
\altaffiltext{6}{Facultad de Ciencias Astron\'{o}micas y Geof\'{i}sicas, Universidad Nacional de La Plata, 
Instituto de Astrof\'{i}sica de La Plata (IALP), CONICET, Paseo del Bosque S/N, B1900FWA La Plata, 
Argentina}
\altaffiltext{7}{Department of Astronomy and Astrophysics, University of 
Chicago, 5640 South Ellis Avenue, Chicago, IL 60637, USA}
\altaffiltext{8}{Department of Physics, Florida State University, Tallahassee, 
FL 32306, USA}
\altaffiltext{9}{Deceased}
\altaffiltext{10}{GMTO Corporation, Avenida Presidente Riesco 5335, Suite 501
Las Condes, Santiago, Chile}
\altaffiltext{11}{Leiden Observatory, Leiden University, PO Box 9513, NL-2300 RA Leiden, 
The Netherlands}
\altaffiltext{12}{Cerro Tololo Inter-American Observatory, Casilla 603, La Serena, Chile}
\altaffiltext{13}{SOAR Telescope, Casilla 603, La Serena, Chile}
\altaffiltext{14}{Department of Astronomy, University of California,
  Berkeley, CA 94720-3411, USA}
\altaffiltext{15}{Miller Senior Fellow, Miller Institute for Basic Research 
in Science, University of California, Berkeley, CA 94720, USA}
\altaffiltext{16}{Infrared Processing and Analysis Center, Caltech/Jet
  Propulsion Laboratory, Pasadena, CA 91125, USA}
\altaffiltext{17}{Department of Astronomy, Ohio State University, Columbus, OH 43210, USA}

\begin{abstract}
  \noindent 


We present final natural system optical ($ugriBV$) and near-infrared ($YJH$)
photometry of \numsne\ supernovae (SNe) with probable white dwarf progenitors
that were observed in 2004--2009 as part
of the first stage of the Carnegie Supernova Project (\mbox{\csp}).  The sample
consists of 123 Type~Ia SNe, 5 Type~Iax SNe, 2 super-Chandrasekhar SN candidates, 2 Type~Ia
SNe interacting with circumstellar matter, and 2 SN~2006bt-like events.  
The redshifts of the objects range from $z = 0.0037$ to 0.0835; the median
redshift is 0.0241. For \numsneir\ (90\%) of these SNe, 
near-infrared photometry was obtained.
Average optical extinction coefficients and color terms are derived 
and demonstrated to be stable during the five \mbox{\csp} observing campaigns. 
Measurements of the \mbox{\csp} near-infrared bandpasses are also described, and 
near-infrared color terms are estimated through synthetic photometry of stellar 
atmosphere models.  Optical and near-infrared magnitudes of local
sequences of tertiary standard stars for each supernova are given, and a new calibration of $Y$-band
magnitudes of the \citet{Per_etal98} standards in the \mbox{\csp} natural
system is presented.
\end{abstract}

\section{Introduction}
\label{sec:intro}

Type~Ia supernovae (SNe, singular SN) are generally agreed to be the result of a 
carbon-oxygen white dwarf that undergoes a thermonuclear runaway \citep{hoyle60} 
owing to mass accretion in a binary system \citep{wheeler71}. The mechanism for the 
ignition of the degenerate material is thought to be tied to the interplay 
between the exploding white dwarf and its companion star. Potential progenitor 
systems are broadly categorized as ``single degenerate'' where the companion star 
is a main sequence, red giant, or helium star, or ``double degenerate'' where the 
system consists of two white dwarfs. Within this scheme, several triggering 
mechanisms have been proposed.  The thermonuclear explosion can be triggered by the 
heat created during the dynamical merger of two white dwarfs after expelling 
angular momentum via gravitational radiation \citep[e.g.,][]{webbink84,Iben_Tut84}. 
The explosion can also be triggered by compressional heating as the white dwarf 
accretes material from a degenerate or nondegenerate companion to close to the 
Chandrasekhar limit \citep[e.g.,][]{whelan73}.  A third mechanism involves the 
explosion of a sub-Chandrasekhar-mass white dwarf triggered by detonating a thin 
surface helium layer which, in turn, triggers a central detonation front 
\citep[e.g.,][]{nomoto82}.  A fourth mechanism might be a collision of two C-O white
dwarfs in a triple-star system \citep{Kus_etal13}.

Currently, it is unclear whether 
the observed SN~Ia population results from a combination of these explosion mechanisms or is largely
dominated by one. The power-law dependence of 
the delay time between the birth of the progenitor system and the explosion 
as a SN Ia (the ``delay-time distribution''; \citealt{maoz10}) and the 
unsuccessful search for evidence of the companions to normal Type~Ia SNe 
\citep[e.g., see][]{Li_etal11,Sch_Pag12,olling15} would seem to favor the double-degenerate 
model, but some events, such as SN~2012cg \citep{marion16} and SN~2017cbv 
\citep{Hos_etal17} show a blue excess in their early-time light curves, indicative 
of nondegenerate companions.  The rare SNe~Ia that interact with 
circumstellar matter (CSM), such as SNe 2002ic \citep{hamuy03} and
PTF~11kx \citep{dilday12}, also favor a single-degenerate system. 

Type~Ia SNe are important for their role in the chemical
enrichment of the Universe \citep[e.g.,][and references 
therein]{nomoto13}.  They also play a fundamental role in observational 
cosmology as luminous standardizable candles in the optical bands 
\citep[e.g.,][]{Phi93,Ham_etal96,riess96,Phi_etal99} and as (essentially) standard candles at 
maximum light in the near-infrared (NIR) \citep[][and references 
therein]{kris04,Kri12, Phi12}.  The most precise current estimates for the value 
of the Hubble constant are 
based on SNe~Ia \citep[][and references therein]{Rie_etal16}; moreover, 
\citet{Rie_etal98} and \citet{Per_etal99} used them to find that the Universe
is currently expanding at an accelerating rate. 



In this age of precision cosmology, observations of SNe~Ia continue 
to play a crucial role \citep[e.g., see][]{sullivan11}. Ironically, we are 
still faced with the situation that many more events have well-observed 
light curves at high redshifts ($z > 0.1$) than at low redshifts 
\citep{betoule14}. Since the SN~Ia results are derived from a 
comparison of the peak magnitudes of distant {\em and} nearby events, the 
relatively heterogeneous quality of the low-redshift data directly affects 
the precision with which we are able to determine the nature of dark 
energy.  Moreover, there are still legitimate concerns about systematic 
errors arising from the conversion of instrumental magnitudes into a 
uniform photometric system, calibration errors, the treatment of host galaxy 
dust reddening corrections, and evolutionary effects caused by differing ages 
or metallicities \citep{Woo_etal07,Fre_etal09,conley11}.


The Carnegie Supernova
Project-I \citep[\mbox{\csp};][]{Ham_etal06} was initiated to address these problems by 
creating a new dataset of low-redshift optical/NIR light curves of 
SNe~Ia in a well-understood and stable photometric system.
The use of NIR data provides several major advantages over optical wavelengths alone.  First, color corrections caused by dust and any 
systematic errors associated with these are up to a factor of five smaller than at optical 
wavelengths \citep{kris00,Fre_etal09}.  The combination of optical and NIR photometry 
also provides invaluable information on the shape of the host-galaxy dust reddening curve 
\citep{Fol_etal10,mandel11,Bur_etal14}.
Finally, both theory and observations indicate that the rest-frame peak NIR magnitudes 
of SNe Ia exhibit a smaller intrinsic scatter \citep{kasen06,kattner12,mandel09}
and require only minimal luminosity vs. decline-rate corrections.

The \mbox{\csp} was a five-year (2004--2009) project funded 
by the National Science Foundation (NSF). It consisted of low-redshift ($z \lesssim 
0.08$) and high-redshift ($0.1 < z < 0.7$) components.  \citet{Ham_etal06} 
presented an overview of the goals of the low-redshift portion of the project, 
the facilities at Las Campanas Observatory (LCO), and details of photometric 
calibration.  It should be noted that the \mbox{\csp} also obtained 
observations of more than 100 low-redshift core-collapse SNe.

\citet[][hereafter Paper 1]{Con_etal10} presented \csp\ photometry of 35 
low-redshift SNe~Ia, 25 of which were observed in the NIR. Analysis 
of the photometry of these objects is given by \citet{Fol_etal10}. 
\citet[][hereafter Paper 2]{Str_etal11} presented \csp\ photometry of 50 
more low-redshift SNe~Ia, 45 of which were observed in the NIR.  This 
sample included two super-Chandrasekhar candidates \citep{howell06} and 
two SN~2006bt-like objects \citep{Foley_etal10b}.  The high-redshift 
objects observed by the \csp\ in the rest-frame $i$ band are discussed by 
\citet{Fre_etal09}.

In this paper, we present optical and NIR photometry of the final 49 SNe in 
the \mbox{\csp} low-redshift sample, including five members of the 
SN~2002cx-like subclass, also referred to as Type~Iax SNe 
\citep[see][]{Foley_etal13}, and two examples of the Type~Ia-CSM subtype 
\citep{silverman13}.  We provide updated optical and NIR photometry of the 
85 previously published low-redshift SNe in the \mbox{\csp} sample since, 
in several cases, we have eliminated bad data points, improved the 
photometric calibrations, and obtained better host-galaxy reference 
images. This combined dataset represents the definitive version of the 
\mbox{\csp} photometry for low-redshift white dwarf SNe, and supersedes 
the light curves published in Papers~1 and~2, as well as those published 
for a few individual objects by \citet{Pri_etal07}, \citet{Phi_etal07}, 
\citet{Sch_etal08}, \citet{Str_etal10}, \citet{taddia12a}, 
\citet{Str_etal14}, and \citet{Gal_etal17}.  Other useful optical and near-IR
observations of Type Ia SNe includes the photometry obtained by 
the Center for Astrophysics group \citep{Hic_etal09, Hic_etal12, Fri_etal15}.

\section {Supernova Sample}
\label{sec:sample}

In Figure~\ref{fig:fcharts} we present finder charts for the \numsne 
~SNe~Ia comprising the low-redshift \mbox{\csp} white dwarf SN 
sample, indicating the positions of the SN and the local sequence of 
tertiary standard stars in each field (see \S\ref{sec:std_stars}). General 
properties of each SN and host galaxy are provided in 
Table~\ref{tab:snproperties}. Two ``targeted'' searches, the Lick 
Observatory SN Search \citep[LOSS;][]{filippenko01,leaman11,li11a} 
with the 0.76~m Katzman Automatic Imaging Telescope (KAIT) and the Chilean 
Automatic Supernova Search \citep{pignata09}, accounted for 55\% of the 
SNe selected for follow-up observations. Another 36\% of the SNe in the sample were 
discovered by amateur astronomers, and the remaining 19\% were drawn from 
two ``untargeted'' (sometimes referred to as ``blind'') searches: the Robotic Optical Transient Search Experiment 
\citep[ROTSE-III;][]{akerlof03} and the Sloan Digital Sky Survey-II 
Supernova Survey \citep[SDSS-II;][]{frieman08}.


Note that the host galaxies of five SNe in the sample are somewhat ambiguous: 


\begin{itemize}
\item The field of SN~2006bt lies at large angular distances from any
potential hosts in a field 
rich in galaxies. Redshift measurements taken from the 
NASA/IPAC Extragalactic Database (NED) show that the majority of the galaxies 
within 10$\arcmin$ of the SN are members of a cluster at $z_{\rm helio} = 0.0482 \pm 
0.0026$. The closest galaxy to the SN, 2MASX~J15562803+2002482, is 35$\arcsec$ 
distant and has $z_{\rm helio} = 0.0463$.  However, the second-closest galaxy, CGCG~108-013, 
which lies 50$\arcsec$ from the SN and has $z_{\rm helio} = 0.0322$, was determined by 
\citet{Foley_etal10b} to be the most likely host using the SuperNova IDentification code 
\citep[SNID;][]{BloTon07}. As discussed by these authors, SN~2006bt displayed 
unusual photometric and spectroscopic properties compared to typical SNe~Ia.
By chance, another object with very similar characteristics whose host galaxy is 
unambiguous, SN~2006ot (see Figure~\ref{fig:fcharts}), was discovered in 
the \mbox{\csp} sample by \citet{Str_etal14}.  After deredshifting our spectra of
SN~2006ot, we used them as templates to determine at what redshift they best matched
spectra of SN~2006bt taken at comparable epochs.  We derive $z_{\rm helio} = 0.0325 \pm 0.0005$ 
for SN~2006bt, confirming that CGCG~108-013 is the likely host. 

\item SN~2007mm exploded in the midst of a compact group of galaxies, five of 
whose members are within 33$\arcsec$ of the SN position.  The redshift listed in 
Table~\ref{tab:snproperties} corresponds to the average of these five galaxies. 

\item SN~2008bf appeared between three galaxies in the NGC~4065 group, the 
closest being 2MASX~J12040495+2014489 which has $z_{\rm helio} = 0.0224$.  However, 
any of the three galaxies could be the host, and so we adopt the NGC~4065 group 
redshift of $z_{\rm helio} = 0.0235$.  A pre-SN image by the Sloan Digital Sky Survey 
(SDSS) shows an unresolved source at the position of the SN.  Host-galaxy reference 
images obtained by the \mbox{\csp} after the SN faded show the same unresolved source.
The colors of this object are most consistent with a star, and so we assume it is
unrelated to the SN.

\item SN~2008ff is 32$\arcsec$ from 2MASX J20135726-4420540, whose redshift is 
$z_{\rm helio} = 0.0194$, and is 40$\arcsec$ from ESO 284-G032 which has $z_{\rm helio} = 
0.0192$.  We assume the average of these redshifts for the SN. 

\item NGC 3425 (also known as NGC 3388), the supposed host of SN~2009al at 
$z_{\rm helio} = 0.0221$, is 66$\arcsec$ distant from the SN.  Although a second galaxy
(SDSS J105124.64+083326.7) with $z_{\rm helio} = 0.0232$ is located 85$\arcsec$ from 
the SN, we adopt $z_{\rm helio} = 0.0221$.

\end{itemize}

The top panel of Figure~\ref{fig:histograms} shows a histogram of the 
heliocentric radial velocities of the host galaxies of the \numsne ~SNe in our sample.  The 
redshifts range from $z = 0.0037$ (for SN~2010ae) to 0.0835 (for SN~2006fw).  The 
median redshift is 0.0241, corresponding to a distance of 100 Mpc for a Hubble 
constant of 72 km s$^{-1}$ Mpc$^{-1}$ \citep {Fre_etal01}.


Table~\ref{tab:spec_details} summarizes spectroscopic classifications for the sample. 
The spectral subtype is listed, along with the epoch of the spectrum 
(relative to the time of $B$-band maximum) used to determine the spectral 
subtype. Also given are classifications in the \citet{branch06} and 
\citet{wang09} schemes using the same criteria as \citet{Fol_etal13}.  Photometric
parameters for the subset of 123 SNe~Ia are provided in
Table~\ref{tab:lcfit_details}.  See \S \ref{sec:Ia} for details.

\section{Imaging}
\label{sec:obs}

Between 2004 and 2010, five 9-month \mbox{\csp} observing campaigns were 
carried out, each running from approximately September through May.  During these 
campaigns, the vast majority of the optical imaging in the $ugriBV$ bandpasses 
was obtained with the SITe3 CCD camera attached to the Las Campanas Observatory 
(LCO) Swope 1~m telescope.  A limited amount of optical imaging was also taken 
with the Tek5 CCD camera on the LCO 2.5~m du~Pont telescope.

NIR imaging of the \mbox{\csp} SNe during the first observing campaign was obtained 
exclusively with the Wide-Field IR Camera (WIRC) on the du~Pont 2.5~m telescope 
\citep{Per_etal02}, and some additional WIRC observations were carried out during 
campaigns 2--5. However, beginning with the second \mbox{\csp} campaign, a new imager 
built especially for the \csp, RetroCam, went into use on the Swope 1~m telescope and 
became the workhorse NIR camera for the remaining four campaigns.

Basic reductions of the optical and NIR images are discussed in detail in 
Paper~1.  For the optical data, these consisted of electronic bias subtraction, 
flat-fielding, application of a linearity correction appropriate for the CCD, and 
an exposure-time correction that corrects for a shutter time delay.  The 
individual dithered NIR images were corrected for electronic bias, detector linearity, 
pixel-to-pixel variations of the detector sensitivity, and sky 
background, and were then aligned and stacked to produce a final image.

Host-galaxy reference images were obtained a year or more after the last follow-up 
image. As described in Paper~1, most of the optical $ugriBV$ reference images were 
obtained with the du~Pont telescope and the Tek5 CCD camera using the same 
filters employed to take the original science images\footnote[18]{If 
the set of \csp\ $ugri$ filters used at the Swope telescope was not available,
a duplicate set produced by Asahi Spectral Company, Ltd. in the same run as the Swope
filters was employed.  Likewise, if the $B$ and/or $V$ filters were not available, 
filters from the du~Pont with similar throughput curves were used.}.  
A smaller set of reference
images was also taken with the du~Pont telescope using a second CCD camera, SITe2, 
and a few were obtained using the Swope~+~SITe3 camera under good 
seeing conditions.  For a small number of objects located far outside their host 
galaxies, subtraction of a reference image was unnecessary.

NIR $YJH$ host-galaxy reference images were obtained exclusively with WIRC on the du Pont telescope
using similar filters as in RetroCam.

\section{Filters}
\label{sec:filters}

Precision photometry requires knowledge of the filter throughputs as a function 
of wavelength \citep[e.g.,][]{bessell90,stubbs06}, so we devised an 
instrument incorporating a monochromator and calibrated detectors to precisely 
determine the response functions (telescope~+~filter~+~instrument) of the \mbox{\csp} 
bandpasses \citep{Rhe_etal14}.   
Paper~2 provides a detailed account of 
the measurement of the optical bandpasses, and in 
Appendix~\ref{sec:nir_bandpasses} we describe the calibration of the NIR 
bandpasses using the same instrument and similar techniques.

Repeated scans of the  \mbox{\csp} $ugriBV$ bandpasses 
show that the relative measurement errors in transmission are $\sim1$\% or less. 
That is, the ratios as a function of wavelength of repeated scans of each 
individual filter fall within an envelope that is $\pm$1\%.
Repeated scans of the $YJH$ bandpasses (both for the Swope~+~RetroCam
and the du~Pont~+~WIRC) indicate that each of these filters has been determined in a relative
sense to a precision of 2--3\%. Unfortunately, the throughput of the WIRC $K_s$ filter is 
highly uncertain beyond 2200~nm (2.2~$\mu$m) owing to the low power of the monochromator light 
source at these wavelengths and the rising thermal contamination at 2.3 $\mu$m.
Nearly all of the $K_s$-band observations made by 
\mbox{\csp} were obtained during the first observing campaign, and were published 
in Papers~1 and~2.  However, due to the uncertainty in the $K_s$ filter response 
function, we have elected not to include any $K_s$-band observations in 
this final data release paper.  {\it Those wishing to employ the \mbox{\csp} photometry for
precision cosmology applications are advised not to use the $K_s$-band measurements 
given in Papers~1 and~2.}

Figure~\ref{fig:filters} displays the optical and NIR bandpasses employed by the \csp\ after including 
atmospheric transmission typical of LCO.  In constructing the optical filter
bandpasses, we have assumed an airmass of 1.2, a value which corresponds to the mode 
of the airmasses of the standard-star observations used to calibrate the data. 
In Appendix~\ref{sec:opt_cterms} we test the validity of the final optical bandpasses 
by reproducing the measured color terms (see \S\ref{sec:swope_opt}) via synthetic 
photometry performed on spectra of Landolt standards.

\section{Photometric Reductions: Overview}
\label{sec:phot}

In this and the following section we define the \csp\ natural photometric system and describe the
methodology used to calibrate it. While this has been described in previous
\mbox{\csp} publications, several changes have been made in our definitions
and procedures. These changes affect the entire \mbox{\csp} sample, and as this is the
final data release, we seek to make the procedure as clear as possible.

\subsection{The \csp\ Natural System}
\label{sec:phot_defined}
Because of differences in instrument throughputs, photometry measured by different
facilities will not agree. These differences are a strong function of color of
the object and can therefore be taken into account through the use of color terms
\citep[e.g., see][]{harris81}. These color terms are typically
measured empirically by observing a set of standard stars with a large range in
color, and allow the observer to transform their instrumental photometry into
the system in which the standards were measured.

The primary difficulty in dealing with {\em supernova} photometry is the fact 
that their spectral energy distributions (SEDs) are significantly different than
those of the stars we use to calibrate. Supernova spectra also evolve 
significantly with time. Consequently, the color terms cannot be used on the SN 
magnitudes to transform them to a standard system. Instead, we adopt a {\em 
natural} system, in which the standard-star magnitudes are transformed to what we 
would measure through our own telescopes/instruments. There are several 
advantages to working in the natural system, as follows.

\begin{itemize}

\item If the system is stable (i.e., color terms do not vary significantly with time), 
nightly calibration of each filter does not rely on other filters to measure colors. This 
can be advantageous if time is short.

\item Working in the natural system requires fewer standard-star measurements to obtain the nightly zero 
points, as the equations have one fewer unknown.  In fact, the equations can be 
reduced to only one unknown, the nightly zero point (see \S \ref{sec:photdetails}).

\item Photometry in the natural system is the ``purest'' form of the data and, given precise
bandpass response function measurements, allows the \mbox{\csp} observations to be more readily
combined with photometry in other photometric systems using S-corrections \citep{Str_etal02,
Kri_etal03}.

\end{itemize}

To compare photometry of SNe in host galaxies at different redshifts,
precision K-corrections must be calculated with the transmission functions 
used in the observations and not that of the standard system. Thus, one 
must back out the standard system color transformation to the natural 
system in order to do the K-correction.

Having introduced the natural system, we now proceed to describe in general terms the
procedure used to measure and calibrate the photometry using standard stars.

\subsection{Standard Stars}
\label{sec:std_stars}

Observations of standard stars are required in order to calibrate the SN 
photometry.  In this paper, we adopt the following nomenclature in 
referring to the different types of standard stars used by the 
\mbox{\csp}.

\begin{itemize}

\item {\bf Primary standards.}  We use this term to refer to Vega ($\alpha$~Lyr),
the F~subdwarf BD+17$\arcdeg$4708, and the two 
CALSPEC \citep{bohlin2014}
solar-analog standards P177D and P330E.


\item {\bf Secondary standards.}  We employed observations of  \citet{Lan92} and
 \citet{Smi_etal02} standard stars to provide the fundamental calibration of the \mbox{\csp} optical 
 photometry.   The Landolt and Smith et al.~stars are considered 
``secondary standards'' since they were calibrated with respect to the primary standards 
Vega and BD+17$\arcdeg$4708, respectively.  In the NIR, the \csp\ photometry is calibrated with respect
to the \citet[][hereafter P98]{Per_etal98} secondary standards, which are tied to Vega.

\item {\bf Tertiary standards.}  A ``local sequence'' of stars was established in each SN field 
in order to allow relative photometry of the SN to be measured.
We refer to the local sequence stars as ``tertiary standards'' because they were calibrated 
via observations of secondary standards.


\end{itemize}

\subsection{Supernova Photometry and Calibration}

In order to measure photometry of the SNe accurately, the underlying
host-galaxy light is first subtracted from each SN image
using the host-galaxy reference images obtained after the SN has 
disappeared. The details of this procedure are discussed in Papers~1 and~2.
DAOPhot \citep{Stetson87}  is then used to measure counts in our CCD frames for 
both the SN and the local sequence of tertiary standards using
point-spread-function (PSF) photometry. For each 
tertiary standard, $i$, we measure a differential magnitude with respect to
the SN,
\begin{equation}
\Delta m_{{\rm SN},i} = -2.5 \log_{10}\left(\frac{e^-_{\rm SN}}{e^-_i}\right) \; ,
\label{eq:deltam}
\end{equation}
where $e^-_{\rm SN}$ and $e^-_i$ are the photoelectrons measured for the SN and 
tertiary standards, respectively. The uncertainty $\sigma\left(\Delta m_{{\rm SN},i}\right)$
is computed assuming Poisson statistics. 

Once the tertiary standards have been
calibrated to the natural system, the final magnitude of the SN can be 
computed as a weighted average,
\begin{equation}
m_{\rm SN} = \frac{\sum_i \left(\Delta m_{{\rm SN},i} + m_i\right)w_i}{\sum_i w_i} \; ,
\label{eq:msn}
\end{equation}
where $m_i$ are the calibrated magnitudes of the tertiary standards,
and the weights $w_i$ are the inverse variance
\begin{equation}
w_i = \left[\sigma^2\left(\Delta m_{{\rm SN},i}\right) + \sigma^2\left(m_i\right)\right]^{-1} \; .
\end{equation}
The uncertainty in the SN photometry is therefore 
\begin{equation}
\label{eq:error_msn}
\sigma\left(m_{\rm SN}\right) = \left[\sum_i 
    \left(\sigma^2\left(\Delta m_{{\rm SN},i}\right) + \sigma^2\left(m_i\right)\right)^{-1}\right]^{-1/2} \; ,
\end{equation}
which contains a variance term for the statistical uncertainty from photon counts as well as a
systematic variance term that describes the uncertainty in each tertiary standard's absolute flux.
This procedure is applied for each SN field in each filter.
The remainder of this section deals with the determination of the calibrated magnitudes
of the tertiary standard stars, $m_i$.

\subsection{Tertiary Standard Calibration}
In the natural system, the calibrated magnitudes of the tertiary standards are
determined relative to the natural magnitudes of secondary standards observed on
photometric nights. For a set of photometric nights (${j}$), on which a set of secondary standards
($k$) is measured, the estimate of the magnitude of the tertiary standard ($i$) in a particular filter ($\lambda$) is
\begin{equation}
m_{i,\lambda} = \frac{\sum_{j,k} \left(m^\prime_{{\rm nat},k,\lambda} + \Delta m_{i,j,k,\lambda}
      - k_\lambda \left(X_{i,j,\lambda}-X_{j,k,\lambda}\right)\right)w_{j,k,\lambda}}{\sum_{j,k} w_{j,k,\lambda}} \; ,
\label{eq:ter_calib}
\end{equation}
where $m^\prime_{{\rm nat},k,\lambda}$ is the natural-system magnitude of the secondary
standard $k$, $\Delta m_{i,j,k,\lambda}$ is the differential magnitude (see Eq.~\ref{eq:deltam})
between the tertiary standard $i$
and the secondary standard $k$ on night $j$, $X_{i,j,\lambda}$ and $X_{j,k,\lambda}$ are the respective
airmasses of the tertiary and secondary standards, and $k_\lambda$ is the 
extinction coefficient.  The weights ($w_{j,k,\lambda}$) are the inverse variances
\begin{equation}
w_{j,k,\lambda} = \left[\sigma^2\left(\Delta m_{i,j,k,\lambda}\right) + 
\sigma^2\left(m^\prime_{{\rm nat},k,\lambda}\right)\right]^{-1}\; ,
\end{equation}
where $\sigma\left(\Delta m_{i,j,k,\lambda}\right)$ is calculated assuming Poisson errors and 
$\sigma\left(m^\prime_{{\rm nat},k,\lambda}\right)$ is taken from the published standard photometry.
The uncertainty, $\sigma\left(m_{i,\lambda}\right)$, is analogous to Eq.~\ref{eq:error_msn}:
\begin{equation}
\label{eq:error_mi}
\sigma\left(m_{i,\lambda}\right) = \left[\sum_i
    \left(\sigma^2\left(\Delta m_{i,j,k,\lambda}\right) + \sigma^2\left(m^\prime_{{\rm nat},k,\lambda}\right)\right)^{-1}\right]^{-1/2} \; .
\end{equation}
Note that because this is a natural system, there is
no color term in Eq.~\ref{eq:ter_calib} and hence no dependence on the 
color of the tertiary standards. The
differential magnitudes, $\Delta m_{i,j,k,\lambda}$, are measured using aperture photometry
as this was found to be more robust for the secondary standard stars, which tend to be
significantly brighter than the tertiary standards. In the optical we used an aperture
of 7\arcsec, while for the NIR, we used an aperture of 5\arcsec. A sky annulus of inner
radius 9\arcsec\ and 2\arcsec\ width was used to estimate the sky level for both the optical and NIR.

The final ingredients are the natural-system magnitudes for the secondary standards,
$m^\prime_{{\rm nat},\lambda}$. As discussed in \S\ref{sec:phot_defined}, color terms are 
used to transform the standard magnitudes of these stars into the natural-system
magnitudes that would be measured through our telescopes.
The form of these transformations is assumed to be linear with color,
\begin{equation}
   m^\prime_{{\rm nat},\lambda} = m^\prime_{\lambda} - \epsilon_\lambda \times C^\prime_\lambda \; ,
   \label{eq:nat_trans}
\end{equation}
where $\epsilon_\lambda$ is the color term and $C^\prime_\lambda$ is the associated color
based on the standard magnitudes. As an example, for $\lambda = B$, we choose
$C^\prime_\lambda = (B - V)$. 
It is important to emphasize that because these color terms are only ever used to
compute $m^\prime_{{\rm nat},\lambda}$, it is the range of colors of the secondary 
standards used for calibration that determines their relative importance.
In other words, we are forcing the zero point of the natural 
system and standard system to be the same at zero color.

Technically, each telescope/instrument used by the \csp\
will have its own color terms and hence its own set of natural magnitudes for
the secondary standards. In the next section, we describe each of these
in detail.

\section{Photometric Reductions: Details}
\label{sec:photdetails}

Our natural system is defined by Eq.~\ref{eq:nat_trans}. If all published 
secondary standards had zero color, then the definition of the natural system would be 
trivial. However, the published standard stars have a range of colors.
To use all the published standards to define the natural system we must 
calculate color terms, as given in Eq.~\ref{eq:nat_trans}, which transform 
the  published standard system into a table of the same stars with 
natural-system photometry.

If something goes wrong with the photometric system and the transmission 
functions change, then the color terms in Eq.~\ref{eq:nat_trans} will 
change. However, for program objects that are stars, the natural system will 
remain well defined because stars were used to define the standard 
to natural systems.

This is not true for SNe because they have different SEDs. Thus, an 
important sanity check on our reductions is to see if the color terms vary 
over time.  Provided that the transmission functions do not change, the color 
terms should never change. However, we must keep track of any variations of 
the color terms to verify that the natural system is stable.

\subsection{Optical Photometry}
\label{sec:opt_phot}

\subsubsection{Swope~+~SITe3}
\label{sec:swope_opt}

We define the transformation of the instrumental $ugribv$ magnitudes into the 
natural system through the following equations:


\begin{equation}
u_{\rm nat}~=~u'~-~\epsilon_u \times (u'~-~g') \; ,
\label{eq:u_nat}
\end{equation}
\begin{equation}
g_{\rm nat}~=~g'~-~\epsilon_g  \times (g'~-~r') \; ,
\label{eq:g_nat}
\end{equation}
\begin{equation}
r_{\rm nat}~=~r'~-~\epsilon_r \times (r'~-~i') \; ,
\label{eq:r_nat}
\end{equation}
\begin{equation}
i_{\rm nat}~=~i'~-~\epsilon_i \times (r'~-~i') \; ,
\label{eq:i_nat}
\end{equation}
\begin{equation}
B_{\rm nat}~=~B~-~\epsilon_b \times (B~-~V) \; ,~{\rm and}
\label{eq:B_nat}
\end{equation}
\begin{equation}
V_{\rm nat}~=~V~-~\epsilon_v \times (V~-~i') \; ,
\label{eq:V_nat}
\end{equation}

\noindent
where $u'g'r'i'BV$ correspond to magnitudes in the standard system. The color 
terms ($\epsilon_{\lambda}$) are measured in the manner described below. The 
magnitudes of the secondary photometric standards of \citet{Lan92} and 
\citet{Smi_etal02} are thus used to calculate new magnitudes of these stars in 
the natural photometric system of the Swope telescope using the above 
equations.

On photometric nights, we can solve for these color terms based on observations of
the secondary standards. To do this, we fit the instrumental magnitudes with the following
equations:

\begin{equation}
u~=~u'~+k_uX~-~\epsilon_u \times (u'~-~g')-~\zeta_u \; ,
\label{eq:u_inst}
\end{equation}
\begin{equation}
g~=~g'~+k_gX~-~\epsilon_g  \times (g'~-~r')-~\zeta_g \; ,
\label{eq:g_inst}
\end{equation}
\begin{equation}
r~=~r'~+k_rX~-~\epsilon_r \times (r'~-~i')-~\zeta_r \; ,
\label{eq:r_inst}
\end{equation}
\begin{equation}
i~=~i'~+k_iX~-~\epsilon_i \times (r'~-~i')-~\zeta_i \; ,
\label{eq:i_inst}
\end{equation}
\begin{equation}
b~=~B~+k_bX~-~\epsilon_b \times (B~-~V)-~\zeta_b \; ,~and 
\label{eq:B_inst}
\end{equation}
\begin{equation}
v~=~V~+k_vX~-~\epsilon_v \times (V~-~i')-~\zeta_v \; .
\label{eq:V_inst}
\end{equation}

\noindent
Note that these equations differ slightly 
from those defined in Eqs.~1--6 of \citet{Ham_etal06} in that the colors on 
the right-hand side of the equations are in the standard system and not the 
instrumental system.

The calibration strategy adopted by the 
\mbox{\csp} for the optical imaging obtained with the Swope telescope was to observe a 
minimum of eight secondary standard stars over a range of airmass during one photometric 
night every week.  During the course of the \csp, different team members would use
IRAF\footnote[19]{IRAF is distributed by the National Optical Astronomy Observatory, which 
is operated by the Association of Universities for Research in Astronomy, Inc., under 
cooperative agreement with the NSF.} tools and procedures
to fit these observations to Eqs.~\ref{eq:u_inst}--\ref{eq:V_inst} to obtain the 
nightly extinction coefficients, color terms, and zero points for each band. 
For this final data release, we have 
redone the nightly measurements of the extinction coefficients, color terms, and 
zero points in a uniform manner using a more sophisticated, noninteractive 
method that accounts for outliers and provides more realistic error bars. In 
detail, we used a Mixture Model Markov Chain Monte Carlo (MCMC) fitting procedure 
\citep[as in][]{hogg10}, which includes a photometric model, a Gaussian model for 
the outliers, an extra variance term, and a $q$ parameter accounting for the 
fraction of the data points that fit the photometric model. 
The MCMC modeling is specified as follows:

\begin{equation}
f_1(k_{\lambda},\epsilon_{\lambda},\zeta_{\lambda}) = m'_{\lambda} + k_{\lambda} X 
- \epsilon_{\lambda}  \times C^\prime_\lambda - \zeta_{\lambda}  
\end{equation}

\noindent
is the photometric model for the observed instrumental magnitudes, and 

\begin{equation}
f_2(\mu,\sigma) = N(\mu,\sigma) 
\end{equation}

\noindent
is the Gaussian normal distribution model for outliers. The seven-parameter likelihood 
function,
$\mathcal{L}(k_{\lambda},\epsilon _{\lambda},\zeta _{\lambda},\sigma_e^2,
q,\mu,\sigma)$, 
is then expressed as
\begin{equation}
\mathcal{L}= \prod_{i=1}^{N} \left[ q\frac{{\rm exp}(-0.5(f_1-m_i)^2/(\sigma_i^2 + \sigma_{\rm extra}^2))}{\sqrt{2\pi (\sigma_i^2 + \sigma_{\rm extra}^2)}} + 
(1-q)\frac{{\rm exp}(-0.5(f_2-m_i)^2/(\sigma_i^2 + \sigma^2))}{\sqrt{2\pi (\sigma_i^2 + \sigma^2)}}\right] \; ,
\end{equation}

\noindent
where $N$ is the number of standard-star observations in one photometric night in one filter.

In this model, $m_i$ and $\sigma_i$ correspond to instrumental magnitude and error 
bars; {$k_{\lambda}$}, $\epsilon_{\lambda}$, and $\zeta_{\lambda}$ are the nightly extinction coefficients, color 
terms, and zero points (respectively) for filter ${\lambda}$; $C^\prime_\lambda$ is the
color associated with the color term; and $\sigma$ is the standard 
deviation of the Gaussian error distribution (for outliers) centered on $\mu$. The 
extra variance, $\sigma_{\rm extra}^2$, is an additional error term added to every 
single measurement.  This is necessary because a single bright secondary standard star 
typically has an uncertainty due to photon statistics of only a few millimagnitudes, 
while the zero-point dispersion for a good night is no better than 0.01~mag.  
Finally, $q$ represents the fraction of the data that fits the 
photometric model, while $1-q$ is the fraction that can be considered as outliers.  
A handful of nights with values of $q < 0.8$ in different filters was discarded as 
likely to have been nonphotometric.

Figure~\ref{fig:extinction} displays nightly values of the atmospheric extinction 
coefficients in $ugriBV$ for the Swope~+~SITe3 camera 
derived with this MCMC model over the five \mbox{\csp} campaigns.  Histograms of the 
collected extinction-coefficient measurements are 
shown in the right-hand part of each panel.  Figure~\ref{fig:CT} shows a similar plot 
of the color terms over the five campaigns.  In neither of these figures is there 
evidence for significant secular changes in the extinction coefficients or color 
terms.

The nightly photometric zero points for the Swope~+~SITe3 camera are shown in 
Figure~\ref{fig:ZP}. The obvious zig-zag pattern arises from the accumulation of 
dust and aerosols 
between the two washings of the primary mirror 
(marked by the red arrows) that occurred during the \csp\ observing campaigns. 
Smaller dips in sensitivity are observed around mid-February 2006 (JD 2,453,780) 
and mid-March 2008 (JD 2,454,540). Similar dips are visible during the summer 
months in the zero points measured by \citet{Bur_etal95} between November 1975 
and August 1994 at the neighboring La Silla Observatory, and we speculate that 
these are associated with an increase in atmospheric haze that occurs due to the 
inversion layer being generally higher at that time of the year.  
Interestingly, these dips do not appear to be accompanied by significant changes in the 
extinction coefficients and color terms.

The demonstrated stability of the nightly extinction coefficients and color terms over
all five \mbox{\csp} campaigns justifies adopting average color terms 
and extinction coefficients for the final photometric reductions.
This reduces the problem to solving for nightly zero points only.
In this way, just a handful of secondary standard
star observations is needed to calibrate the natural
photometry for the local sequence tertiary standards observed during the same night. 
The final mean extinction coefficients and color terms adopted for the five \mbox{\csp} campaigns 
are given in Table~\ref{tab:ec_ct}. Using these mean color terms, natural-system
magnitudes for the \citet{Smi_etal02} and \citet{Lan92} secondary standards were calculated
via Eqs.~\ref{eq:u_nat}--\ref{eq:V_nat}.  These, in turn, were used to derive
magnitudes in the natural system of the local sequences of tertiary
standards in each of the SN fields.

Final $u'g'r'i'BV$ magnitudes of the local sequences of tertiary standards
for all \numsne~SNe are listed in Table~\ref{tab:opt_local_stds}.  
Note that these are given in the {\em standard} system  (i.e., as calculated using
Eqs.~\ref{eq:u_nat}--\ref{eq:V_nat}) in order to facilitate their usage by 
others.
In all cases, these magnitudes are based on observations made on at least 
three different photometric nights, and the 
accompanying uncertainties are 
weighted averages of the errors computed from these multiple 
measurements.\footnote[20]{In
this paper, in the tables of field-star magnitudes or SN photometry, any
entry given as ``0.000(000)'' indicates missing data.}



As discussed in Paper~2, on 14~January~2006 (unless otherwise noted, UT dates are used throughout
this paper; JD~2,453,749) the original $V$ filter 
(``LC-3014'') used at the Swope telescope was broken and subsequently replaced by another $V$ filter 
(``LC-3009'').  However, after a few nights of use, it was determined that the 
replacement filter had a significantly different color term compared to the 
original. This filter was  replaced on 25~January~2006 (JD~2,453,760) with a third 
filter (``LC-9844''), which was used for the remainder of the \mbox{\csp} campaigns.  
Although the bandpass of the LC-9844 filter is slightly broader than that of the 
LC-3014 filter (see Figure~\ref{fig:filters}), observations at the telescope as well 
as synthetic photometry showed the color terms to be the same to within 
$\sim0.002$.  Hence, we adopted the same natural magnitudes of the local 
sequences of tertiary standards for observations made in both of these filters.  
However, the color 
term of the LC-3009 filter was sufficiently different that we have treated separately the 
reduction of the smaller number of observations obtained with this filter.
Table~\ref{tab:ec_ct} gives the mean color term for $V$-band 
transformations for the LC-3014 and LC-9844 filters, whose value is $-0.058$.
For the small amount of $V$-band photometry obtained with the LC-3009 filter,
we assume the color term of $-0.044$ derived in Paper~2.

\subsubsection{du~Pont~+~Tek5}
\label{sec:dupont_opt}

As already mentioned, owing to its larger aperture and better delivered image quality, the 2.5~m du~Pont 
telescope was used during the \mbox{\csp} to obtain host-galaxy reference images in 
$ugriBV$ using the facility Tek5 CCD camera.  A small amount of SN follow-up 
imaging was also obtained with this telescope/instrument combination.  
Unfortunately, precise measurements of the filter response functions with the 
Tek5 camera were not carried out, and this camera has since been decommissioned.  
Nevertheless, it is possible to estimate the color terms of this system using 
the local sequences of tertiary standards (established with the Swope~+~SITe3 camera) in the fields 
of the SNe observed with the du~Pont~+~Tek5 camera.

To carry out this experiment we chose two objects, SNe 2007ab and 2008O, that were 
observed in both the Swope~+~SITe3 and du~Pont~+~Tek5 systems.  Both SNe are at 
relatively low Galactic latitudes with many foreground stars in their fields.  
Natural-system magnitudes in the $ugriBV$ bandpasses were measured for the 100 
brightest stars in each field using all of the Swope~+~SITe3 images calibrated by the 
respective local sequence of tertiary standards.  The range in $(B-V)$ colors covered by these stars was 
$+0.2$ to $+1.5$~mag for SN~2007ab, and $+0.4$ to $+1.2$~mag for SN~2008O.

SN~2007ab was
observed on one night with the du~Pont~+~Tek5 camera, and SN~2008O on four nights.
Instrumental magnitudes were measured for the same 100 field stars in each of the 
images taken on these nights, and differences ($\Delta m$) were calculated with respect 
to the Swope+SITe3 natural system magnitudes:

\begin{equation}
\Delta m = m({\rm Swope\; SITe3})_{\rm nat} - m({\rm du~Pont\; Tek5})_{\rm ins} \; .
\end{equation}

\noindent
If the response functions for a given filter are identical between the Swope SITe3
and du~Pont Tek5 cameras, we would expect $\Delta m$ to be a constant.
On the other hand, if the response functions are significantly different, we 
would expect to detect a relative color term as well.  We therefore analyzed 
the observations by fitting the model

\begin{equation}
\Delta m_{\lambda} =  \epsilon_{\lambda} \times C_{\lambda} + \zeta_{\lambda} \; .
\end{equation}

\noindent
Here, the color $C_{\lambda}$ is in the natural system and depends on the filters 
as per Eqs.~\ref{eq:u_inst}--\ref{eq:V_inst}. For example,
$\Delta B~=~\epsilon_b~(B-V)~+~\zeta_b$ for the $B$ band.

For the $griBV$ filters, we find that the color term is within 1--2$\sigma$ of zero. The color 
term for the $u$ filter is also consistent with zero to $\sim2\sigma$, but these results 
are of lower confidence since this filter was utilized only one night for each SN.

Based on these results, it is justified to assume that the SN photometry obtained 
in the $griBV$ filters with the du~Pont~+~Tek5 camera is on the same natural system as the 
Swope~+~SITe3 camera.  It also seems likely that any difference between the $u$ bandpasses
is small.  We have therefore opted to calibrate the SN photometry obtained with the
du~Pont~+~Tek5 camera using the natural system 
tertiary standard star magnitudes, mean extinction coefficients, and mean color terms 
measured with the Swope~+~SITe3 camera.

\subsection{NIR Photometry}
\label{sec:nir_phot}

\subsubsection{Swope~+~RetroCam}
\label{sec:swope_nir}

The Swope~+~RetroCam $YJH$ bandpasses are shown in Figure~\ref{fig:filters}.
On the night of 2008~December~8 (JD~2,454,808), the observer
detected a change in the $J$-band dome flat-field images, suggesting either
contamination or that the filter might be starting to delaminate. The decision was 
taken to replace the suspect filter, and this was accomplished approximately a 
month later.  The last observations made with the original filter, which we will refer 
to as ``$J_{\rm RC1}$'', were obtained on 2009~January~2 (JD~2,454,833).
Observations with the replacement filter, which we call ``$J_{\rm RC2}$'', began on 
2009~January~15 (JD~2,454,846).  It was eventually determined that the 
change in the $J_{\rm RC1}$ filter was due to contamination, and this problem
affected the $J_{\rm RC1}$ observations obtained between 
JD~2,454,808 and 2,454,833.  Although we have no evidence that the contamination
significantly changed the bandpass of the $J_{\rm RC1}$ filter and have therefore 
included these observations in this paper, we caution the reader that the
reliability of these observations is less certain than that of the other $J$-band
photometry published in this paper.

In Papers~1 and~2, we neglected any color terms that might exist 
in transforming $J$ and $H$ measurements made 
by the \csp\ to the P98 photometric system.  In order to check this assumption, we have 
reproduced the P98 bandpasses by combining 
the filter transmission data and typical NICMOS3 quantum-efficiency curve given 
by these authors with two aluminum reflectivity curves (one for the primary and 
another for the secondary mirror) and an atmospheric transmission spectrum 
typical of LCO.  The resulting response functions are plotted in red in 
Figure~\ref{fig:filters}.

The $(J-H)$ colors of the P98 secondary standard stars used by the \csp\ to calibrate 
both the Swope and du~Pont NIR observations range from only $+0.19$ to $+0.35$~mag.  
This is too small of a color range to measure the NIR color terms, and so we must
resort to synthetic photometry of model atmospheres
to estimate these. We downloaded the \citet{castelli03} 
atmosphere models for a range of stellar parameters. For each model spectrum, we 
then computed synthetic photometry for a range of reddenings ($E(B-V) = 
0.0$~to~2.5 mag), and plotted the differences between the P98 magnitudes and the 
RetroCam and WIRC $J$ and $H$ magnitudes as a function of the $(J-H)_{\rm P98}$ color (see  
Figure~\ref{fig:castelli_cts}). Linear fits to these data yield the following:

\begin{equation}
J_{\rm P98}~=~J_{\rm RC1}~+~0.039 \times (J~-~H)_{\rm P98}+~\zeta_j \; ,
\label{eq:j_rc1}
\end{equation}
\begin{equation}
J_{\rm P98}~=~J_{\rm RC2}~+~0.016 \times (J~-~H)_{\rm P98}+~\zeta_j \; ,~{\rm and}
\label{eq:j_rc2}
\end{equation}
\begin{equation}
H_{\rm P98}~=~H_{\rm RC}~-~0.029 \times (J~-~H)_{\rm P98}+~\zeta_h \; .
\label{eq:h_rc}
\end{equation}

\noindent In Appendix~\ref{sec:nir_cterms}, we present du~Pont~+~RetroCam observations of P98 standards
covering a much wider range of colors  that validate the accuracy of this procedure. 

 Although the effect of the 
color terms in Eqs.~\ref{eq:j_rc1}--\ref{eq:h_rc} is less than 0.01~mag
over the small range of color of the P98 standards, it is a systematic effect and so we use them to 
transform the P98 magnitudes to the natural system.

The $Y$ photometric band was introduced by \citet{hillenbrand02}.  
\citet{Ham_etal06} calculated synthetic $(Y-K_s)$ and $(J-K_s)$ colors from 
Kurucz model atmosphere spectra using the estimated filter response functions for the Magellan 
6.5~m Baade telescope ``PANIC'' NIR imager.  These values were fitted with a 
fifth-order polynomial with the requirement that $(Y-K_s) = 0.0$ when $(J-K_s) = 
0.0$~mag, consistent with the definition that $\alpha$~Lyr (Vega) has magnitudes of 
zero at all NIR wavelengths \citep{elias82}.  This relation was then used to 
compute $Y$-band magnitudes from $J$ and $K_s$ for all of the P98 secondary standards.
In Appendix~\ref{sec:y_phot}, we repeat this exercise using the {\em measured} Swope~+~RetroCam 
$Y$-band response function along with the $J$ and $K_s$ filter response functions 
we have derived for the P98 standards. 

\subsubsection{du~Pont~+~WIRC}
\label{sec:dupont_nir}

Color terms for the du~Pont~+~WIRC system were calculated from synthetic
photometry of the \citet{castelli03} stellar
atmosphere models in the manner described previously for the 
Swope~+~RetroCam.  We find

\begin{equation}
J_{\rm P98}~=~J_{\rm WIRC}~+~0.015 \times (J~-~H)_{\rm P98}+~\zeta_j \; ,~{\rm and}
\label{eq:j_wirc}
\end{equation}
\begin{equation}
H_{\rm P98}~=~H_{\rm WIRC}~-~0.031 \times (J~-~H)_{\rm P98}+~\zeta_h \; .
\label{eq:h_wirc}
\end{equation}

\noindent We note that these color terms are nearly identical to those for the 
$J_{\rm RC2}$ and $H_{\rm RC}$ filters (cf. Eqs.~\ref{eq:j_rc2} and \ref{eq:h_rc}).

Figure~\ref{fig:filters} shows that the du~Pont~+~WIRC $Y$ bandpass cuts off more
rapidly at blue wavelengths than is the case for the Swope~+~RetroCam $Y$ filter.
We again employ synthetic photometry to evaluate the color term required to transform the 
du~Pont~+~WIRC $Y$-band secondary standard star observations to the Swope~+~RetroCam
system.  This gives

\begin{equation}
Y_{\rm RetroCam}~=~Y_{\rm WIRC}~-~0.042 \times (J~-~H)_{\rm P98}+~\zeta_y \; .
\label{eq:y_wirc}
\end{equation}

\noindent 
Over the range of $(J-H)$ colors of the local sequence tertiary standards, 
this color term is too large to be ignored.  This means that for
the $Y$ band, we must work in two different natural systems:  that of the 
Swope~+~RetroCam, which we adopt as the ``standard'' system, and that of the
du~Pont~+~WIRC.

\subsubsection{NIR Natural-System Photometry}
\label{sec:nir_nat_sys}

From the above, we conclude that the color terms for the $J_{\rm RC2}$ and $J_{\rm WIRC}$
filters are sufficiently similar that we can average them and, therefore, create a single 
natural system for all of the tertiary standards and SNe observed. 
Likewise, the color terms for the $H_{\rm RC}$ and $H_{\rm WIRC}$ filters are nearly
identical, and the photometry obtained with them can also be considered on the same
natural system.  However, the $J_{\rm RC1}$ color term differs considerably with respect 
to those of the other two $J$ filters, and therefore defines its own natural
system.  Likewise, the color term for the $Y_{\rm WIRC}$ filter compared to $Y_{\rm RC}$ is 
too large to be ignored.

We therefore adopt the following equations to transform the secondary standard
star magnitudes to the natural systems in $YJH$
for the Swope~+~RetroCam:

\begin{equation}
Y_{\rm nat,RC}~\equiv~Y_{\rm RC} \; ,
\label{eq:y_nat_rc}
\end{equation}
\begin{equation}
J_{\rm nat,RC1}~=~J_{\rm P98}~-~0.039 \times (J~-~H)_{\rm P98} \; ,
\label{eq:j_nat_rc1}
\end{equation}
\begin{equation}
J_{\rm nat,RC2}~=~J_{\rm P98}~-~0.015 \times (J~-~H)_{\rm P98} \; ,~{\rm and}
\label{eq:j_nat_rc2}
\end{equation}
\begin{equation}
H_{\rm nat,RC}~=~H_{\rm P98}~+~0.030 \times (J~-~H)_{\rm P98} \; .
\label{eq:h_nat_rc}
\end{equation}

\noindent For the du Pont + WIRC system, the transformation equations for
$J_{\rm nat,WIRC}$ and $H_{\rm nat,WIRC}$ are identical to Eqs.~\ref{eq:j_nat_rc2} and \ref{eq:h_nat_rc},
while for $Y_{\rm nat,WIRC}$ the equation is

\begin{equation}
Y_{\rm nat,WIRC}~=~Y_{\rm RC}~+~0.042 \times (J~-~H)_{\rm P98} \; ,
\label{eq:y_nat_wirc}
\end{equation}

\noindent
where the $Y_{\rm RC}$ magnitudes are taken from Appendix~\ref{sec:y_phot}, and the 
$J_{\rm P98}$ and $H_{\rm P98}$ magnitudes are from P98. 

For each photometric night where secondary standard stars were observed, the NIR
photometric equations are then simplified to

\begin{equation}
Y_{\rm nat}~=~y~-k_yX~+~\zeta_y \; ,
\label{eq:y_nat}
\end{equation}
\begin{equation}
J_{\rm nat}~=~j~-k_jX~+~\zeta_j \; ,~{\rm and}
\label{eq:j_nat}
\end{equation}
\begin{equation}
H_{\rm nat}~=~h~-k_hX~+~\zeta_h \; ,
\label{eq:h_nat}
\end{equation}

\noindent
where $y$, $j$, and $h$ correspond to the instrumental magnitudes; and
$k_y$, $k_j$, and $k_h$ are extinction coefficients.  
Figure~\ref{fig:nir_extinction} displays nightly values of the atmospheric 
extinction coefficients in $YJH$ over the five \mbox{\csp} campaigns for both the
Swope~+~RetroCam and du~Pont~+~WIRC systems. These were  derived using the
MCMC fitting procedure described in \S\ref{sec:swope_opt} from 
observations of typically 2--10  secondary standards per night.  Histograms of the 
collected extinction-coefficient measurements are shown at the right-hand side of each 
panel.  No significant difference is observed between the two telescope~+~camera
systems, so we can combine the observations.  The resulting
mean extinction coefficients are given in Table~\ref{tab:ec_ct}.  
As was found to be the case for the optical bandpasses, the stability of the extinction 
coefficients during the five \mbox{\csp} observing  campaigns is such that these average
values can be adopted, leaving only the nightly zero points in
Eqs.~\ref{eq:y_nat}--\ref{eq:h_nat} to be determined.

Thirteen SN fields  were not observed for the requisite minimum of three photometric nights.
In order to improve the photometric calibration of the tertiary standards for these fields, 
we devised a ``hybrid'' 
calibration  whereby calibrated tertiary standards from one field are used to calibrate
the tertiary standards in another field that is observed on the same night
under photometric conditions, but when secondary
standards were not observed.  In this case, we use a modified version of Eq.~\ref{eq:ter_calib}:

\begin{equation}
m_{i,\lambda} = \frac{\sum_{j,k} \left(m_{{\rm nat},k,\lambda} + \Delta m_{i,j,k,\lambda}
      - k_\lambda \left(X_{i,j,\lambda}-X_{j,k,\lambda}\right)\right)w_{j,k,\lambda}}{\sum_{j,k} w_{j,k,\lambda}} \; ,
\label{eq:ter_calib_hybrid}
\end{equation}
where $k$ now refers to calibrated tertiary standards of different SN fields observed 
in filter $\lambda$ on the same photometric night ($j$).

In brief, this procedure worked as follows.

\begin{itemize}

\item A catalog of tertiary standard stars was produced from 126 SN fields 
(90 SNe~Ia and 36 SNe of other types) calibrated on a minimum of 
four photometric nights in each of the three NIR filters.

\item This catalog of tertiary standards was then used to measure an alternative set of
zero points for each night of NIR imaging during the five \mbox{\csp} campaigns.

\item These new zero points were then filtered to include only those nights where
(1) a minimum of three SN fields with calibrated tertiary standards was observed, 
(2) a minimum continuous span of three hours of imaging was obtained, and
(3) a maximum dispersion of 0.08~mag in the nightly zero point as calculated from
the observations of the tertiary standards was observed that night.  This latter criterion
is similar to that used in filtering the photometric nights chosen for calibrating the
tertiary standard stars using the P98 secondary  standards, but it should be noted
that the typical dispersion in zero points for photometric nights was significantly
less (0.02--0.03~mag).

\end{itemize}

In Figures~\ref{fig:wirc_zero_pts} and \ref{fig:rc_zero_pts}, 
the zero points calculated using only the P98 secondary standard stars are plotted 
as a function of time for the du~Pont~+~WIRC and the Swope~+~RetroCam, respectively.  
Shown for comparison are
the zero points obtained using the hybrid method described
above.  Note that the agreement is generally excellent, although the 
uncertainties in the zero points derived in the hybrid method are 
generally larger since the local sequence stars are typically 3--4~mag 
fainter than the P98 secondary standards. 
The hybrid method provides potential photometric zero points for an additional 52 nights in $Y$, 
44 nights in $J$, and 40 nights in $H$ for the du~Pont~+~WIRC observations, and
an additional 154 nights in $Y$, 139 nights in $J$, and 123 nights in $H$ for the 
Swope~+~RetroCam data.   Nevertheless, we used only those nights
that allowed us to improve the calibration of the thirteen SN fields.

\subsubsection{Filter Contamination}
\label{sec:contamination}

Close inspection of Figure~\ref{fig:rc_zero_pts} reveals a faster-than-expected change 
in the zero-point evolution of the RetroCam $Y$ and $H$ filters during campaign~3, 
producing large breaks between the end of campaign~3 and the beginning of 
campaign~4.  These discontinuities do not correspond to when the primary mirror was 
washed (indicated by the vertical gray lines in Figure~\ref{fig:rc_zero_pts}).  A 
similar problem is observed with the $H$ band during campaign~4, where the zero point 
decreases by nearly 1~mag between the washing of the primary mirror and the 
end of the campaign, as opposed to the much smaller changes observed for the $Y$ and 
$J$ filters over the same period.  This behavior suggests slowly increasing 
contamination of the filters.  To test this hypothesis, we plot with dashed blue lines 
in Figure~\ref{fig:rc_zero_pts} the dates that the RetroCam dewar was warmed up, 
pumped, and then cooled down again.  The recovery of the $Y$ and $H$ zero-point values 
in campaign~3, and the $H$ zero point in campaign~4 is seen to coincide with when the 
dewar was pumped, consistent with the contamination hypothesis.  Comparison of flat 
fields taken during campaigns~3 and~4 provide further evidence for slowly changing 
contamination seen as a radial pattern of increasing counts from the center to the 
edges of the filter that disappears when the dewar is pumped.

To examine the effect of this changing contamination on the photometry, we used 
observations of stars in the fields of several SNe at low Galactic latitudes.  The 
observations of the Type IIn SN 2006jd \citep{Str_etal12} from both 
campaigns 3 and 4 were used, supplemented by observations 
of the Type~Ia SNe 2007hj and 2007on, and the Type~II SNe~2008M and 2008ag carried out 
during campaign~4.  Figure~\ref{fig:phot_diff_campaign4} shows an example of the 
magnitude differences in the $Y$, $J$, and $H$-band photometry of stars in the field 
of SN~2008ag between images obtained on 20~October~2007 and 5~April~2008.  For the 
$Y$ and $J$ bands in campaign 4, the magnitude differences are consistent with zero over the entire 
detector.  In contrast, the $H$ filter shows clear evidence of a radial gradient 
amounting to $\sim0.014$~mag per 100 pixels as measured from the center of the 
detector.  However, as the lower-right panel of Figure~\ref{fig:phot_diff_campaign4} 
shows, there is no evidence that the filter bandpass itself was changed significantly 
by the contamination since both blue and red stars show the same radial-gradients.
$Y$ and $H$-band observations in campaign 3 show a similar radial gradient effect.

Unfortunately, these contamination problems were not recognized during the course of 
the \csp\ campaigns.  For most SNe, the error in the photometry due to the 
contamination is relatively small (0.02--0.03~mag), but systematic; we must therefore 
correct for the effect.  Fortunately, the growth of the contamination was nearly 
linear in time.  This is illustrated in Figure~\ref{fig:mag_diff_slopes} where 
magnitude differences in $H$-band photometry of stars in the field of SN 2008ag are 
plotted at five epochs between 20~February 2008 and 20~June 2008 with respect to 
observations made on 16~February 2008. Fitting these trends with straight lines 
provides a recipe for correcting the SN and tertiary standard photometry, with the 
correction being a function of both time and the ($x,y$) coordinates of the SN or 
standard on the RetroCam detector. Specifically, we fit the slope measurements as a 
function of time by the relation

%
%
%
%

\begin{equation}
p(\rm JD) =  p({\rm JD}_{\rm end}) \times ({\rm JD} - {\rm JD}_{\rm start}) / 365.25,
\label{eq:slope_vs_time}
\end{equation}

\noindent  where JD is the Julian date of the observation and
$p({\rm JD}_{\rm end})$ is the slope  (measured in units of mag per 100~pixels)  
on the Julian date at the end of the period of contamination, ${\rm JD}_{\rm end}$.
The formula for calculating the correction to the photometry of the tertiary standards and SN 
in an image take on any Julian date during the period of contamination is then

\begin{equation}
\Delta m_{\rm corr}({\rm JD}) =  p({\rm JD}) \times (d_{\rm radial} / 100 - 256/100),
\label{eq:corr_vs_time}
\end{equation}

\noindent where $d_{\rm radial}$ is the radial distance in pixels from of the star or SN from the
center of the image, and the constant $256/100$ makes the average of 
the magnitude corrections for each image approximately zero. 

These corrections were applied to photometry obtained with the Swope~+~RetroCam
as follows:

\begin{itemize}

\item Campaign~3, $Y$ band: ${\rm JD}_{\rm start} = 2,453,980.0$, ${\rm JD}_{\rm end} = 2,454,260.0$, $p({\rm JD}_{\rm end}) = 0.027$

\item Campaign~3, $H$ band: ${\rm JD}_{\rm start} = 2,453,980.0$, ${\rm JD}_{\rm end} = 2,454,260.0$, $p({\rm JD}_{\rm end}) = 0.021$

\item Campaign~4, $H$ band: ${\rm JD}_{\rm start} = 2,454,509.3$, ${\rm JD}_{\rm end} = 2,454,646.0$, $p({\rm JD}_{\rm end}) = 0.110$

\end{itemize}

\subsubsection{Final Photometry}
\label{sec:final_nir_phot}

Final $YJH$ magnitudes of the tertiary standards for all \numsne~SNe are listed in 
Table~\ref{tab:nir_local_stds}.  Note that the $J$ and $H$ magnitudes are given in 
the {\em standard} P98 system,
whereas the $Y$ magnitudes are in 
the {\em natural} system of the Swope~+~RetroCam (which we have adopted
as the ``standard'' system).  The accompanying uncertainties 
are the dispersions of the multiple measurements of each sequence star.


\section{Final Light Curves}
\label{sec:lightcurves}

Final optical and NIR photometry of the \numsne\ SNe in the \mbox{\csp} sample is given in 
Tables~\ref{tab:opt_phot_site3}--\ref{tab:Ia_nir_phot_wirc}. 
Tables~\ref{tab:opt_phot_site3} and \ref{tab:opt_phot_tek5} give the $ugriBV$ 
photometry in the natural systems of the Swope~+~SITe3 and du~Pont~+~Tek5
(respectively), and Table~\ref{tab:lc3009_v_phot} gives the small amount of 
$V$-band photometry obtained in the natural system of the LC-3009 filter at the 
Swope.  NIR photometry of \numsneir ~SNe in the natural systems of the 
Swope~+~RetroCam and du~Pont~+~WIRC is found in Tables~\ref{tab:Ia_nir_phot_rc1},
\ref{tab:Ia_nir_phot_rc2}, 
and \ref{tab:Ia_nir_phot_wirc}. On those occasions when more than one NIR 
measurement is given for an object on a given night, it is because the WIRC
used on the du Pont telescope images the SN location on more than 
one chip.  Rather than averaging the measurements, we give the individual values.

\subsection{Type~Ia SNe}
\label{sec:Ia}


Plots of the individual light curves of the Type~Ia SNe\footnote[21]{By ``Type~Ia 
SNe,'' we  mean those classified as ``normal,'' ``SN~1991T-like,'' ``SN~1986G-like,'' and 
``SN~1991bg-like'' in Table~\ref{tab:spec_details}.} in the \mbox{\csp} sample are displayed 
in Figure~\ref{fig:sample_lcs} along with fits (solid red lines) using SNooPy 
\citep{burns11}.  Photometric parameters derived from the SNooPy fits are 
provided in Table~\ref{tab:lcfit_details}.  In some cases we can directly measure 
the epoch of $B$-band maximum, $T_{\rm max}(B)$, and the $B$-band decline rate, \dmb.  
The latter is defined as the number of magnitudes the object faded in the first 
15 days since the time of $B$-band maximum, and has long been known to 
correlate with the absolute magnitudes of SNe~Ia at maximum light 
\citep{Phi93}.  Often, however, photometric coverage is not optimal for direct 
measurements, and it is more robust to estimate $T_{\rm max}(B)$ and the 
decline rate using a family of light-curve templates. Hence, for each object in 
Table \ref{tab:lcfit_details}, we have used the ``max model'' method of SNooPy to 
calculate template-derived estimates of the epoch of $B$-band maximum and the 
decline-rate parameter, \dmb, that we denote as $T_{\rm max}$(template) and \dm(template), 
respectively. We also give the dimensionless ``stretch $BV$ parameter,'' 
$s_{BV}$, which is equal to $\Delta T_{BV}$/(30~d), where $\Delta T_{BV}$ is 
the number of days since $T(B_{\rm max})$ that a supernova's $(B-V)$ color reaches 
its maximum value \citep{Bur_etal14}.  Burns et al. discuss the advantages of this new 
parameter over the usual decline-rate parameter, especially for rapidly 
declining light curves. In particular, plots of reddening-corrected colors vs. 
$s_{BV}$ show low root-mean-square scatter, allowing a more definitive characterization of the 
photometric properties of SNe~Ia.

One ``normal'' SN~Ia listed in Table ~\ref{tab:lcfit_details} that we cannot
fit using just \csp\ photometry is SN~2006dd, because the \csp\ data cover only the
post-maximum linear decline.  We refer the reader to
\citet{Str_etal10}, which contains pre-maximum, maximum-light, and post-maximum
photometry of this SN obtained with the CTIO 1.3~m telescope
using its dual optical/NIR imager ANDICAM.  

The middle panel of Figure \ref{fig:histograms} shows a histogram of the values of 
the $B$-band decline rate, \dmb, as obtained from the template fits.  The bottom panel 
of this figure shows a histogram of ``stretch $BV$'' values. 

\subsection{Type~Iax SNe}
\label{sec:Iax}


Type~Iax SNe are spectroscopically similar to Type~Ia SNe that are more 
luminous than average because they show high-ionization lines such as Fe~III, 
but have lower maximum-light velocities and fainter absolute magnitudes for 
their light-curve decline rates \citep{Foley_etal13}.  The prototype of this 
subclass is SN~2002cx \citep{Filippenko03,Li_etal03}. Plots of the 
individual light curves 
of the five SNe~Iax SNe in the \mbox{\csp} sample (SNe~2005hk, 2008ae, 
2008ha, 2009J, and 2010ae) are displayed in Figure~\ref{fig:Iax_lcs}.
Preliminary photometry of SN 2005hk was published by \citet{Phi_etal07}.
\citet{Str_etal14, Str_etal15} have published updated 
photometry of SN 2005hk and new photometry of SNe 2008ha and 2010ae.

\subsection{Other Subtypes}
\label{sec:other}

Two objects observed by the \mbox{\csp}, SNe 2007if and 2009dc, are candidates for the
super-Chandrasekhar (``SC'') subtype \citep{howell06}.  SNe~2006bt and 2006ot are 
members of the SN~2006bt-like subclass \citep[][Paper~2]{Foley_etal10b}. Two other 
events (SNe~2005gj and 2008J) belong to the rare Type~Ia-CSM subtype
\citep{silverman13}.
The light curves of these six SNe are shown in Figure~\ref{fig:other_lcs}.

\section{Conclusions}
\label{sec:conclusion}

In this paper we have presented the third and final data release of optical and 
NIR photometry of the 134 nearby ($0.004 \la z \la 0.08$) white dwarf SNe 
observed during the \mbox{\csp}.  This sample consists of 123 Type~Ia SNe, 5 
Type~Iax SNe, 2 super-Chandrasekhar candidates, 2 Type~Ia-CSM SNe, and 2 SN~2006bt-like events.  
NIR photometry was obtained for 90\% of these SNe.  Optical spectroscopy has 
already been published for approximately two-thirds of the SNe~Ia in the 
sample, and the remaining spectra are currently being prepared for publication.  
In addition to providing a new set of light curves of low-redshift SNe~Ia 
in a stable, well-characterized photometric system for cosmological studies, 
the combined \mbox{\csp} dataset is allowing us to improve dust extinction 
corrections for SNe~Ia \citep{Bur_etal14}.  The excellent precision and 
high cadence of the \csp\ observations also facilitate detailed analysis of the 
light curves, leading to a deeper understanding of the physics of thermonuclear 
events \citep[e.g.,][]{hoeflich10, hoeflich17}.

Over the course of the \mbox{\csp}, more than 100 core-collapse SNe were 
observed. Photometry of 7 SNe~IIn has already been presented by 
\citet{Str_etal12} and \citet{Tad_etal13}. In 
an accompanying paper to this one \citep{Str_etal17}, the final data release of 
optical and NIR photometry of 34 stripped-envelope core-collapse SNe is 
presented.  Publication of optical and NIR photometry of 83 SNe~II 
observed during the course of the \mbox{\csp} is also in preparation. 
Preliminary $V$-band light curves for this sample have already been published 
by \citet{anderson14}, and \csp\ observations of two SN~1987A-like events were 
presented by \citet{taddia12b}. Extensive optical spectroscopy of many of these 
core-collapse SNe was also obtained, and is currently being prepared for 
publication.

In 2011, we began a second phase of the CSP to obtain optical and NIR observations of
SNe~Ia in the smooth Hubble flow.  Over a four-year period, light curves were obtained for
nearly 200 SNe~Ia, $\sim100$ of which were at $0.03 \la z \la 0.10$.
NIR spectra were also obtained of more than 150 SNe~Ia. This dataset, which we plan
to publish over the next three years, in combination with the \mbox{\csp} light curves published
in the present paper, should provide a definitive test of the ultimate precision
of SNe~Ia as cosmological standard candles.

\section{Electronic Access}
\label{sec:access}

To obtain an electronic copy of the photometry of any of the SNe included in 
this paper, the reader is directed to the CSP website at 
\url{http://csp.obs.carnegiescience.edu/}.  (At the time of the posting of this
preprint the tar ball of the photometry was not yet posted at this website.)
Also available at this website are 
the optical spectra of \mbox{\csp} SNe~Ia published by \citet{Fol_etal13}.

\acknowledgments 

This paper is dedicated to the memory of our dear colleague, Wojtek Krzeminski
(1933--2017), who played an important role in the early history of Las Campanas
Observatory and who, during his retirement, obtained many of the observations
presented in this paper.

The CSP particularly thanks the mountain staff of the Las Campanas Observatory 
for their assistance throughout the duration of our observational program, and 
to Jim Hughes and Skip Schaller for computer support. Special thanks are due to 
Allyn Smith and Douglas Tucker for allowing us to publish their $u'g'r'i'$ 
magnitudes of P177D and P330E (and to Dan Scolnic for leading us to Allyn and 
Douglas). This project was supported by NSF under grants AST--0306969, 
AST--0908886, AST--0607438, and AST--1008343.  M. Stritzinger, C. Contreras, and E. Hsiao 
acknowledge generous support from the Danish Agency for Science and Technology 
and Innovation through a Sapere Aude Level 2 grant. 
M. Stritzinger is supported by a research grant (13261) from VILLUM FONDEN. 
M. Hamuy acknowledges support by 
CONICYT through grants FONDECYT Regular 1060808, Centro de Astrof\'{i}sica FONDAP 
15010003, Centro BASAL CATA (PFB--06), and by the Millennium Center for 
Supernova Science (P06--045-F). A. V. Filippenko is 
grateful for the financial support of the NSF, the Richard and Rhoda Goldman
Fund, the TABASGO Foundation, 
the Christopher R. Redlich Fund, and the Miller Institute for Basic Research 
in Science (U.C. Berkeley). This 
research has made use of the NASA/IPAC Extragalactic Database (NED) which is 
operated by the Jet Propulsion Laboratory, California Institute of Technology, 
under contract with the National Aeronautics and Space Administration.
N. Suntzeff is grateful for the support provided by the Mitchell/Heep/Munnerlyn
Chair in Observational Astronomy at Texas A\&M University.  The CSP thanks the
Mitchell Foundation and Sheridan Lorenz for sponsoring our group meetings at
Cook's Branch Nature Conservancy.

We thank the Lick Observatory staff for their assistance with the
operation of KAIT. LOSS, which discovered many of the SNe studied
here, has been supported by many grants from the NSF (most recently
AST-0908886 and AST-1211916), the TABASGO Foundation, US Department of
Energy SciDAC grant DE-FC02-06ER41453, and US Department of Energy
grant DE-FG02-08ER41563. KAIT and its ongoing operation were made
possible by donations from Sun Microsystems, Inc., the Hewlett-Packard
Company, AutoScope Corporation, Lick Observatory, the NSF, the
University of California, the Sylvia \& Jim Katzman Foundation, and
the TABASGO Foundation.  We give particular thanks to Russ Genet, who
made KAIT possible with his initial special gift, and the TABASGO
Foundation, without which this work would not have been completed.    


\appendix

\section{\csp\ Near-Infrared Bandpasses}
\label{sec:nir_bandpasses}


Paper 2 provided a detailed description of the calibration of the 
\mbox{\csp} optical bandpasses.  The setup consisted of a  
monochromator that allowed the throughput of the entire telescope 
plus detector system to be measured {\em in situ} without having to rely on 
multiple calibrations (filters, windows, aluminum reflections, detector quantum 
efficiency) multiplied together.  Here we provide a summary of the calibration of 
the NIR bandpasses using the same monochromator.

The calibration of the two NIR cameras used for \csp\ was carried out in 
late-July 2010 (Swope~+~RetroCam) and early-August 2010 (du~Pont~+~WIRC).  
The measurements were made on 
at least two different nights for each filter to ensure that the method was 
repeatable. The monochromator system uses a fiber that splits, sending 90\% 
of the power to the dome-flat screen and 10\% to a ``witness screen.'' The 
witness screen was placed in a dark box to prevent ambient light from reaching 
it.

Two germanium detectors were used, each 10 mm in diameter, which were calibrated 
in the lab at Texas A\&M University using a NIST traceable Gentec calibrated 
photodiode.  The Ge detectors are sensitive from 900 to 1600~nm, and were 
used only for the calibration of the $Y$-band filter (900 to 1100~nm). To calibrate 
the three longer-wavelength \csp\ bandpasses ($J$, $H$, and $K_s$), a 2~mm diameter 
InGaAs detector was purchased and shipped to the National Research Council of 
Canada for calibration prior to its use in Chile.

Due to the lower light levels produced by this system in the NIR and because of 
the smaller area of the InGaAs detector, all of the actual calibrations were done 
using the 35~cm by 35~cm witness screen made of the same material 
as the dome-flat screen. When taking $Y$-band data, one Ge detector detected photons 
from the dome-flat screen while the other Ge detector simultaneously detected 
photons from the witness screen in the dark box.  The two detectors in the box 
were placed about 10~cm from the witness screen.

The $Y$-band signal at the witness screen was about 50 times stronger than the 
signal from the dome-flat screen.  The voltage of the Ge photodiode (as a function 
of wavelength) measured from the dome-flat screen and scaled by a factor of 
$\sim50$ matches, within the errors of measurement, the voltage of the other Ge 
photodiode used to measure the witness screen.  Thus, we are confident that  
the witness screen can give reliable results for the longer-wavelength bandpasses.

A challenge inherent to IR measurements is the presence of background
thermal drifts occurring on timescales of seconds. Also, during the day and on
nights with moonlight, even with the dome closed there is some ambient light
in the dome, which is relevant to the $Y$-band calibration described here.

To minimize the problem of thermal drifts and residual light in the dome, we took 
data using the following method. For each wavelength we obtained two 30~s ``Dark'' 
images and two 30~s ``Light'' images.  A ``Dark'' image is taken with the light 
off and a ``Light'' image is taken with the light on.  We subtract one ``Dark'' 
image from one ``Light'' and average the two net values to get a single measurement.  
Additionally, RetroCam on the Swope telescope exhibited severe image persistence, 
so we took a short (6~s) ``Dark'' exposure after each ``Light'' one to clear the 
camera detector.

The photodiode detectors are not temperature stabilized, so the amplifier
offsets and the background drift significantly during the 30~s required
for an exposure.  They are relatively stable on a 1~s timescale.
The error caused by the drift is much more important than the noise in the
detector.  It is then better to take shorter integration times ($\sim1$~s)
to avoid drift problems even if we sacrifice a bit of averaging of the noise.

Since the output of the lamp is very stable ($<1$\%) over a period
of hours, we assumed that the output was constant over the 30~s
exposure time and only measured the amplitude just before opening the shutter
to take a camera image.  Before each exposure we cycled the light on and
off 5 times for a 10~s period to obtain 5 values of the amplitude, which we
averaged to calculate the power at the detector.

The normalized transmission curves for each filter were obtained using a 
subsection ($x$ and $y$ from 60 to 964 pixels) of chip 2 in WIRC 
on the du~Pont telescope.  For RetroCam on the Swope telescope we used the same 
subsection of its chip. In general, we are confident that the measured transmission 
is accurate to 2--3\% of the peak transmission.

We have normalized each filter for each camera separately.  We estimate that the 
transfer from the dome-flat screen to the witness screen and the relative 
photodiode calibration uncertainties are 1\% and 2\%, respectively, for the 
Swope and du~Pont cameras.  For the Swope~+~RetroCam, the relative photodiode noise 
level is 0.5\% for the $YJH$ filters, as is the relative noise level of the 
 camera.  The total uncertainty in each filter is obtained from the square root 
of the sum of the individual components added in quadrature, or 2.3\%. For 
the du~Pont~+~WIRC, the relative photodiode noise level and the relative noise 
level of the camera is 1--2\% for $YJH$.  For these filters the total 
uncertainty is 2.6\% in $Y$ and $J$, and 3.2\% in $H$.

We have also investigated the focal-plane uniformity of the filters in the Swope 
and du~Pont cameras.  Each filter was scanned twice on at least two separate 
nights. The scans were performed with a wavelength step of 5 or 10 nm.  The 
analysis involved dividing each chip into four quadrants and comparing the 
relative response curves for each quadrant.  At worst there is a 1.5 nm shift in 
the filter cutons and cutoffs as a function of quadrant and chip.  This effect 
would be negligible unless a very narrow emission line happens to fall at that 
exact wavelength.

\section{Optical Color Terms}
\label{sec:opt_cterms}

To test the accuracy of the response functions of the \mbox{\csp} optical bandpasses 
shown in Figure~\ref{fig:filters}, we computed color terms using a subset of the 
stars from the spectrophotometric atlas of Landolt standards published by 
\citet{stritzinger05}.  Eighteen of the stars in this atlas are in the list of 
\citet{Smi_etal02} $u'g'r'i'$ standards.  Of these, one star (SA98-653) does not 
have sufficient wavelength coverage to include the $u$ and $g$ bands, and we 
suspect that another (SA104-598) is variable.  The results of synthetic 
photometry of the sixteen remaining stars are shown in 
Figure~\ref{fig:opt_syn_cts}.  In each plot, the abscissa is the color from the 
published standard-star magnitudes, and the ordinate is the difference of the 
magnitude calculated via synthetic photometry using the bandpasses in 
Figure~\ref{fig:filters} and the published standard-star magnitude. The red 
triangles correspond to the 16 stars in the \citet{stritzinger05} atlas, and the 
dashed lines are the best fits to these points.  The slopes of these fits 
correspond to the color terms, and the values are indicated in red.  The slopes 
of the observed color terms are indicated by the solid black lines, with the 
values shown in black. The histograms in each plot correspond to the colors of 
the \citet{Smi_etal02} and \citet{Lan92} standard stars observed routinely by 
\mbox{\csp}.

In general, the agreement between the measured and computed color terms is good.  
For the $u$, $r$, $i$, $B$, and $V$ bands, the color terms agree to better than 
$1\sigma$.  For the $g$ band, the agreement is within $2\sigma$.  
Considering the relatively small number of stars used for this test, and the fact 
that these cover a somewhat smaller range of color than the actual standards used 
routinely by \mbox{\csp}, we consider the results of this test to be consistent with 
the observations.

\section{NIR Color Terms}
\label{sec:nir_cterms}

In principle, we can check the color terms derived in \S\ref{sec:swope_nir} for 
the $J_{\rm RC2}$ and $H$ filters of the Swope~+~RetroCam through observations of the red 
stars listed in Table~3 of P98. Unfortunately, although some of the red stars 
were observed on a few nights during the \mbox{\csp} campaigns, these data were not 
reduced at the time they were taken, and a subsequent disk crash made it 
impossible to recover them.  However, since mid-2011, RetroCam has been in use on 
the du~Pont telescope, and observations of several of the P98 red stars were made 
in late-December 2015.


Figure~\ref{fig:nir_ct_dupont_rc} shows synthetic synthetic photometry in $J_{\rm RC2}$ and $H$ 
of the \citet{castelli03} atmosphere models for a range of stellar parameters and 
reddenings ($E(B-V) = 0.0$~to~2.5 mag).  Here the differences between the P98 
magnitudes and the RetroCam magnitudes are plotted as a function of the 
$(J-H)_{\rm P98}$ color. The filter response functions used for these calculations 
were measured in 2013 November for RetroCam on the du~Pont telescope using the 
same monochromator and setup described in Appendix~\ref{sec:nir_bandpasses}. 
The predicted color terms of RetroCam 
on the du~Pont telescope and on the Swope telescope are very similar, as might be 
expected.  The black symbols in Figure~\ref{fig:nir_ct_dupont_rc} are our observations 
of several P98 standards, along with three stars with $(J-H)_{\rm P98} > 1.5$~mag taken 
from the P98 red-star list that correspond to reddened M giants (typically 
M3~III) in Bok globules in the Coal Sack.  The three red stars plotted with red 
symbols are young stellar objects (YSOs) from the P98 list. The latter stars are 
not representative of the \citet{castelli03} models used for calculating the 
synthetic photometry, and are also often variable.\footnote[22]{Our observations of 
the YSO IRAS-537-S, which is included in the P98 list of red stars, showed a 
brightness change of $\sim0.2$~mag in a space of two nights.}  Hence, we do not 
include these stars in the fits shown in Figure~\ref{fig:nir_ct_dupont_rc}.

Comparing the synthetic photometry of the models with the observations 
(excluding the YSOs), we find excellent agreement between the predicted
and observed color terms in $J$ and $H$.  This gives us confidence in the
reliability of the color terms calculated for the Swope~+~RetroCam in 
Eqs.~\ref{eq:j_rc2} and ~\ref{eq:h_rc} of \S\ref{sec:swope_nir}.

\section{$Y$-Band Photometric Standards}
\label{sec:y_phot}

\citet[][Appendix C]{Ham_etal06} describe how we derived $Y$-band magnitudes 
for the NIR standards of P98. In brief, we used \citet{Kur91}
model atmospheres, the P98 filter functions, $J_{\rm P98}$
and $K_{\rm P98}$, and our best estimate of 
the transmission of our $Y$ filter to compute 
synthetic $(Y-K_{s,{\rm P98}})$ colors as a function of
synthetic $(J_{\rm P98}-K_{s,P98})$ colors. Since then, we have scanned our
NIR filters (see Appendix~\ref{sec:nir_bandpasses}) and improved 
stellar atmosphere models \citep{castelli03} have become available.
Hence, in this Appendix we rederive $Y$-band magnitudes 
for the NIR standards of P98.  Note that the photometric zero points for the 
NIR filters are computed assuming
Vega colors are all zero. 

Figure \ref{fig:Y_calibration} shows the results for the Swope~+~RetroCam $Y$ band. 
The grey circles correspond to synthetic photometry
of model dwarf-star atmospheres with nearly solar metallicity 
($\log(g) > 4.0$, $-0.1 < \log(Z/{\rm Z}_{\odot}) < 0.1$, where $g$ is the local acceleration
of gravity in cm~s$^{-2}$ and $Z$ is the abundance of elements heavier
than helium). The gray region indicates the color range spanned by the P98
standards used by the \csp. 
The green solid line is a fifth-order polynomial fit to the data,


\begin{multline}
   Y - K_{s,P98} = 2.393\left(J_{\rm P98}-K_{s,{\rm P98}}\right) -
                   4.473\left(J_{\rm P98}-K_{s,{\rm P98}}\right)^2 +
                   10.715\left(J_{\rm P98}-K_{s,{\rm P98}}\right)^3 - \\
                   13.011\left(J_{\rm P98}-K_{s,{\rm P98}}\right)^4 +
                   6.232\left(J_{\rm P98}-K_{s,{\rm P98}}\right)^5 \; .
\end{multline}

\noindent
The blue dashed line is the fit from \citet{Ham_etal06}.
As with \citet{Ham_etal06}, we do not allow a constant offset, forcing the
polynomial through $Y-K_{s,{\rm P98}} = J_{\rm P98}-K_{s,{\rm P98}} = 0$~mag. 
However, we
note that the synthetic colors of the \citep{castelli03} 
model atmospheres do not have zero NIR colors for an A0~V star. This can be
seen in the lower panel of Figure \ref{fig:Y_calibration}, where we have
subtracted a linear fit (solid red line) to the points in order to better 
visualize the difference between the old and new fits. We could adjust
the NIR zero points to make all synthetic NIR colors zero at the expense
of Vega acquiring nonzero colors, but choose not to do this in order to
be more consistent with our previous natural system.

In Table \ref{tab:Y_stds}
we give the final $Y$-band photometric values of most of the P98 standard stars.
We include values for stars considerably further north than we can
observe at Las Campanas Observatory, in case northern hemisphere
observers require them.

\section{Photometric Zero Points}
\label{sec:zeropts}

The zero points of a photometric system are necessary for computing
transformations to other photometric systems (commonly referred to as
S-corrections) as well as computing cross-band K-corrections within the same
photometric system. The definition of the magnitude of a
source with SED $f_\lambda$ measured by an
instrument with response $F_{\lambda}$ is given by

\begin{equation}
m = m^{*}-2.5\log_{10}\left(\frac{\frac{1}{ch}\int F_{\lambda} f_\lambda \lambda d\lambda}
                             {\frac{1}{ch}\int F_{\lambda} f^*_\lambda \lambda d\lambda}
                             \right) \; ,
\label{eq:mag_defined}
\end{equation}
where $f^*_\lambda$ is some standard SED (e.g., Vega) and $m^{*}$ is its
magnitude through the instrument defined by $F_{\lambda}$. Here 
and in Eq.~\ref{eq:zero_point_defined} below $c$ and $h$ are
the speed of light and Planck's constant, respectively. The numerator
within the log function is the observed photon flux detected by the CCD, while
the denominator is the photon flux of the standard SED through the same
instrument and is generally {\em not} observed. Defining the zero point as
\begin{equation}
\zeta = m^{*} + 2.5\log_{10}\left(\frac{1}{ch}\int F_{\lambda} f^*_\lambda
\lambda d\lambda\right) \; ,
\label{eq:zero_point_defined}
\end{equation}
we clearly need three pieces of information to compute the zero point: the total
instrument response, the standard SED, and the magnitude $m^{*}$. As mentioned
previously, the \csp\ has directly measured all components of $F_{\lambda}$ except
the atmosphere. This leaves the standard SED and value of $m^{*}$.

The \csp\ used three sets of secondary standards to calibrate our photometry: 
\citet{Lan92} for $BV$, \citet{Smi_etal02} for $ugri$, and P98 for $YJH$. 
These standards, more than anything else, define our photometric zero point. 
However, we also require a high-fidelity SED that covers the wavelength range 
of our filters, and such SEDs generally do not exist for these standards. 
\citet{stritzinger05} give SEDs at optical wavelengths for 18 \citet{Lan92} 
and \citet{Smi_etal02} standards. Ultimately, the \citet{Lan92} and P98 
standards are tied to $\alpha$ Lyr while the \cite{Smi_etal02} standards are 
tied to BD$+17\arcdeg 4708$, both of which have accurate SEDs 
\citep{bohlin04a,bohlin04b}, so we use these to compute our zero points.

Lastly, we need the value of $m^{*}$ for each instrument and SED combination. We
begin with the standard magnitudes of each star in the system for which it was
defined and use our color terms (see Table~\ref{tab:ec_ct}) to compute the magnitudes that would have been
observed through {\em our} instrument. The adopted standard magnitudes and
transformed natural-system magnitudes are listed in Table~\ref{tab:pm_mags}. For
$B$ and $V$, we adopt the standard magnitudes for $\alpha$~Lyr from
\citet{Fuk_etal96}. For $YJH$, we adopt zero magnitudes for $\alpha$~Lyr to be
consistent with \citet{elias82}. For $ugri$, we adopt the standard
magnitudes of BD$+17\arcdeg 4708$ from \citet{Smi_etal02}.

The reader should note that \citet{boh_lan15} present evidence that
BD$+17\arcdeg 4708$ is slightly variable.  From 1986 to 1991 this star
brightened by $\sim0.04$ mag in multiple optical bands.  Following the suggestion
of \citet{boh_lan15}, we have also calculated zero points in $ugriBV$ using the standard and
natural-system magnitudes of the CALSPEC standards P177D and P330E.
For $B$ and $V$, we use the values measured by \citet{boh_lan15}, while for
$ugri$, we adopt unpublished measurements made by Allyn Smith in the
USNO $u'g'r'i'z'$ standard system of \citet{Smi_etal02}.

In Table~\ref{tab:pm_zpts} we give corresponding zero points calculated with the
SEDs of the four primary calibration standards.  The agreement in zero
points between
P177D, P330E, and $\alpha$~Lyr for $B$ and $V$ is excellent, as is also the case
between P177D, P330E, and BD$+17\arcdeg 4708$ for $ugri$.

It is worth pointing out that we have two different networks of standard stars
in the optical. As such, there is no guarantee that the $BV$ zero points will be
consistent with the $ugri$ zero points in an absolute sense. To investigate
this, we compute synthetic photometry of the \citet{pickles98} stellar library
and compare the $(B-g)$ and $(V-g)$ colors with the observed colors of stars common
to \citet{Lan92} and \citet{Smi_etal02}. The results, shown in
Figure~\ref{fig:pickles}, indicate a systematic offset between the synthetic and
observed $(B-g)$ and $(V-g)$ colors of $\sim0.03$ mag. This could easily be
fixed by adding the appropriate offsets to the zero points of the $B$ and $V$
filters, thereby bringing both systems into alignment. However, this introduces a
reliance on the \citet{pickles98} spectra and presupposes that it is the $B$ and
$V$ magnitudes that should be adjusted, when the problem could just as easily be with
the $ugri$ zero points. Hence, we prefer {\em not} to apply a zero-point correction
to either system.  Nevertheless, users of the \mbox{\csp} photometry should be aware
of this inconsistency.

\section{AB Magnitude Offsets}
\label{sec:abmagoffsets}

According to Eq.~7 of \citet{Fuk_etal96}, a broad-band AB magnitude is defined as

\begin{equation}
m_{\mathrm{AB}} = -2.5\log_{10}~\left[ \frac{\int d(\log_{10}~\nu)f_\nu F_\nu}{\int d(\log_{10}~\nu)F_\nu} \right] - 48.6 \; ,
\label{eqn:abmag}
\end{equation}

\noindent where $f_\nu$ is the flux per unit frequency of the object expressed in units of 
ergs~s$^{-1}$~cm$^{-2}$~Hz$^{-1}$, and $F_\nu$ is the response function of the filter.  
Since AB magnitudes are directly related to physical units \citep{oke83}, they offer a straight-forward way
of transforming magnitudes to flux densities.

To convert from natural magnitudes to AB 
magnitudes, we need to solve for an offset for each filter such that

\begin{equation}
m_{\mathrm{AB}} = m_{\mathrm{natural}} + \mathrm{offset} \; .
\label{eqn:aboffset}
\end{equation}

\noindent Combining equations~\ref{eqn:abmag} and \ref{eqn:aboffset} with the definition of our CSP
natural magnitudes,

\begin{equation}
m_{\mathrm{natural}} = -2.5\log_{10} \left[ \frac{1}{ch}\int F_{\lambda} f_\lambda \lambda d\lambda \right] +\zeta \; ,
\label{eq:nat_mag_def}
\end{equation}

\noindent it can be shown that

\begin{equation}
\mathrm{offset} = 16.847 - \zeta + 2.5\log_{10} \left[ \int \frac{F_{\lambda} d\lambda}{\lambda} \right] \; ,
\label{eq:offset_def}
\end{equation}

\noindent where $\zeta$ is the zero point of the filter.
The zero points of the CSP natural magnitudes are derived in Appendix~\ref{sec:zeropts} 
and are listed in Table~\ref{tab:pm_zpts}.
  
Table~\ref{tab:AB_offsets} shows offsets calculated with equation~\ref{eq:offset_def} for all of the CSP-I filters .   
Once the offsets have been applied, the flux in each band is given by:

\begin{equation}
\left<f_\nu\right>_{\nu} = 10^{-0.4 (m_{\mathrm{AB}}+48.6)}\,~\rm{erg}\,~\rm{s}^{-1}\,\rm{cm}^{-2}\,\rm{Hz}^{-1} \; . 
\label{eqn:refname2}
\end{equation}

\noindent 
where $\left<f_\nu\right>_{\nu}$ is the weighted average of $f_\nu$ 
with weight function $F\left(\nu\right)\cdot \nu^{-1}$.  Note that 
equation~\ref{eqn:refname2} is not the proper inverse of 
equation~\ref{eqn:abmag}. One cannot derive a precise monochromatic flux from 
an AB magnitude, especially for objects such as supernovae that have SEDs 
very different from the stars used in the fundamental spectrophotometric 
system.  This is discussed in detail by \citet{Bro_etal16}.

\clearpage

\newpage
\begin{deluxetable} {lcrlllll}
\rotate
\tabletypesize{\scriptsize} 
\tablecolumns{8}
\tablewidth{0pt}
\tablecaption{General Properties of $\numsne$ Type Ia Supernovae\label{tab:snproperties}}
\tablehead{%
\colhead{SN} &
\colhead{SN}   &
\colhead{SN}   &
\colhead{Host} &
\colhead{Morphology\tablenotemark{a}} &
\colhead{$z_{\rm helio}$\tablenotemark{a}} &
\colhead{Discovery} &
\colhead{Discovery Indiv/} \\
\colhead{Name} &
\colhead{$\alpha (2000)$} &
\colhead{$\delta (2000)$} &
\colhead{Galaxy} &
\colhead{}   &
\colhead{}   &
\colhead{Reference} &
\colhead{Group\tablenotemark{b}} }
\startdata
  2004dt &     02:02:12.77 &   $-$00:05:51.5 &                        NGC 799 &           (R$^{\prime}$)SB(s)a & 0.0197 &            IAUC 8386 &                                     LOSS  \\
  2004ef &     22:42:10.02 &     +19:59:40.4 &                      UGC 12158 &                             Sb & 0.0310 &            IAUC 8399 &                         Boles/Armstrong   \\
  2004eo &     20:32:54.19 &     +09:55:42.7 &                       NGC 6928 &                        SB(s)ab & 0.0157 &            IAUC 8406 &                                  Itagaki  \\
  2004ey &     21:49:07.81 &     +00:26:39.2 &                      UGC 11816 &                        SB(rs)c & 0.0158 &            IAUC 8419 &                               Tenagra II  \\
  2004gc &     05:21:49.95 &     +06:40:33.7 &                  ARP 327 NED04 &                            --- & 0.0321 &            IAUC 8442 &                     del Olmo/Tenagra II   \\

  2004gs &     08:38:23.18 &     +17:37:39.8 &                 MCG +03-22-020 &                            S0? & 0.0267 &            IAUC 8453 &                                     LOSS  \\
  2004gu &     12:46:24.72 &     +11:56:56.1 &                       FGC 175A &                            --- & 0.0459 &            IAUC 8454 &                                ROTSE-III  \\
  2005A  &     02:30:43.25 &   $-$02:56:19.8 &                        NGC 958 &                        SB(rs)c & 0.0191 &            IAUC 8459 &                                     LOSS  \\
\enddata
\tablenotetext{a}{Host-galaxy morphology and heliocentric redshift are
from the NASA/IPAC Extragalactic Database (NED) or SDSS unless otherwise indicated.}
\tablenotetext{b}{References/URLs:
CHASE (http://www.das.uchile.cl/proyectoCHASE/);
CSS (http://www.lpl.arizona.edu/css/);
LOSS \citep{filippenko01, filippenko05, leaman11};
Perth \citep{williams97};
ROTSE-III \citep{quimby06}; Puckett (http://www.cometwatch.com);
Tenagra-II (http://www.tenagraobservatories.com/Discoveries.htm);
SDSS-II \citep{frieman08}.}
\tablenotetext{c}{Host-galaxy redshift of SN~2005gj from \cite{Pri_etal07}.
Note that \citet{Ald_etal06} give $z$ = 0.0667.}
\tablenotetext{d}{Most likely host according to \citet{Foley_etal10b}.  See text for further details.}
\tablenotetext{e}{The host of SN~2006is was observed with the Magellan
Baade Telescope and IMACS; the hosts of SNe~2006lu and 2008fr were observed with
the du Pont 2.5 m and WFCCD. Emisssion/absorption lines for radial velocity determination were weighted 
according to their equivalent widths.}
\tablenotetext{f}{Ambiguous host. See text for further details.}
\end{deluxetable}

\clearpage
%
\begin{deluxetable} {lccccrrr}
\tablecolumns{8}
\tablewidth{0pt}
\tablecaption{Photometric and Spectroscopic Properties of the Supernovae\label{tab:spec_details}}
\tablehead{%
\colhead{SN} &
\colhead{$\Delta m_{15}$($B$)\tablenotemark{a}} &
\colhead{} &
\colhead{Wang} &
\colhead{Branch} &
\colhead{$N_{\rm{opt}}$\tablenotemark{e}} &
\colhead{$N_{\rm{IR}}$\tablenotemark{f}} &
\colhead{$t_{\rm sp}$\tablenotemark{g}} \\
\colhead{Name} &
\colhead{(mag)} &
\colhead{Subtype\tablenotemark{b}} &
\colhead{Subtype\tablenotemark{c}} &
\colhead{Subtype\tablenotemark{d}} &
\colhead{} &
\colhead{} &
\colhead{(days)} }
\startdata
  2004dt &    1.067(081) &     normal & HV   & BL   &  49 &   0 &      23.6  \\
  2004ef &    1.371(060) &     normal & HV   & BL   &  49 &   4 &    $-$1.8  \\
  2004eo &    1.339(060) &     normal & N    & CL   &  42 &   9 &   $-$10.3  \\
  2004ey &    0.950(061) &     normal & N    & CN   &  32 &   8 &    $-$0.8  \\
  2004gc &    1.103(084) &     normal & \ldots & \ldots &  29 &   0 &      11.1  \\
  2004gs &    1.587(060) &     normal & N    & CL   &  54 &   9 &    $-$3.4  \\
  2004gu &    0.796(067) &  06gz-like & 91T  & SS   &  31 &   3 &    $-$3.1  \\
  2005A  &    0.977(072) &     normal & HV   & BL   &  36 &   9 &      25.4  \\
\enddata   
\tablenotetext{a}{Values of $\Delta m_{15}$($B$) were determined from the template fits. In 
parentheses are the 1$\sigma$ uncertainties in units of 0.001 mag.}
\tablenotetext{b}{Type Ia subtypes from SNID \citep{BloTon07}.  
Members of Type Iax and Ia+CSM subtypes are listed as ``Iax'' and ``Ia+CSM'', respectively; 
super-Chandrasekhar candidates are denoted by ``SC''; SN~2006bt-like are listed as ``06bt-like''.}
\tablenotetext{c}{\citet{wang09} classification. N = normal, HV = high velocity, 91bg = 
SN 1991bg-like.}
\tablenotetext{d}{CN = core normal, CL = cool, SS = shallow silicon, BL = broad line \citep{branch06}.}
\tablenotetext{e}{Number of nights with optical photometry.}
\tablenotetext{f}{Number of nights with near-IR photometry.}
\tablenotetext{g}{Epoch of the spectrum, in days, with respect to $T$($B_{\rm max}$) 
used to determine spectral subtype. If no epoch is given, the spectral classification 
was taken from the literature.}
\end{deluxetable}

\clearpage
\begin{deluxetable} {lcccccc}
\tabletypesize{\scriptsize}
\tablecolumns{7}
\tablewidth{0pt}
\tablecaption{Supernova Light Curve Parameters\label{tab:lcfit_details}}
\tablehead{%
\colhead{SN} &
\colhead{$T_{max}$($B$)\tablenotemark{a}} &
\colhead{$T_{max}$(template)\tablenotemark{b}} &
\colhead{$\Delta m_{15}$($B$)\tablenotemark{c}} &
\colhead{$\Delta m_{15}$(template)\tablenotemark{d}} &
\colhead{$s_{BV}$\tablenotemark{e}} &
\colhead{$s_{BV}$ (template)\tablenotemark{f}} \\
\colhead{Name} &
\colhead{} &
\colhead{} &
\colhead{(mag)} &
\colhead{(mag)} &
\colhead{} &
\colhead{}}
\startdata
2004dt & \ldots & $53234.72 (0.73)$ & \ldots & $1.067 (0.081)$ & \ldots & $1.189 (0.061)$\\
2004ef & $53263.77 (0.12)$ & $53264.38 (0.77)$ & $1.353 (0.014)$ & $1.371 (0.060)$ & $0.846 (0.007)$ & $0.816 (0.060)$\\
2004eo & $53278.24 (0.08)$ & $53278.53 (0.60)$ & $1.389 (0.010)$ & $1.339 (0.060)$ & $0.835 (0.010)$ & $0.824 (0.060)$\\
2004ey & $53303.73 (0.13)$ & $53304.18 (0.60)$ & $0.954 (0.040)$ & $0.950 (0.061)$ & $1.088 (0.008)$ & $1.010 (0.060)$\\
2004gc & \ldots & $53324.78 (0.98)$ & \ldots & $1.098 (0.084)$ & \ldots & $0.923 (0.064)$\\
2004gs & $53355.93 (0.12)$ & $53356.41 (0.78)$ & $1.628 (0.025)$ & $1.587 (0.060)$ & $0.730 (0.008)$ & $0.702 (0.060)$\\
2004gu & $53362.97 (1.05)$ & $53361.76 (0.66)$ & $0.853 (0.091)$ & $0.796 (0.067)$ & \ldots & $1.248 (0.061)$\\
2005A & $53379.72 (0.85)$ & $53379.67 (0.88)$ & $1.115 (0.111)$ & $0.977 (0.072)$ & \ldots & $0.964 (0.061)$\\
\enddata
\tablenotetext{a}{Modified Julian date of $B$-band maximum, as derived from $B$-band data only.}
\tablenotetext{b}{Modified Julian date of maximum light, as derived from data in all filters.}
\tablenotetext{c}{As derived from the $B$-band data only, the number of $B$-band magnitudes the SN faded
in the first 15 days since the time of $B$-band maximum.}
\tablenotetext{d}{Assigned $B$-band decline rate (mag in 15 days), as derived using $B$-band templates.
For SNe~2007cg, 2007hx, 2008cd, and 2009le, the decline rate is derived from data in all filters.}
\tablenotetext{e}{Stretch $BV$ from data. See text for details.}
\tablenotetext{f}{Stretch $BV$ from templates.  See text for details.}
\end{deluxetable}

\clearpage
\begin{deluxetable} {ccc}
\tablecolumns{3}
\tablewidth{0pt}
\tablecaption{Photometric Reduction Terms\label{tab:ec_ct}}
\tablehead{
\colhead{Filter} &
\colhead{Extinction Coefficient\tablenotemark{a}}   &
\colhead{Color Term\tablenotemark{b}} }
\startdata
\multicolumn{3}{c}{Swope+SITE3} \\
$u$ & 0.511 $\pm$ 0.057 &  \phn0.046 $\pm$ 0.017 \\
$B$ & 0.242 $\pm$ 0.022 &  \phn0.061 $\pm$ 0.012 \\
$g$ & 0.191 $\pm$ 0.021 & $-$0.014 $\pm$ 0.011 \\
$V$ & 0.144 $\pm$ 0.018 & $-$0.058 $\pm$ 0.011 \\
$r$ & 0.103 $\pm$ 0.019 & $-$0.016 $\pm$ 0.015 \\
$i$ & 0.059 $\pm$ 0.020 & $-$0.002 $\pm$ 0.015 \\
\hline
\multicolumn{3}{c}{Swope+RetroCam} \\
\enddata
\tablenotetext{a}{Measured in mag airmass$^{-1}$.  
All uncertainties in this table are the ``standard deviations 
of the distributions,'' not the standard deviations of the means.}
\tablenotetext{b}{See Eqs.~\ref{eq:u_nat}--\ref{eq:V_nat}
or \ref{eq:u_inst}--\ref{eq:V_inst}
for which standard colors are used in combination with these coefficients 
to obtain the color correction terms for the optical photometry.
$V$-band photometry obtained with the LC-3009 filter and given in
Table ~\ref{tab:lc3009_v_phot} is associated with a color term
of $-$0.044.  See Appendix B of Paper 2.
As described in the text, color terms are not needed to transform
the $J$ and $H$ magnitudes between the Swope+Retrocam and duPont+WIRC natural systems.}
\tablenotetext{c}{J-band observations with RetroCam prior to 15 January 2009 require
a color term of +0.039.  Starting 15 January 2009 the appropriate color
term is +0.016.  See Equations 26 and 27.}
\tablenotetext{d}{Color terms estimated from synthetic photometry of 
\citet{castelli03} stellar atmosphere models.}
\end{deluxetable}

\clearpage
\begin{deluxetable} {lcccccccc}
\rotate
\tabletypesize{\scriptsize}
\tablecolumns{9}
\tablewidth{0pt}
\tablecaption{Optical Photometry of Secondary Standards\tablenotemark{a}\label{tab:opt_local_stds}}
\tablehead{
\colhead{ID} &
\colhead{$\alpha$ (2000)}   &
\colhead{$\delta$ (2000)}   &
\colhead{$u'$} &
\colhead{$g'$} &
\colhead{$r'$} &
\colhead{$i'$} &
\colhead{$B$} &
\colhead{$V$} }
\startdata
\multicolumn{9}{c}{SN2004dt}\\
 1 & 02:02:09.95 & -00:08:43.8 & 17.479(008) & 15.477(003) & 14.652(004) & 14.266(004) & 16.053(003) & 15.100(003) \\
 2 & 02:02:05.30 & -00:02:03.4 & 16.387(007) & 15.385(003) & 15.185(003) & 15.136(003) & 15.634(003) & 15.270(003) \\
 3 & 02:02:04.23 & -00:02:31.3 & 18.567(012) & 16.128(002) & 15.068(003) & 14.621(003) & 16.784(003) & 15.667(003) \\
 4 & 02:02:24.14 & -00:08:31.2 &  0.000(000) & 18.819(008) & 17.522(005) & 16.678(004) & 19.291(016) & 18.245(008) \\
 5 & 02:02:16.90 & -00:02:52.3 &  0.000(000) & 19.484(014) & 18.166(007) & 16.986(005) & 20.264(038) & 18.929(014) \\
 6 & 02:02:10.06 & -00:06:50.4 & 18.874(014) & 16.786(003) & 16.021(003) & 15.778(003) & 17.341(004) & 16.429(003) \\
 7 & 02:02:08.30 & -00:03:30.2 & 20.142(043) & 19.317(012) & 18.884(013) & 18.723(019) & 19.599(021) & 19.089(016) \\
\enddata
\tablecomments{Table \ref{tab:opt_local_stds} is published in its entirety in the electronic edition of the journal.
A portion is shown here for guidance regarding its form and content.}
\tablenotetext{a}{All photometry is measured in magnitudes.
Values in parentheses are 1$\sigma$ uncertainties. This photometry of field stars is in the
two standard systems, not the natural system.}
\end{deluxetable}

\clearpage
\begin{deluxetable} {lccccc}
\tabletypesize{\scriptsize}
\tablecolumns{6}
\tablewidth{0pt}
\tablecaption{Infrared Photometry of Secondary Standards\tablenotemark{a}\label{tab:nir_local_stds}}
\tablehead{
\colhead{ID} &
\colhead{$\alpha$ (2000)}   &
\colhead{$\delta$ (2000)}   &
\colhead{$Y$} &
\colhead{$J$} &
\colhead{$H$} }
\startdata
\multicolumn{6}{c}{SN2004ef}\\
101 & 22:42:11.38 & +20:00:57.6 & 14.690(046) & 14.382(052) & 14.008(069)  \\
102 & 22:42:12.06 & +20:00:25.1 & 15.262(066) & 14.940(049) & 14.414(067)  \\
103 & 22:42:11.13 & +19:58:37.6 & 15.698(050) & 15.326(057) & 14.698(051)  \\
104 & 22:42:13.33 & +19:58:32.7 & 16.539(044) & 16.072(081) & 15.396(094)  \\
106 & 22:42:17.65 & +19:59:44.9 & 15.933(035) & 15.573(040) & 14.994(068)  \\
108 & 22:42:12.82 & +20:00:02.8 & 16.496(083) & 16.121(056) & 15.452(038)  \\
109 & 22:42:12.90 & +19:59:18.3 & 16.869(049) & 16.552(096) & 16.248(057)  \\
110 & 22:42:15.27 & +19:59:16.0 & 13.114(042) &  0.000(000) &  0.000(000)  \\
111 & 22:42:13.70 & +19:59:30.1 &  0.000(000) & 16.894(075) &  0.000(000)  \\
112 & 22:42:10.76 & +19:59:03.3 & 17.381(119) & 16.997(095) &  0.000(000)  \\
\enddata
\tablecomments{Table \ref{tab:nir_local_stds} is published in its entirety in the electronic edition of the journal.
A portion is shown here for guidance regarding its form and content.}
\tablenotetext{a}{All photometry is measured in magnitudes.
Values in parentheses are 1$\sigma$ uncertainties. The $JH$ photometry  
presented here is in the P98 system. The $Y$-band photometry is in the natural system.}
\end{deluxetable}

\clearpage
\begin{deluxetable} {lcccccc}
\tabletypesize{\scriptsize}
\tablecolumns{7}
\tablewidth{0pt}
\tablecaption{Optical Photometry of SNe in Natural System of Swope (+SITe3)\tablenotemark{a}\label{tab:opt_phot_site3}}
\tablehead{
\colhead{JD} &
\colhead{$u$} &
\colhead{$g$} &
\colhead{$r$} &
\colhead{$i$} &
\colhead{$B$} &
\colhead{$V$} }
\startdata
\multicolumn{7}{c}{SN2004dt}\\
2453249.79 & 16.665(017) & 15.593(006) & 15.659(006) & 16.079(008) & 15.769(006) & 15.575(006) \\
2453250.81 & 16.804(012) & 15.675(009) & 15.695(008) & 16.112(010) & 15.859(008) & 15.630(009) \\
2453251.83 & 16.912(012) & 15.769(009) & 15.745(010) & 16.111(009) & 15.961(009) & 15.677(008) \\
2453252.86 & 17.069(014) & 15.846(006) & 15.796(006) & 16.117(007) & 16.069(007) & 15.723(006) \\
2453253.78 & 17.158(012) & 15.915(006) & 15.785(006) & 16.091(005) & 16.142(014) & 15.770(006) \\
2453256.75 & 17.637(010) & 16.241(006) & 15.807(006) & 16.024(007) & 16.573(007) & 15.950(006) \\
2453257.73 & 17.787(013) & 16.357(008) & 15.833(012) & 16.029(009) & 16.715(008) & 16.012(008) \\
\enddata
\tablecomments{Table \ref{tab:opt_phot_site3} is published in its entirety in the electronic edition of the journal.
A portion is shown here for guidance regarding its form and content.}
\tablenotetext{a}{All photometry is measured in magnitudes.
Values in parentheses are 1$\sigma$ uncertainties.}
\end{deluxetable}

\clearpage
\begin{deluxetable} {lcccccc}
\tabletypesize{\scriptsize}
\tablecolumns{7}
\tablewidth{0pt}
\tablecaption{Optical Photometry of SNe in Natural System of du Pont (+Tek5)\tablenotemark{a}\label{tab:opt_phot_tek5}}
\tablehead{
\colhead{JD} &
\colhead{$u$} &
\colhead{$g$} &
\colhead{$r$} &
\colhead{$i$} &
\colhead{$B$} &
\colhead{$V$} }
\startdata
\multicolumn{7}{c}{SN2005hj}\\ 
2453678.68 & 18.592(033) & 17.848(006) & 17.748(009) & 18.354(012) & 17.927(009) & 17.820(009) \\ 
\multicolumn{7}{c}{SN2005ke}\\ 
2453967.85 &  0.000(000) & 20.720(149) &  0.000(000) &  0.000(000) & 20.661(143) & 20.867(167) \\ 
\multicolumn{7}{c}{SN2006bh}\\ 
2453828.88 &  0.000(000) & 14.636(007) & 14.672(024) & 15.004(021) & 14.724(006) & 14.748(007) \\ 
2453829.88 & 14.924(011) & 14.523(009) &  0.000(000) &  0.000(000) & 14.618(006) & 14.663(009) \\ 
2453830.89 & 14.880(011) & 14.422(011) & 14.519(014) & 15.030(051) & 14.544(006) & 14.550(008) \\
\enddata
\tablenotetext{a}{All photometry is measured in magnitudes.
Values in parentheses are 1$\sigma$ uncertainties.}
\end{deluxetable}

\clearpage
\begin{deluxetable} {lcccc}
\tabletypesize{\scriptsize}
\tablecolumns{5}
\tablewidth{0pt}
\tablecaption{$V$-band Photometry of SNe in Natural System of Swope (+SITe3)\tablenotemark{a}\label{tab:lc3009_v_phot}}
\tablehead{
\colhead{UT Date} &
\colhead{JD} &
\colhead{Object} &
\colhead{$V$} &
\colhead{$\sigma _V$} }
\startdata
2006-01-16 &  2453751.65  & SN2005eq &  19.654 &  0.071 \\
2006-01-19 &  2453754.60  & SN2005gj &  18.319 &  0.018 \\
2006-01-16 &  2453751.60  & SN2005hc &  20.392 &  0.095 \\
2006-01-20 &  2453755.61  & SN2005hc &  20.534 &  0.065 \\
2006-01-17 &  2453752.54  & SN2005hj &  20.505 &  0.077 \\
2006-01-16 &  2453751.53  & SN2005hk &  18.023 &  0.017 \\
2006-01-15 &  2453750.54  & SN2005iq &  19.640 &  0.090 \\
2006-01-20 &  2453755.55  & SN2005iq &  19.884 &  0.052 \\
2006-01-15 &  2453750.60  & SN2005ke &  16.967 &  0.010 \\
2006-01-20 &  2453755.64  & SN2005ke &  17.111 &  0.009 \\
2006-01-16 &  2453751.79  & SN2005ki &  18.164 &  0.024 \\
2006-01-17 &  2453752.78  & SN2005ki &  18.120 &  0.025 \\
2006-01-17 &  2453752.61  & SN2005lu &  19.058 &  0.028 \\
2006-01-17 &  2453752.74  & SN2005mc &  18.198 &  0.022 \\
2006-01-20 &  2453755.74  & SN2005mc &  18.473 &  0.019 \\
2006-01-16 &  2453751.68  & SN2005na &  16.570 &  0.007 \\
2006-01-19 &  2453754.69  & SN2005na &  16.787 &  0.007 \\
2006-01-20 &  2453755.68  & SN2005na &  16.851 &  0.008 \\
2006-01-16 &  2453751.83  & SN2006D  &  14.693 &  0.006 \\
2006-01-17 &  2453752.86  & SN2006D  &  14.538 &  0.006 \\
2006-01-19 &  2453754.77  & SN2006D  &  14.331 &  0.007 \\
\enddata
\tablenotetext{a}{All photometry is measured in magnitudes.
Data taken using LC-3009 filter.}
\end{deluxetable}

\clearpage
\begin{deluxetable} {lccc}
\tabletypesize{\scriptsize}
\tablecolumns{4}
\tablewidth{0pt}
\tablecaption{Infrared Photometry of SNe in Natural System of Swope (+RetroCam)\tablenotemark{a}\label{tab:Ia_nir_phot_rc1}}
\tablehead{
\colhead{JD} &
\colhead{$Y$} &
\colhead{$J$} &
\colhead{$H$} }
\startdata
\multicolumn{4}{c}{SN2005el}\\
2453640.80 & 15.652(012) & 15.650(014) & 15.707(022) \\
2453640.81 & 15.647(011) & 15.617(013) & 15.730(021) \\
2453643.79 & 15.561(010) & 15.513(013) & 15.733(026) \\
2453643.80 & 15.542(010) & 15.521(015) & 15.739(027) \\
2453648.79 & 16.051(030) & 15.858(021) & 16.044(048) \\
2453648.80 & 16.047(017) & 15.905(029) & 15.928(058) \\
2453656.83 & 16.608(018) & 17.426(065) &  0.000(000) \\
2453656.83 & 16.603(020) &  0.000(000) &  0.000(000) \\
2453657.76 & 16.600(017) & 17.364(040) & 16.442(039) \\
2453657.77 & 16.585(020) & 17.266(038) & 16.371(034) \\
2453660.83 & 16.484(017) & 17.409(058) & 16.277(039) \\
2453660.84 & 16.460(018) & 17.320(044) & 16.401(062) \\
2453664.82 & 16.151(017) & 17.151(041) &  0.000(000) \\
2453664.84 & 16.183(020) & 17.304(058) &  0.000(000) \\
2453668.80 & 15.800(012) & 16.798(040) & 15.936(029) \\
2453668.81 & 15.765(012) & 16.873(034) & 15.916(034) \\
\enddata
\tablecomments{Table \ref{tab:Ia_nir_phot_rc1} is published in its entirety in the electronic edition of the journal.
A portion is shown here for guidance regarding its form and content.}
\tablenotetext{a}{The photometry presented here was obtained with the $J$-band filter RC1, which was used
prior to 15 January 2009.  All photometry is measured in magnitudes.
Values in parentheses are 1$\sigma$ uncertainties.}
\end{deluxetable}

\clearpage
\begin{deluxetable} {lccc}
\tabletypesize{\scriptsize}
\tablecolumns{4}
\tablewidth{0pt}
\tablecaption{Infrared Photometry of SNe in Natural System of Swope (+RetroCam)\tablenotemark{a}\label{tab:Ia_nir_phot_rc2}}
\tablehead{
\colhead{JD} &
\colhead{$Y$} &
\colhead{$J$} &
\colhead{$H$} }
\startdata
\multicolumn{4}{c}{SN2008hv}\\
2454849.78 & 15.764(012) & 16.960(024) & 16.332(036) \\
2454849.79 & 15.769(016) & 16.988(025) & 16.336(033) \\
2454854.78 & 16.090(013) & 17.521(036) & 16.634(044) \\
2454854.79 & 16.088(013) & 17.462(036) & 16.700(046) \\
2454868.66 & 16.971(020) & 18.370(100) & 17.252(047) \\
2454868.67 & 16.985(021) & 18.560(100) & 17.283(048) \\
2454875.65 & 17.339(018) & 18.867(112) & 17.549(065) \\
2454875.66 & 17.333(026) & 19.074(112) & 17.584(074) \\
2454887.66 & 17.970(036) &  0.000(000) &  0.000(000) \\
2454887.66 & 17.996(037) &  0.000(000) &  0.000(000) \\
2454889.66 & 18.131(052) &  0.000(000) &  0.000(000) \\
2454889.67 & 18.046(047) &  0.000(000) &  0.000(000) \\
2454895.58 & 18.212(051) &  0.000(000) &  0.000(000) \\
2454895.59 & 18.283(051) &  0.000(000) &  0.000(000) \\
2454899.62 & 18.339(069) &  0.000(000) &  0.000(000) \\
2454912.69 & 18.774(067) &  0.000(000) &  0.000(000) \\
2454912.70 & 18.710(068) &  0.000(000) &  0.000(000) \\
2454912.70 & 18.710(068) &  0.000(000) &  0.000(000) \\
\enddata
\tablecomments{Table \ref{tab:Ia_nir_phot_rc2} is published in its entirety in the electronic edition of the journal.
A portion is shown here for guidance regarding its form and content.}
\tablenotetext{a}{The photometry presented here was obtained with the $J$-band filter RC2, which was used
after 15 January 2009.  All photometry is measured in magnitudes.
Values in parentheses are 1$\sigma$ uncertainties.}
\end{deluxetable}

\clearpage
\begin{deluxetable} {lccc}
\tabletypesize{\scriptsize}
\tablecolumns{4}
\tablewidth{0pt}
\tablecaption{Infrared Photometry of SNe in Natural System of du Pont (+WIRC)\tablenotemark{a}\label{tab:Ia_nir_phot_wirc}}
\tablehead{
\colhead{JD} &
\colhead{$Y$} &
\colhead{$J$} &
\colhead{$H$} }
\startdata
\multicolumn{4}{c}{SN2004ef}\\
2453269.70 & 17.897(042) & 17.946(056) & 17.817(096) \\
2453269.68 &  0.000(000) & 17.919(068) &  0.000(000) \\
2453277.63 & 17.835(025) & 18.679(049) & 17.764(050) \\
2453277.64 &  0.000(000) & 18.559(047) & 17.829(052) \\
2453281.59 & 17.590(021) & 18.463(036) & 17.747(037) \\
2453281.60 & 17.574(021) & 18.384(036) & 17.732(039) \\
2453328.58 & 19.161(051) &  0.000(000) &  0.000(000) \\
2453328.59 & 19.211(061) &  0.000(000) &  0.000(000) \\
\enddata
\tablecomments{Table \ref{tab:Ia_nir_phot_wirc} is published in its entirety in the electronic edition of the journal.
A portion is shown here for guidance regarding its form and content.}
\tablenotetext{a}{All photometry is measured in magnitudes.
Values in parentheses are 1$\sigma$ uncertainties.}
\end{deluxetable}

\clearpage
\begin{deluxetable} {llccc}
\tablecolumns{5}
\tablewidth{0pt}
\tablecaption{$Y$-band Photometric Standards \label{tab:Y_stds}}
\tablehead{
\colhead{ID\tablenotemark{a}} &
\colhead{HST} &
\colhead{$\alpha$ (2000)}   &
\colhead{$\delta$ (2000)}   &
\colhead{$Y$\tablenotemark{b}} }
\startdata
9101 &  P525-E &    00:24:28.3 &    +07:49:02 &    11.880(011) \\
9103 &  S294-D &    00:33:15.2 &  $-$39:24:10 &    11.163(011) \\
9104 &  S754-C &    01:03:15.8 &  $-$04:20:44 &    11.282(011) \\
9105 &  P530-D &    02:33:32.1 &    +06:25:38 &    11.567(017) \\
9106 &  S301-D &    03:26:53.9 &  $-$39:50:38 &    12.396(015) \\
9107 &  P247-U &    03:32:03.0 &    +37:20:40 &    12.205(011) \\
9108 &  P533-D &    03:41:02.4 &    +06:56:13 &    11.995(016) \\
9109 &  S055-D &    04:18:18.9 &  $-$69:27:35 &    11.759(005) \\
\enddata
\tablenotetext{a}{Star ID in \citet{Per_etal98}.}
\tablenotetext{b}{All photometry is measured in magnitudes.
Values in parentheses are 1$\sigma$ uncertainties.}
\end{deluxetable}

\begin{deluxetable} {ccccccccc}
\tabletypesize{\scriptsize}
\tablecolumns{9}
\tablewidth{0pt}
\tablecaption{Standard and Natural System Magnitudes of Fundamental Standards\label{tab:pm_mags}}
\tablehead{
\colhead{Filter} &
\colhead{Vega} & 
\colhead{Vega} &
\colhead{BD +17$\arcdeg 4708$} &
\colhead{BD +17$\arcdeg 4708$} &
\colhead{P177D} &
\colhead{P177D} &
\colhead{P330E} &
\colhead{P330E} \\
\colhead{} &
\colhead{std} &
\colhead{nat} &
\colhead{std} &
\colhead{nat} &
\colhead{std} &
\colhead{nat} &
\colhead{std} &
\colhead{nat} }
\startdata
$u$          & \ldots & \ldots & 10.560 & 10.518 & 15.128 & 15.066 & 14.539 & 14.481 \\
$g$          & \ldots & \ldots &  9.640 &  9.644 & 13.771 & 13.777 & 13.287 & 13.293 \\
$r$          & \ldots & \ldots &  9.350 &  9.352 & 13.316 & 13.318 & 12.848 & 12.850 \\
$i$          & 0.382  & \ldots &  9.250 &  9.250 & 13.170 & 13.170 & 12.701 & 12.701 \\
\hline
$B$          & 0.030  & 0.030  & \ldots & \ldots & 14.138 & 14.099 & 13.658 & 13.620 \\
$V$(LC-3014) & 0.030  & 0.010  & \ldots & \ldots & 13.492 & 13.511 & 13.028 & 13.047 \\
$V$(LC-3009) & 0.030  & 0.015  & \ldots & \ldots & 13.492 & 13.506 & 13.028 & 13.042 \\
$V$(LC-9844)  & 0.030  & 0.010  & \ldots & \ldots & 13.492 & 13.511 & 13.028 & 13.047 \\
\hline
$Y$        & 0.000 & 0.000 & \ldots & \ldots & \ldots & \ldots & \ldots & \ldots \\
$J$        & 0.000 & 0.000 & \ldots & \ldots & \ldots & \ldots & \ldots & \ldots \\
$H$        & 0.000 & 0.000 & \ldots & \ldots & \ldots & \ldots & \ldots & \ldots \\

\enddata
\tablecomments{The $V$(LC-3014) filter was in use until JD~2453749 (2006 January 14 UT).
The $V$(LC-3009) filter was employed from JD~2453750--2453759 (2006 January 15-24 UT).
From JD~2453760 (2006 January 25 UT) onward, the $V$(LC-9844) filter was used.}
\end{deluxetable}

\begin{deluxetable} {ccccccc}
\tabletypesize{\scriptsize}
\tablecolumns{7}
\tablewidth{0pt}
\tablecaption{Photometric Zero Points Derived via Fundamental Standards\tablenotemark{a}\label{tab:pm_zpts}}
\tablehead{
\colhead{Filter} &
\colhead{Vega} & 
\colhead{BD +17$\arcdeg 4708$} &
\colhead{P177D} &
\colhead{P330E} &
\colhead{Average} &
\colhead{Average} \\
\colhead{} &
\colhead{} &
\colhead{} &
\colhead{} &
\colhead{} &
\colhead{with BD+17$\arcdeg 4708$} &
\colhead{without BD+17$\arcdeg 4708$} }
\startdata
$u$          &  \ldots & 12.98 & 12.98 & 12.98 & 12.98     & 12.98 \\
$g$          &  \ldots & 15.11 & 15.12 & 15.12 & 15.12     & 15.12 \\
$r$          &  \ldots  & 14.90 & 14.90 & 14.90 & 14.90     & 14.90 \\
$i$          &  \ldots & 14.54 & 14.53 & 14.52 & 14.53     & 14.53 \\
\hline
             &        &       &       &       & Average   & Average \\
             &        &       &       &       & with Vega & without Vega \\
$B$          &  14.33 &  \ldots & 14.32 & 14.32 & 14.32     & 14.32 \\
$V$(LC-3014)          &  14.44 &  \ldots & 14.43 & 14.44 & 14.44     & 14.44 \\
\enddata
\tablecomments{The $V$(LC-3014) filter was in use until JD~2,453,749 (2006 January 14 UT).
The $V$(LC-3009) filter was employed from JD~2,453,750--2,453,759 (2006 January 15-24 UT).
From JD~2,453,760 (2006 January 25 UT) onward, the $V$(LC-9844) filter was used.
The $J_{\rm{RC1}}$ filter was in use through 2009~January~2 (JD~2,454,833),
and was not used thereafter.
All observations obtained from 2009~January~15 (JD~2,454,846) onward were 
obtained with the $J_{\rm{RC2}}$ filter.}
\tablenotetext{a}{See Equation~\ref{eq:zero_point_defined} of Appendix~\ref{sec:zeropts}.}
\end{deluxetable}

\begin{deluxetable} {ccccc}
\tabletypesize{\small}
\tablecolumns{5}
\tablewidth{0pt}
\tablecaption{AB Magnitude Offsets for CSP-I Filters\label{tab:AB_offsets}}
\tablehead{
\colhead{Filter} &
\colhead{AB Offset} &
\colhead{   } &
\colhead{Filter} &
\colhead{AB Offset}
} 
\startdata
$u$          &  $-$0.06  & & $Y_{RC}$       &  +0.63 \\
$g$          &  $-$0.02  & & $J_{RC1}$      &  +0.91 \\
$r$          &  $-$0.01 & & $J_{RC2}$      &  +0.90 \\
$i$          &  \phn0.00  & & $H_{RC}$       & +1.34 \\
$B$          &  $-$0.13 & &     &    \\
$V$(LC-3014)          &  $-$0.02  & & $Y_{WIRC}$      &  +0.64 \\
$V$(LC-3009)          &  $-$0.02  & & $J_{WIRC}$     &  +0.90 \\
$V$(LC-9844)          &  $-$0.02  & & $H_{WIRC}$     &  +1.34 \\
\enddata
\end{deluxetable}

\clearpage


\figcaption[]{A mosaic of $V$-band CCD images of \numsne ~Type~Ia SNe observed by 
\mbox{\csp}. The location of each SN is indicated by a blue circle. The 
positions of secondary standards used for calibrating the optical photometry are 
indicated by red squares.  For uniformity, each finder chart is 5\arcmin\ 
$\times$ 5\arcmin\ in size.  
In some cases, a few of the local standard stars are outside 
the boundaries of the chart. {\sc Note} -- all \numsne\ finder charts will be 
reproduced in the electronic edition of the journal.
\label{fig:fcharts}}


\figcaption[]{{\em Top:} Histogram of values of heliocentric redshift of host galaxies
of the \numsne\ SNe included in this paper. {\em Middle:} Histogram
of values of the $B$-band decline rate, \dmb, of the
SNe~Ia, as determined from the template fits. {\em Bottom:}
Histogram of values of the color stretch parameter $s_{BV}$
for the SNe~Ia.\label{fig:histograms}}


\figcaption[] {\csp\ optical and NIR filter response functions.  The optical
bandpasses shown in the upper portion of the figure 
were measured using a monochromator as described in Paper~2, 
and have then been multiplied by an atmospheric
absorption and extinction spectrum typical of LCO for an airmass of 1.2.
The NIR bandpasses shown in the lower half of the figure were determined 
with the same monochromator (see Appendix~\ref{sec:nir_bandpasses} 
for details), and also include atmospheric absorption appropriate for LCO. 
Note that three different $V$ filters were used 
with the Swope telescope~+~SITe2 CCD camera during the course of 
the \csp\ (see \S\ref{sec:swope_opt} for details), and two $J$ filters were
utilized at the Swope with RetroCam (see \S\ref{sec:swope_nir}).
The dashed red lines show the response functions for the 
$J$ and $H$ filters employed by P98.  These were derived by combining the 
filter transmission and detector quantum efficiency data given by these 
authors with two aluminum reflections and an
atmospheric transmission function appropriate to LCO. 
\label{fig:filters}}



\figcaption[]{Optical broadband atmospheric extinction values 
(mag airmass$^{-1}$) measured at LCO from September 2004 through November 2009. 
Histograms for the entire five years are shown at the right of each panel.
\label{fig:extinction}}


\figcaption[]{Optical broadband color terms from LCO 1~m photometry. 
Histograms for the entire five years are shown at the right of each panel.\label{fig:CT}}


\figcaption[]{Nightly photometric zero points derived from observations of secondary
standard stars with the LCO Swope telescope~+~SITe3 camera over the course of \mbox{\csp}.
The vertical gray lines indicate dates on which the primary mirror was washed.  
\label{fig:ZP}}


\figcaption[]{{\em Top:} Differences between synthetic 
photometry in the natural systems of the two Swope~+~RetroCam 
filters, $J_{\rm RC1}$ and $J_{\rm RC2}$, and the P98 $J$ bandpass versus 
the $(J-H)_{\rm P98}$ color (green and red points, respectively).
The differences between synthetic 
photometry in the natural system of the du~Pont~+~WIRC $J$ filter  
and the P98 $J$ bandpass are also shown (blue points).  The slopes 
give the $J$-band color terms (CT) for the Swope~+~RetroCam
and the du~Pont~+~WIRC systems using synthetic photometry derived from 
\citet{castelli03} atmosphere models covering a range of stellar parameters and
reddenings ($E(B-V) = 0.0$~to~2.5~mag).  
{\em Bottom:} Same as top panel but for $H$-band magnitude differences vs.
$(J-H)_{\rm P98}$ color.  The slopes of the solid and dashed lines give the
$H$-band color terms derived for the Swope~+~RetroCam and du~Pont~+~WIRC $H$ filters.
\label{fig:castelli_cts}}


\figcaption[]{NIR broadband extinction values (mag airmass$^{-1}$) measured at
LCO from September 2004 through November 2009. Blue points correspond to
observations made with the du~Pont~+~WIRC, and red points show measurements
made with the Swope~+~RetroCam.  Histograms for the entire five years are shown 
at the right of each panel.\label{fig:nir_extinction}}
 

\figcaption[]{Nightly NIR photometric zero points for the du~Pont~+~WIRC
derived from observations of both
secondary and tertiary standard stars over the course of the \mbox{\csp}. 
The blue circles denote
zero points calculated from secondary standards, and the red squares indicate
zero points derived from tertiary standards.
The vertical gray lines indicate dates on which the primary mirror was aluminized.  
\label{fig:wirc_zero_pts}}


\figcaption[]{Nightly NIR photometric zero points for the Swope~+~RetroCam
derived from observations of both
secondary and tertiary standard stars over the course of the \mbox{\csp}.  
The blue circles denote
zero points calculated from secondary standards, and the red squares indicate
zero points derived from tertiary standards.
The vertical gray lines indicate dates on which the primary mirror was washed.  
\label{fig:rc_zero_pts}}


\figcaption[]{Magnitude differences in the $Y$ (upper-left plot), $J$ (upper-right
plot), and $H$ (lower-left plot) shown as a function of radial distance in pixels
from the detector center of field stars in RetroCam images obtained on
20~October~2007 and 5~April~2008.  The ``magnitude difference'' is the arithmetic
difference of the instrumental magnitudes of a given star observed on two different
nights.  The lower-right plot shows the
$H$-band magnitude differences dividing the field stars by color into two
subsamples, with the inset plot showing the $(J-H)$ color distribution.
The blue symbols correspond to stars with $(J-H) \leq 0.45$~mag and the red
symbols are stars with $(J-H) > 0.45$~mag. 
\label{fig:phot_diff_campaign4}}


\figcaption[]{Magnitude differences in the $H$ band plotted as a function
of radial distance in pixels from the detector center of field stars in RetroCam
images obtained on five different epochs during campaign~4.  The slope of 
each fit is given in the sub-plots. The dates shown
in the individual plots are the dates of sunset, not the UT date at midnight.
\label{fig:mag_diff_slopes}}


\figcaption[]{Multiband light curves of the Type~Ia SNe observed by \mbox{\csp}. 
For better intercomparison each subplot has an abscissa ($x$-axis) range of 
100 {\em observer-frame} days, and an ordinate ($y$-axis) range of 6 mag. Note
that $x = 0$ corresponds to the time of 
$B$-band maximum.  Best-fitting SNooPy fits using the ``max model'' mode are shown 
for each SN. As the $Y$-band photometry obtained with two different cameras has 
different color terms, we color code the $Y$-band photometry obtained with the du 
Pont telescope and WIRC with orange points.  Starting 15 January 2009 the $J$-band 
photometry was obtained with the Swope telescope and RetroCam using the $J_{\rm RC2}$ 
filter.  Those data points are color coded orange. {\sc Note} -- all 123 light 
curves for this figure will be reproduced in the electronic edition of the 
journal.
\label{fig:sample_lcs}} 


\figcaption[]{Same as Figure \ref{fig:sample_lcs}, except for Type~Iax SNe. 
\label{fig:Iax_lcs}} 


\figcaption[]{Multiband light curves of the Type~Ia-CSM object SN~2005gj, SN~2008J, two
SN~2006bt-like objects, and two super-Chandrasekhar candidates
observed by \mbox{\csp}.  The axes are laid out as in Figures \ref{fig:sample_lcs}
and \ref{fig:Iax_lcs}.\label{fig:other_lcs}} 


\figcaption[]{Magnitude differences vs. colors for $ugriBV$, 
derived synthetically from the scanned \mbox{\csp} bandpasses
and the \citet{stritzinger05} spectrophotometric atlas of Landolt 
standards. The dashed lines are the best fits to the red triangles; 
the slope of each fit is listed numerically in the sub-plots using a red font.
The slopes corresponding to the measured mean color terms 
(see Table~~\ref{tab:ec_ct}) are indicated by the solid black lines; the 
numerical values of these are reproduced in the sub-plots using a black font. 
The histograms in each plot correspond to the colors of the 
standard stars observed routinely by the \mbox{\csp}. \label{fig:opt_syn_cts}}


\figcaption[]{
Observations of P98 standards and six red stars with $(J-H)_{\rm P98} > 1.5$~mag from Table~3 
of P98 obtained with RetroCam on the duPont telescope.  The differences between the P98 
and observed magnitudes are plotted versus the $(J-H)_{\rm P98}$ color. The red stars 
plotted with black symbols correspond to reddened M~giants in the Coal Sack, while 
those plotted with red symbols are YSOs. The solid and dashed lines are fits to the 
observations excluding and including (respectively) the YSOs. ({\em Top:}) The swath 
of grey points shows the differences between synthetic photometry in the 
natural system of the Swope~+~RetroCam $J_{\rm RC2}$ bandpass and 
the P98 $J$ filter for \citet{castelli03} 
atmosphere models covering a range of stellar parameters and reddenings ($E(B-V) = 
0.0$~to~2.5~mag), plotted versus the $(J-H)_{\rm P98}$ color. ({\em Bottom:}) Same, but for the
$H$ band. \label{fig:nir_ct_dupont_rc}}


\figcaption[]{Synthetic $(Y-K_{s,P98})$ colors plotted as a function of synthetic 
$(J_{\rm P98}-K_{s,{\rm P98}})$ colors.  The gray circles correspond to model dwarf star
atmospheres with nearly solar metallicity.  The red and green solid lines are
linear and fifth-order polynomial fits to
these data, while the blue dashed line is the relation given by \citet{Ham_etal06}.
The gray shaded region indicates the color range of the P98 standards used
by the \mbox{\csp}.
\label{fig:Y_calibration}}


\figcaption[]{Comparison of synthetic photometry of  the \citet{pickles98} stellar
library with the $(B-g)$ and $(V-g)$ colors of stars common to \citet{Lan92} and
\citet{Smi_etal02}.
\label{fig:pickles}}

\clearpage


\begin{figure}[t]
\epsscale{.54}
\plottwo{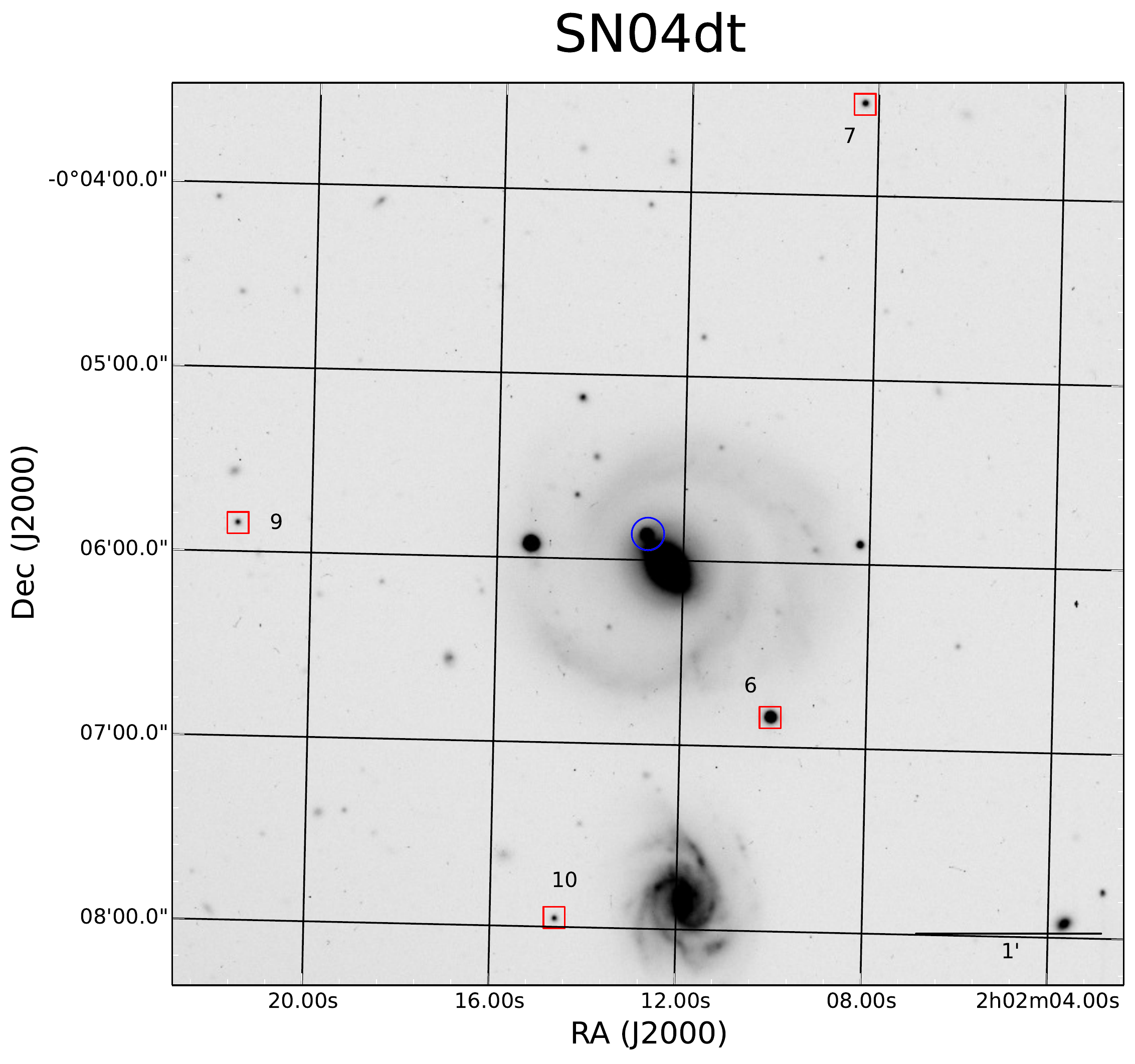}{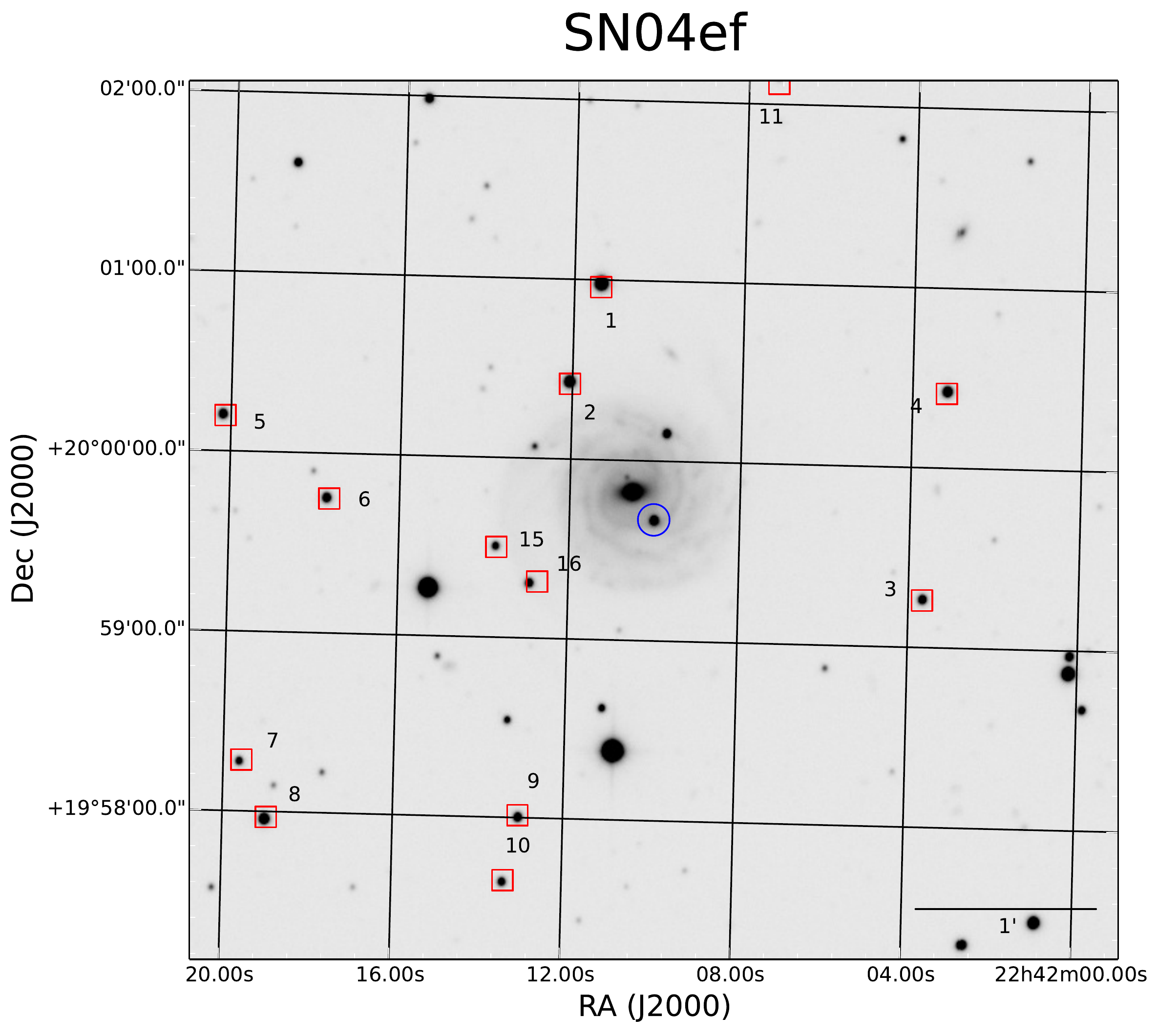}
\plottwo{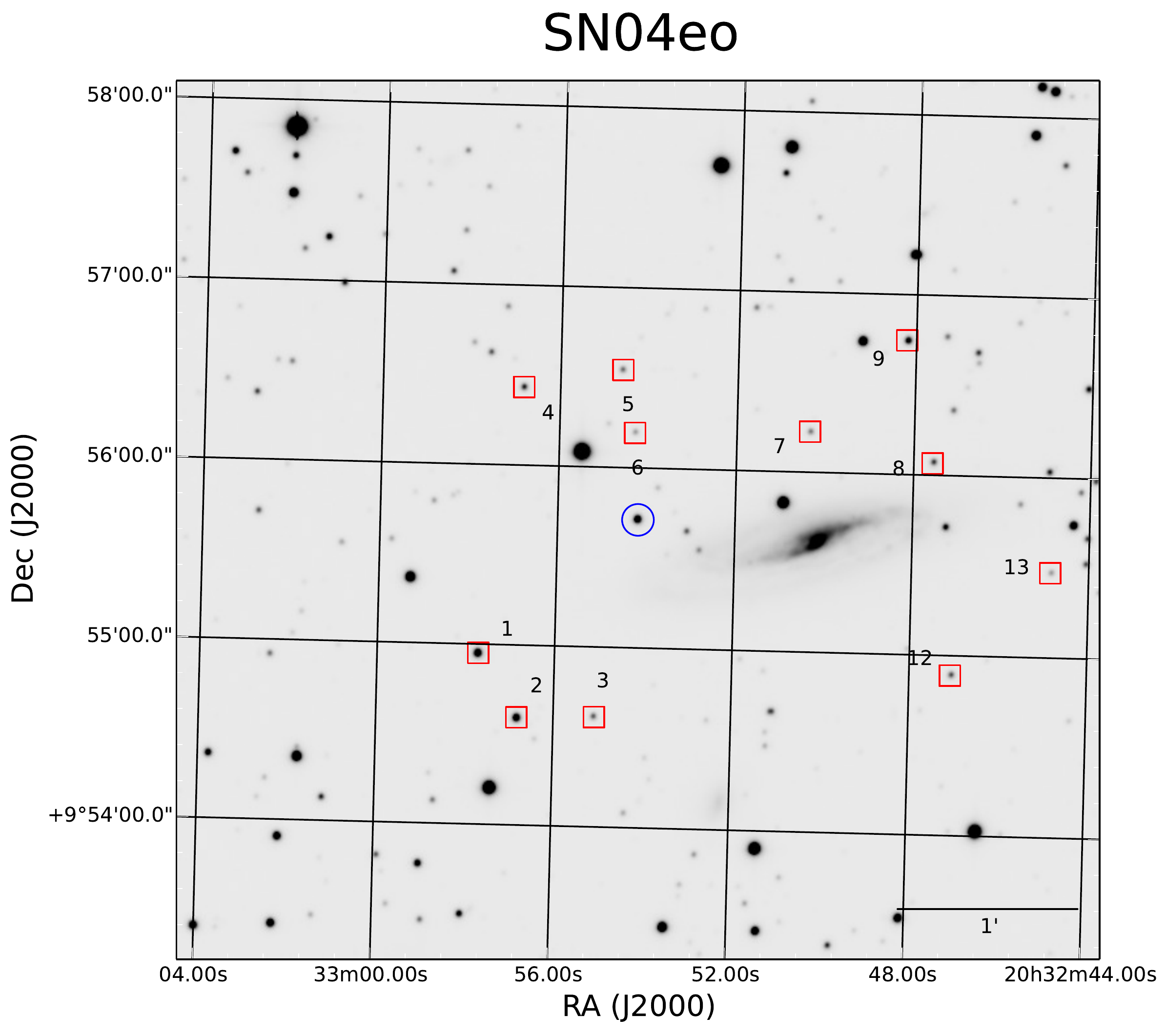}{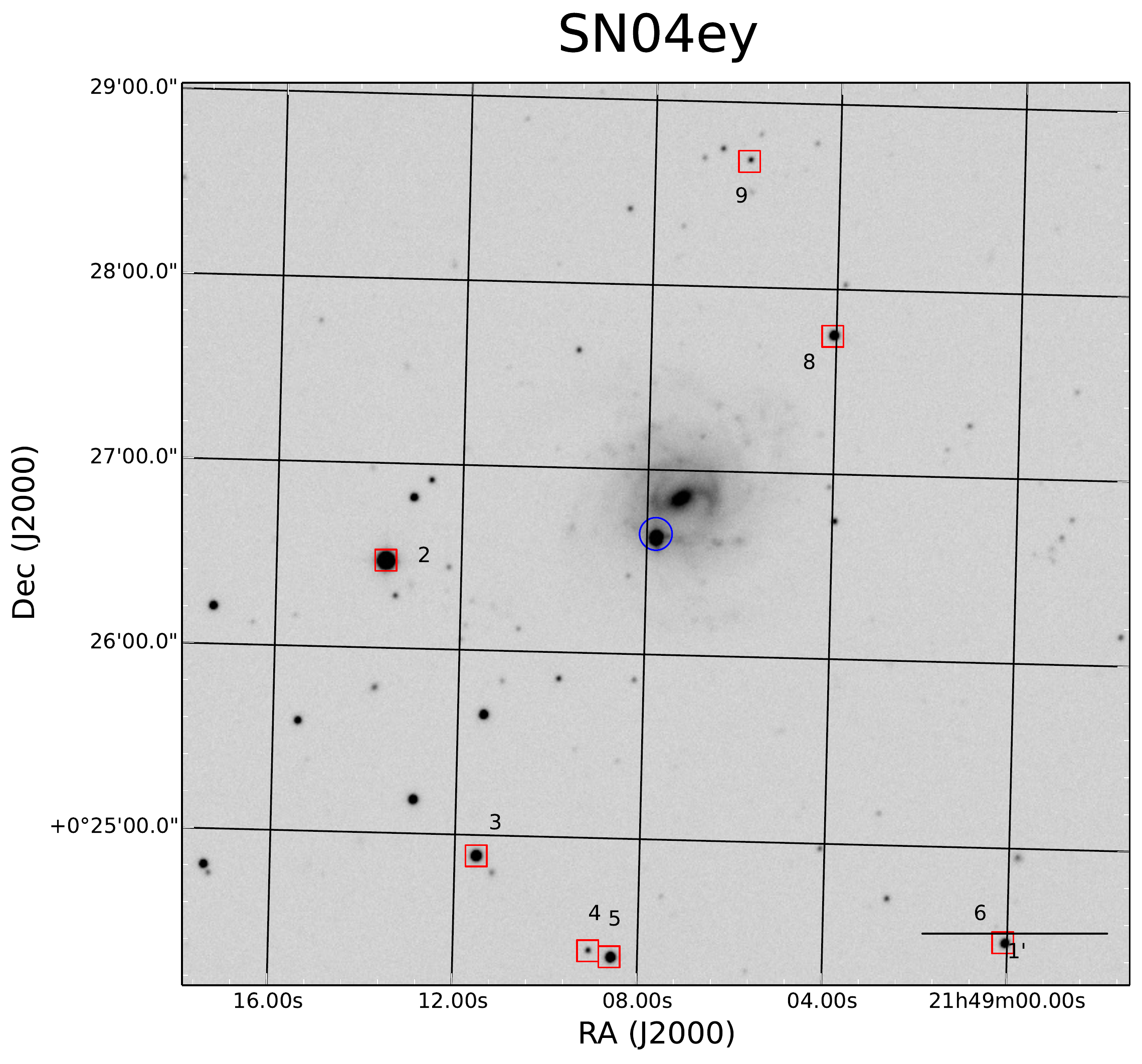}
\newline                                                                     
\plottwo{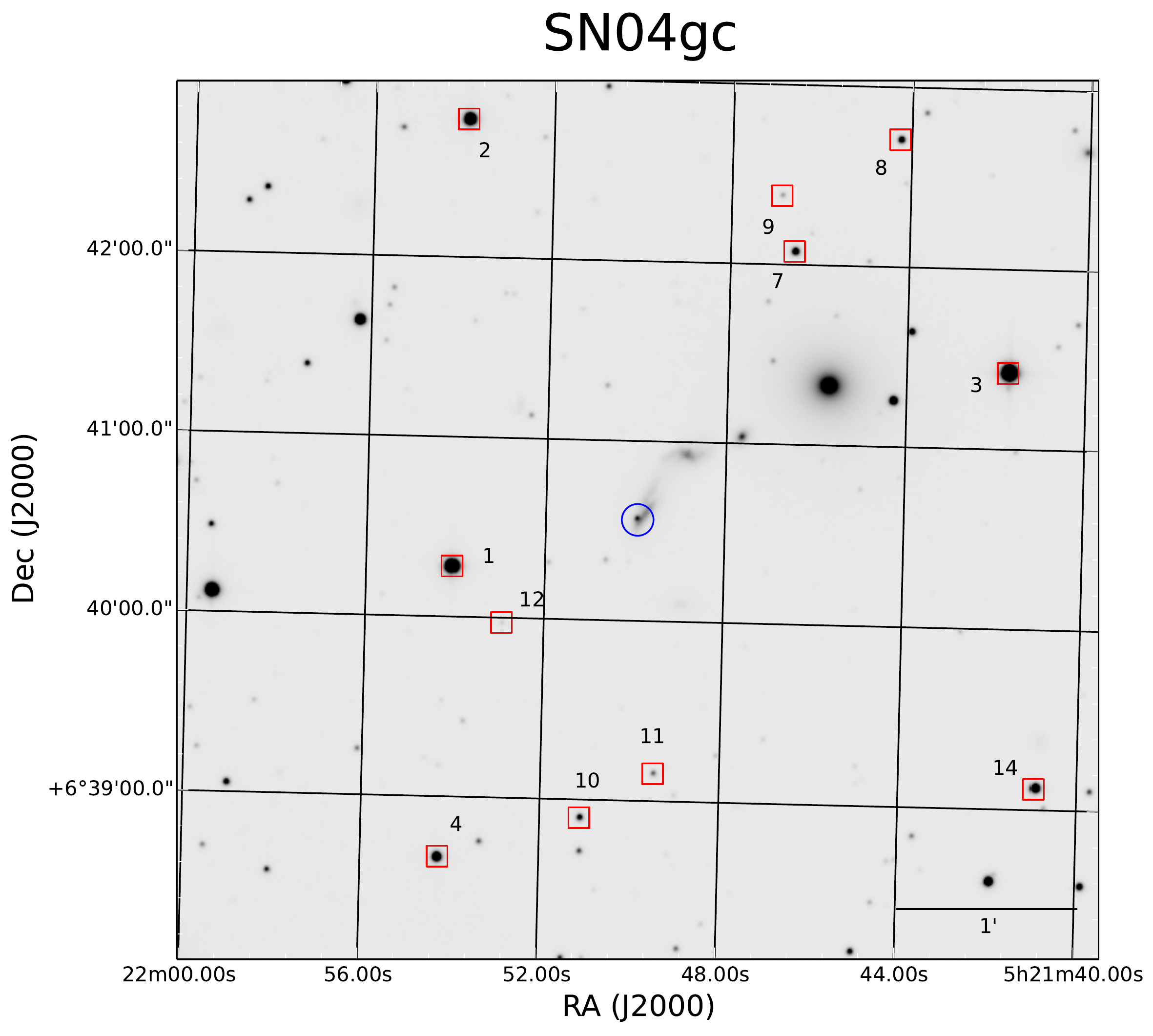}{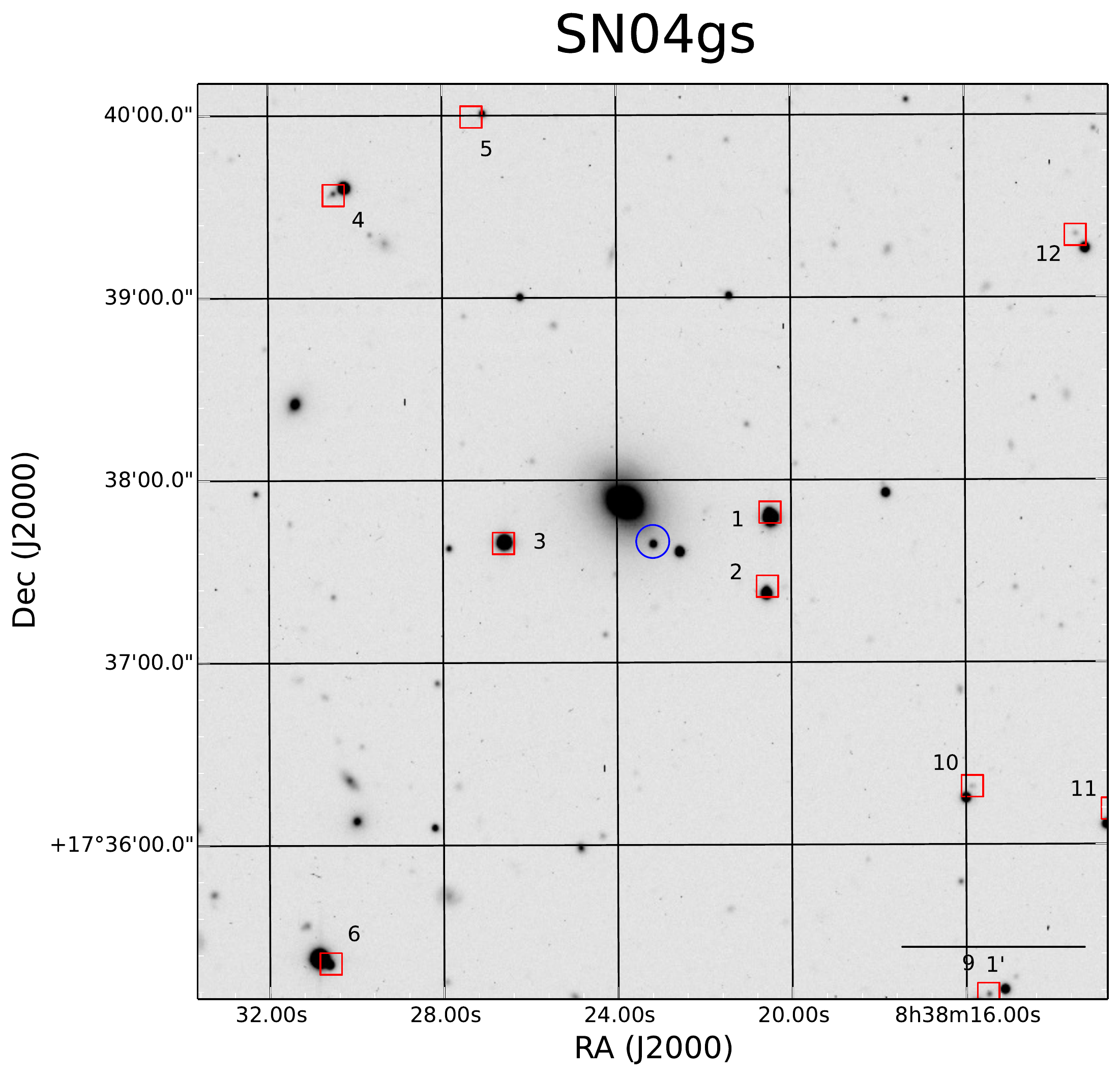}
\plottwo{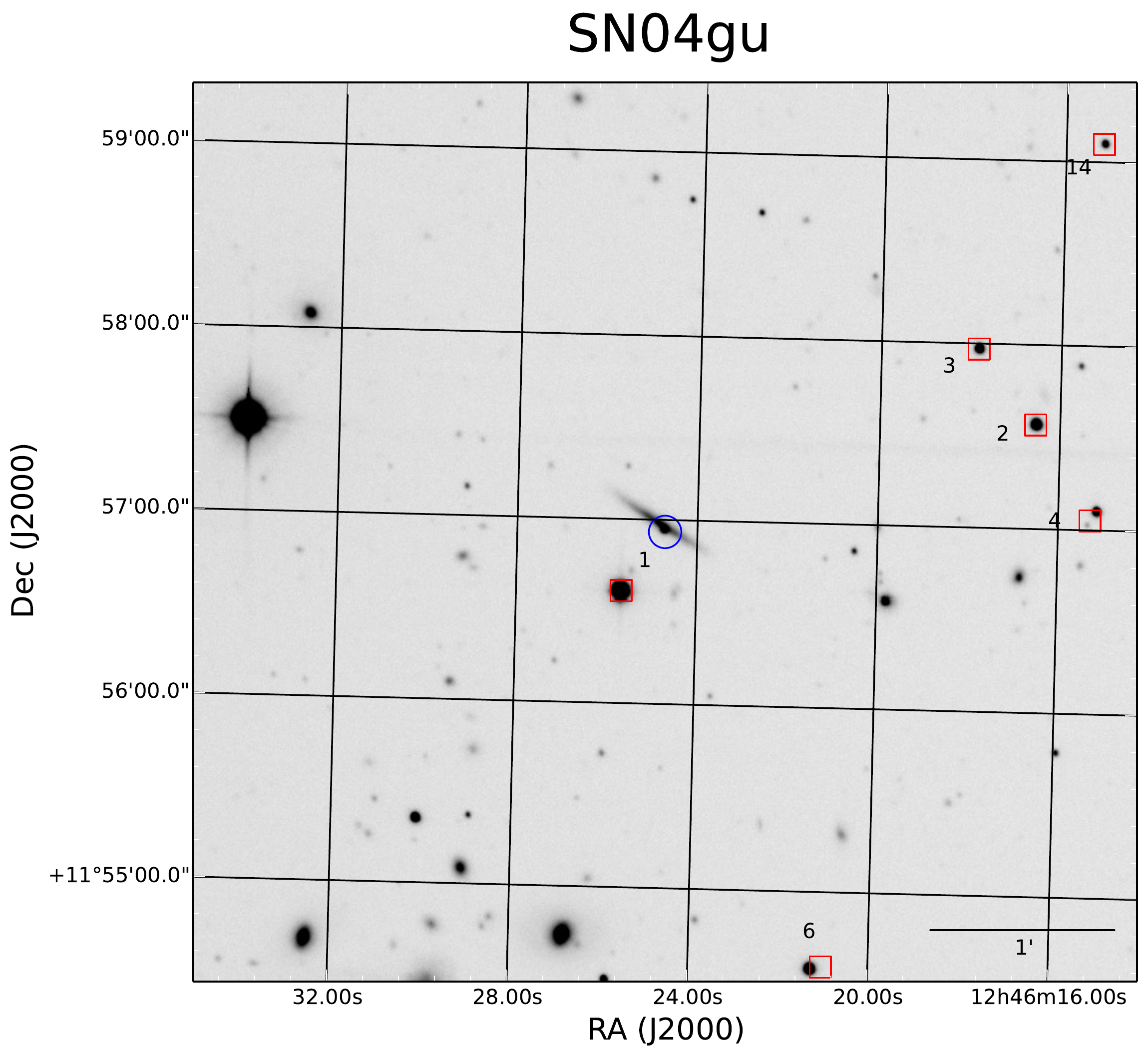}{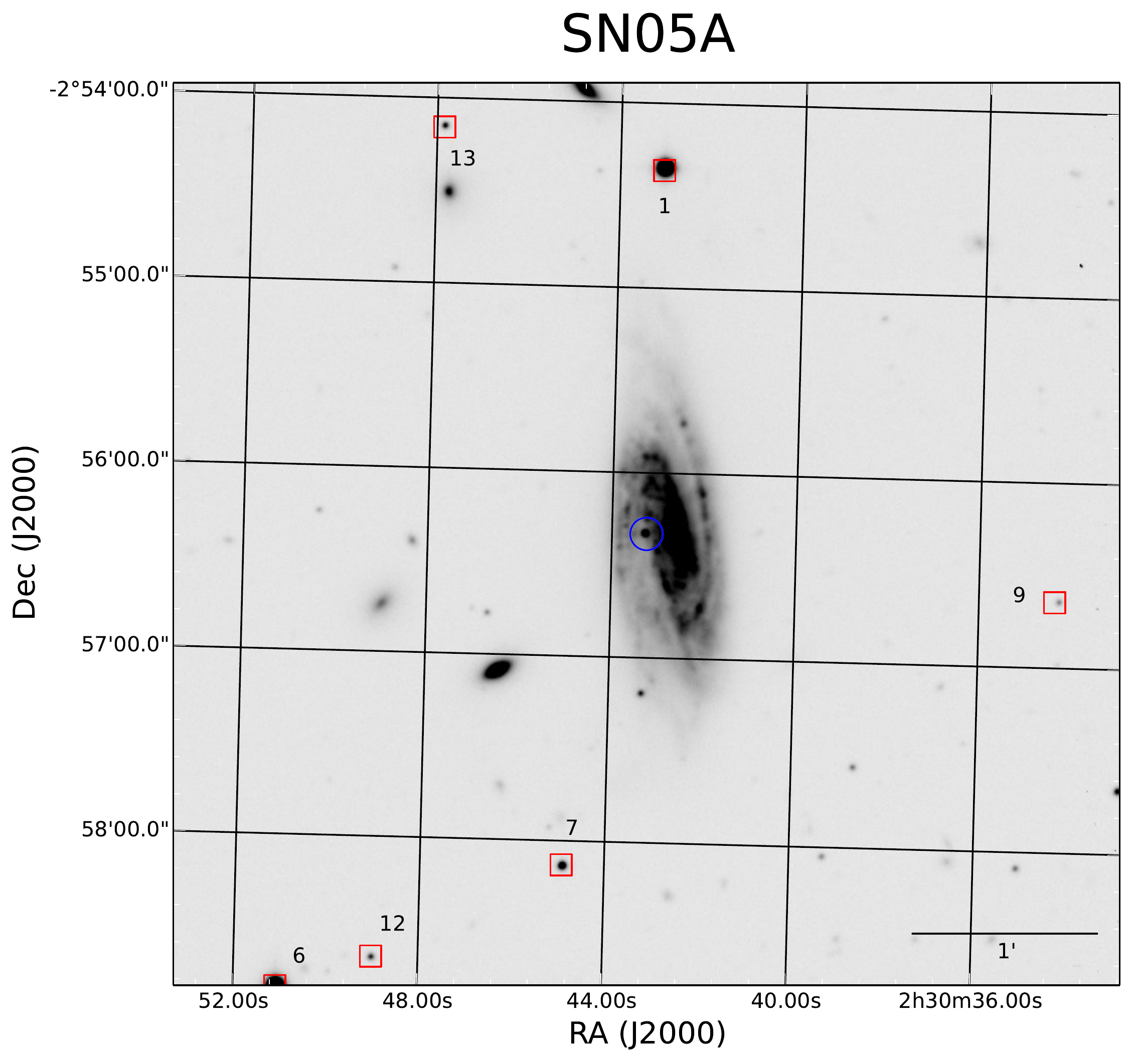}
\newline                                                                     
\plottwo{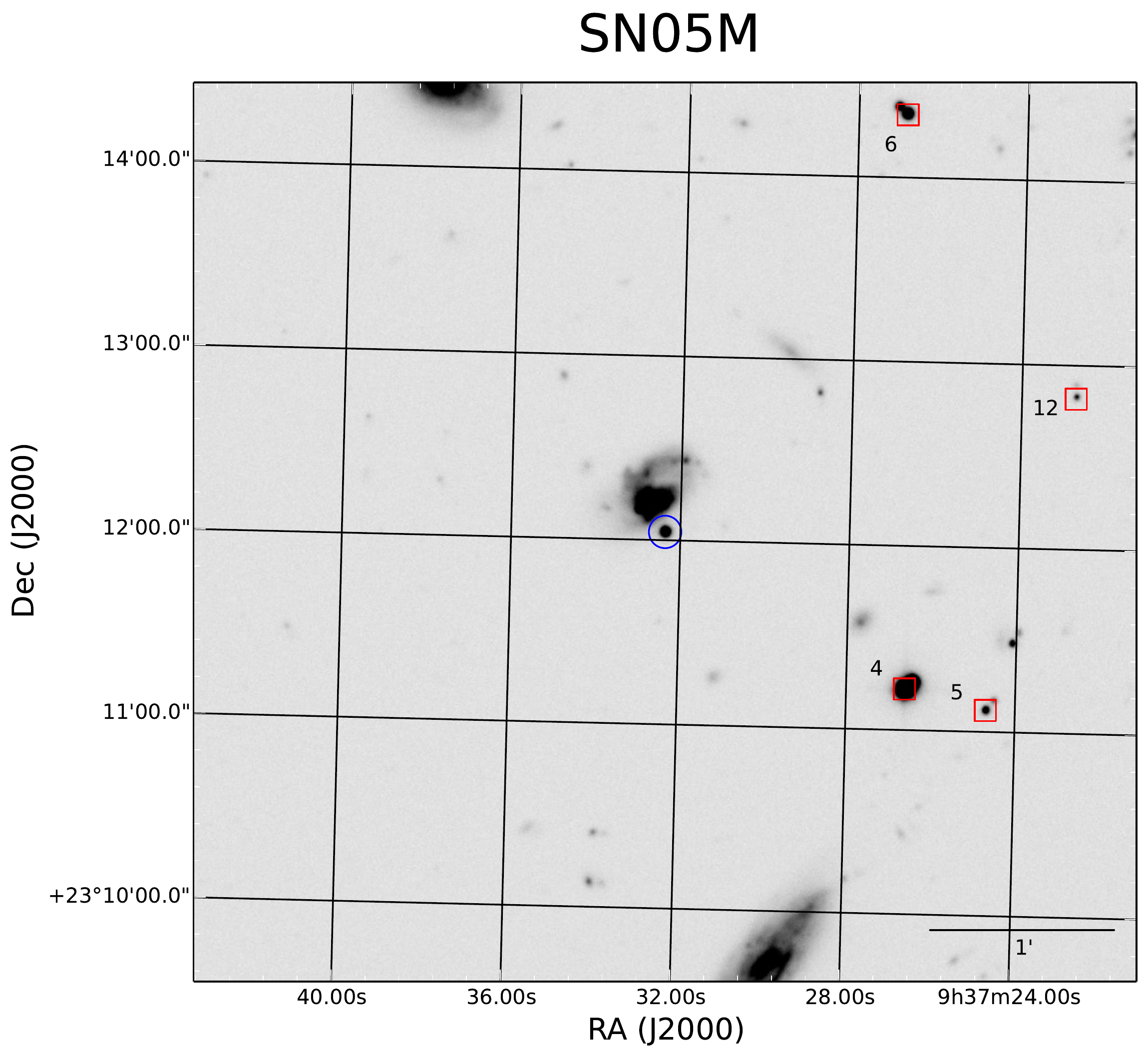}{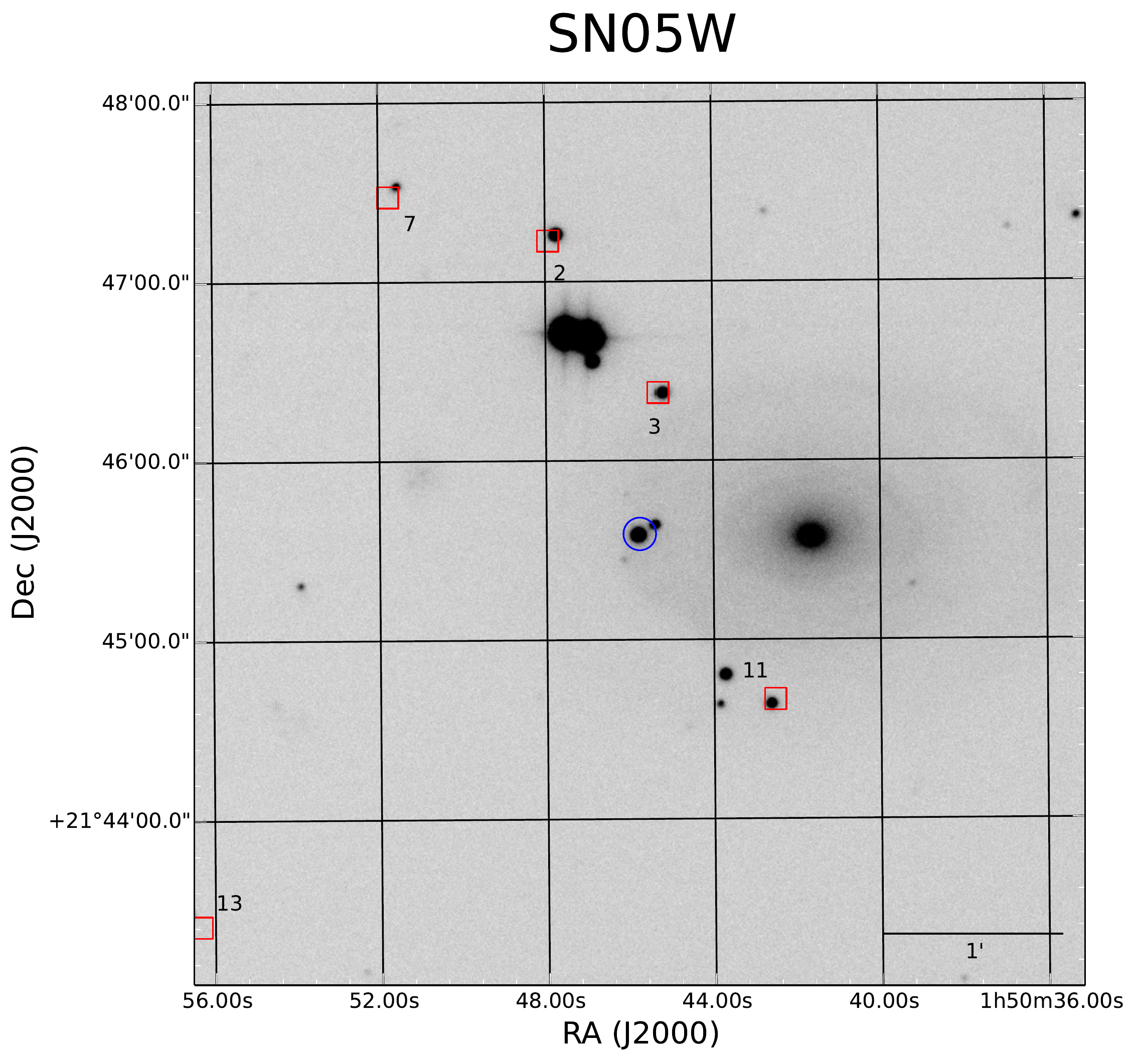}
\plottwo{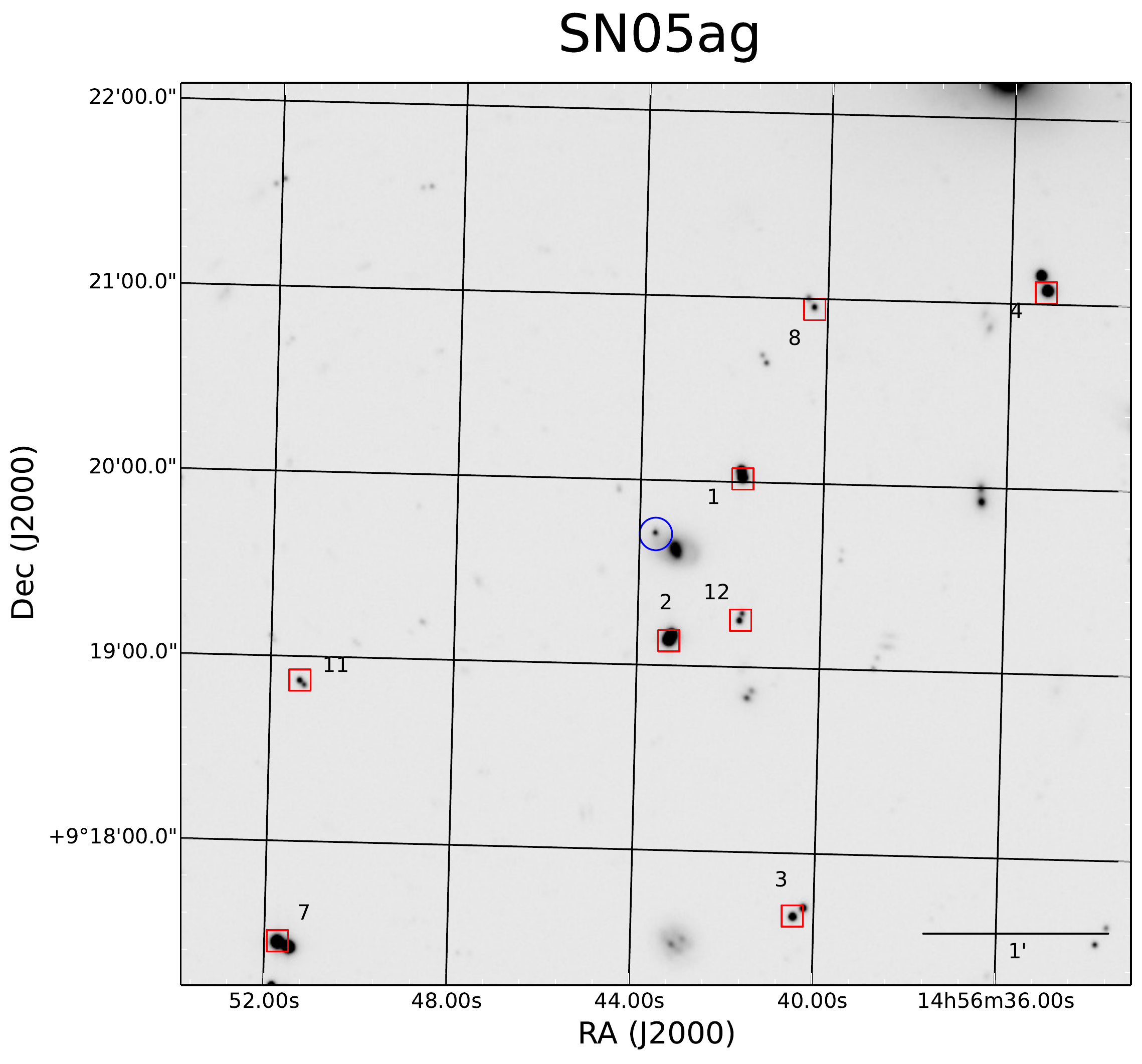}{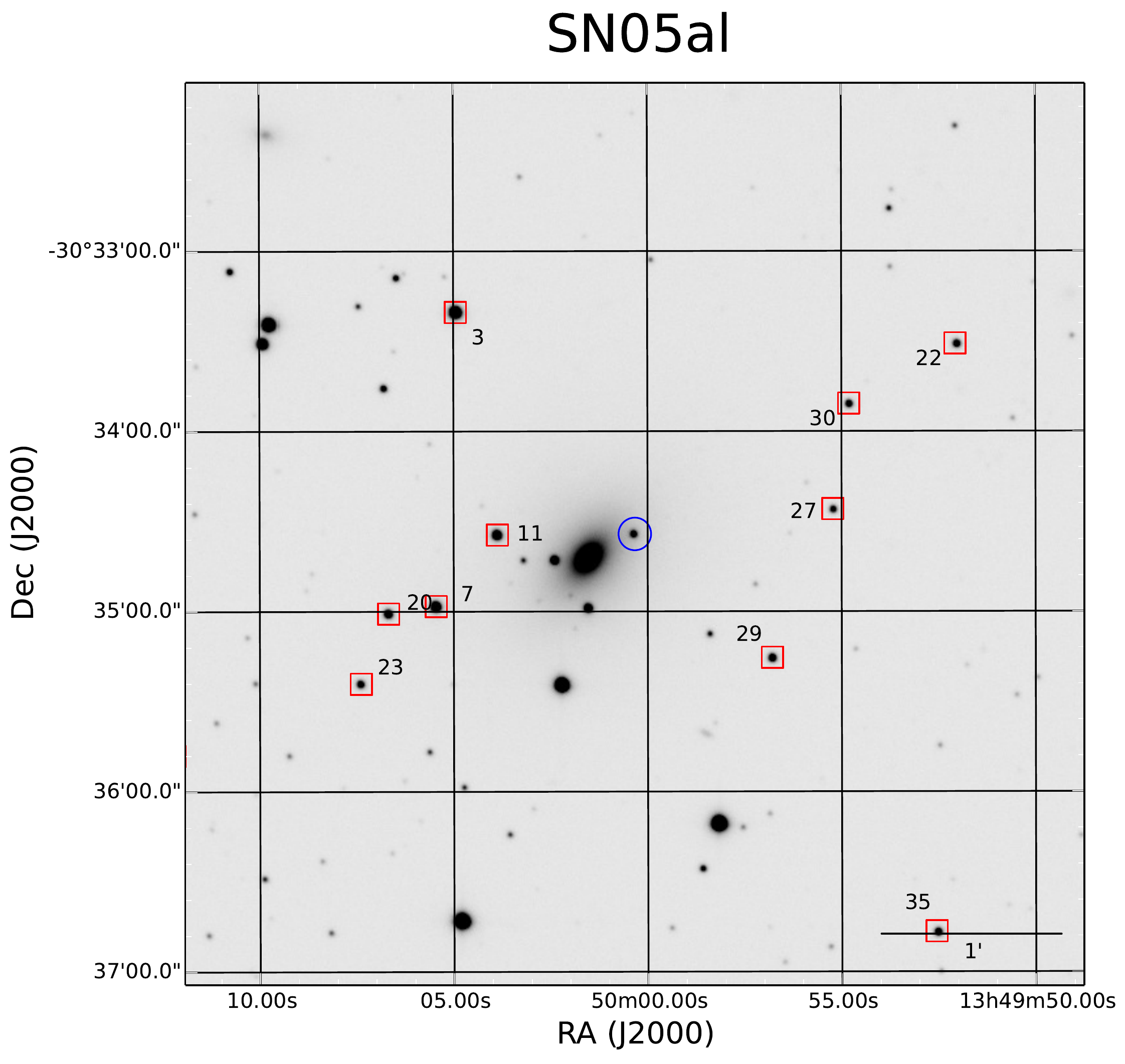}
\newline
\plottwo{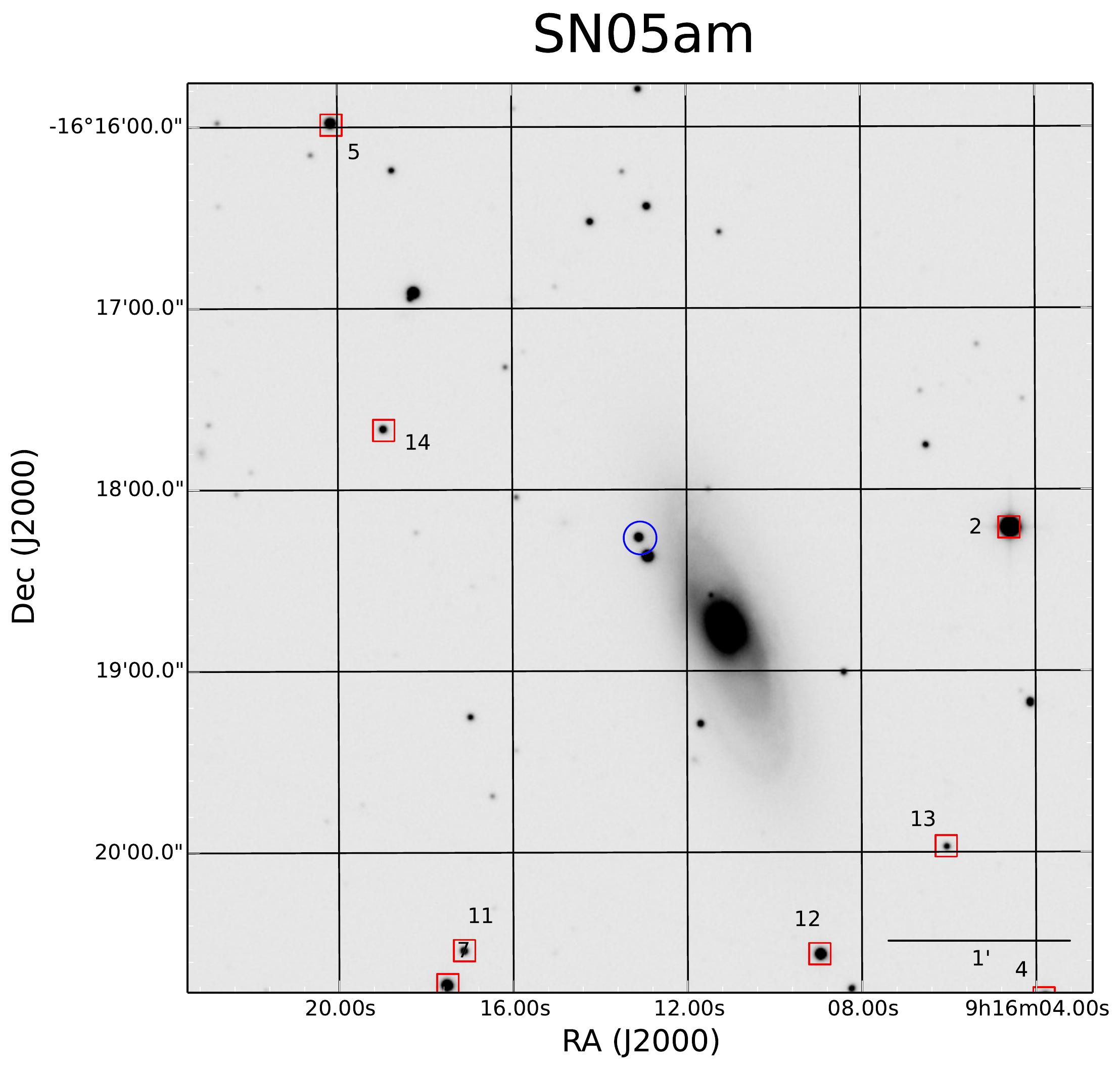}{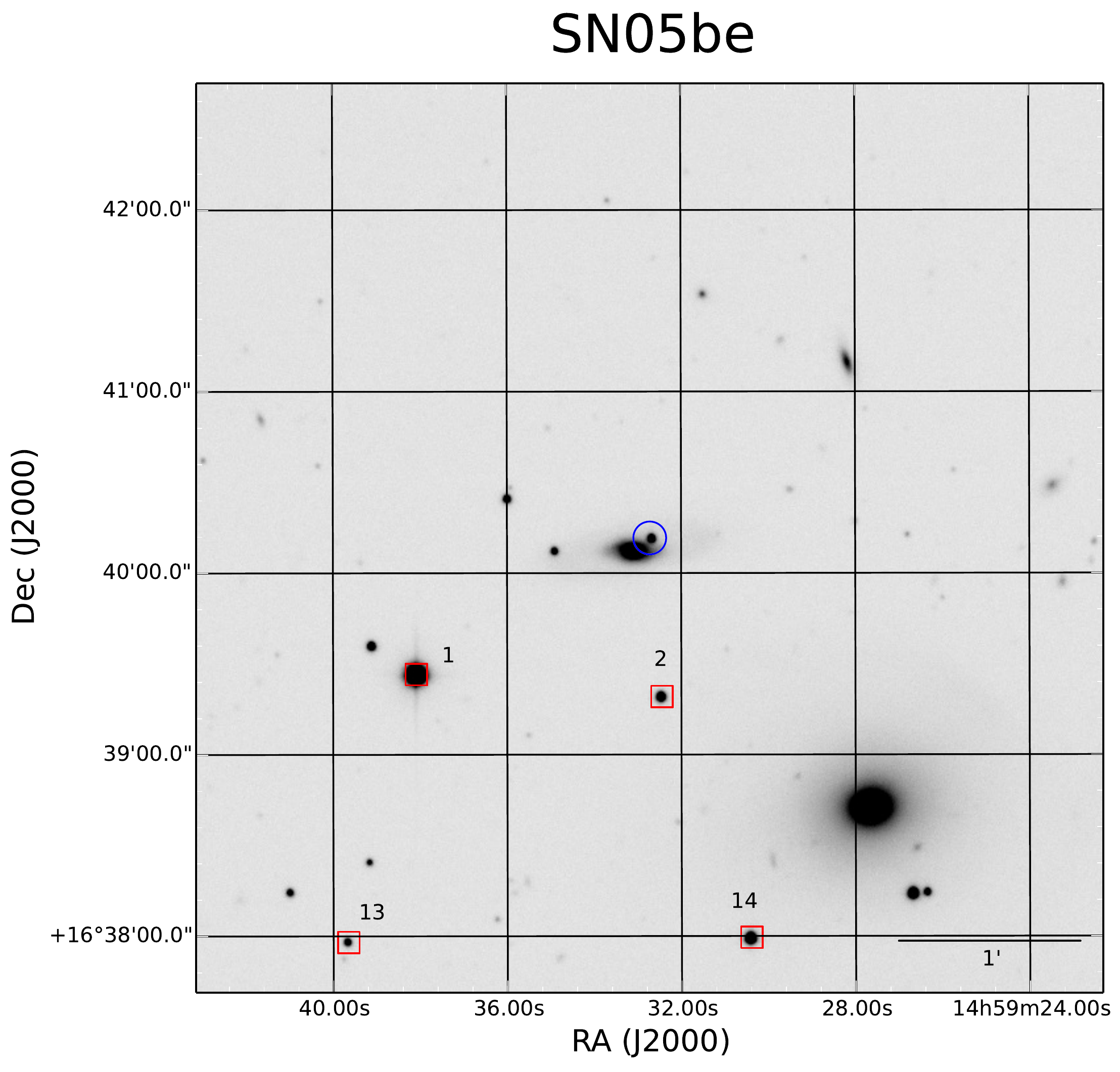}
\plottwo{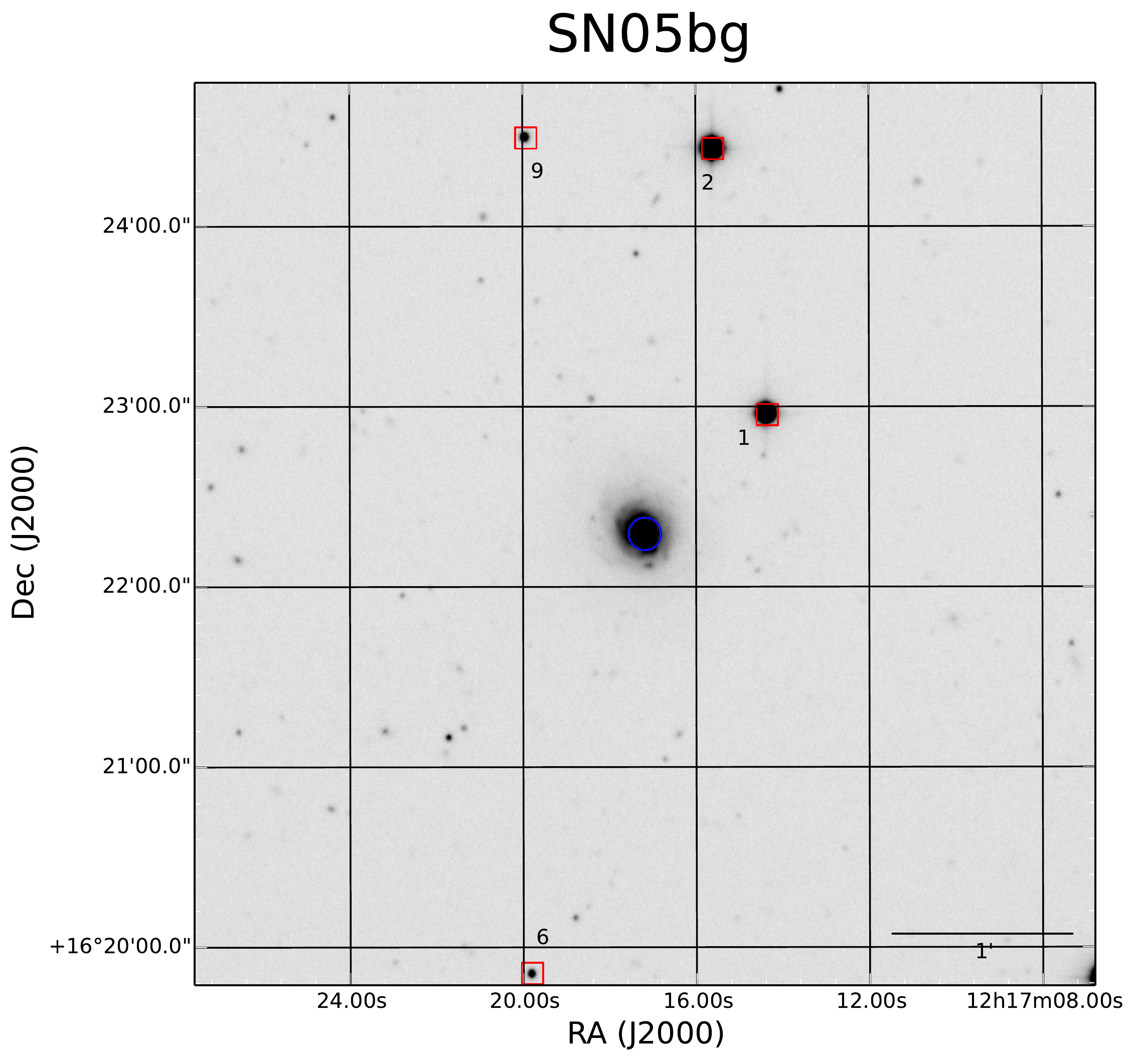}{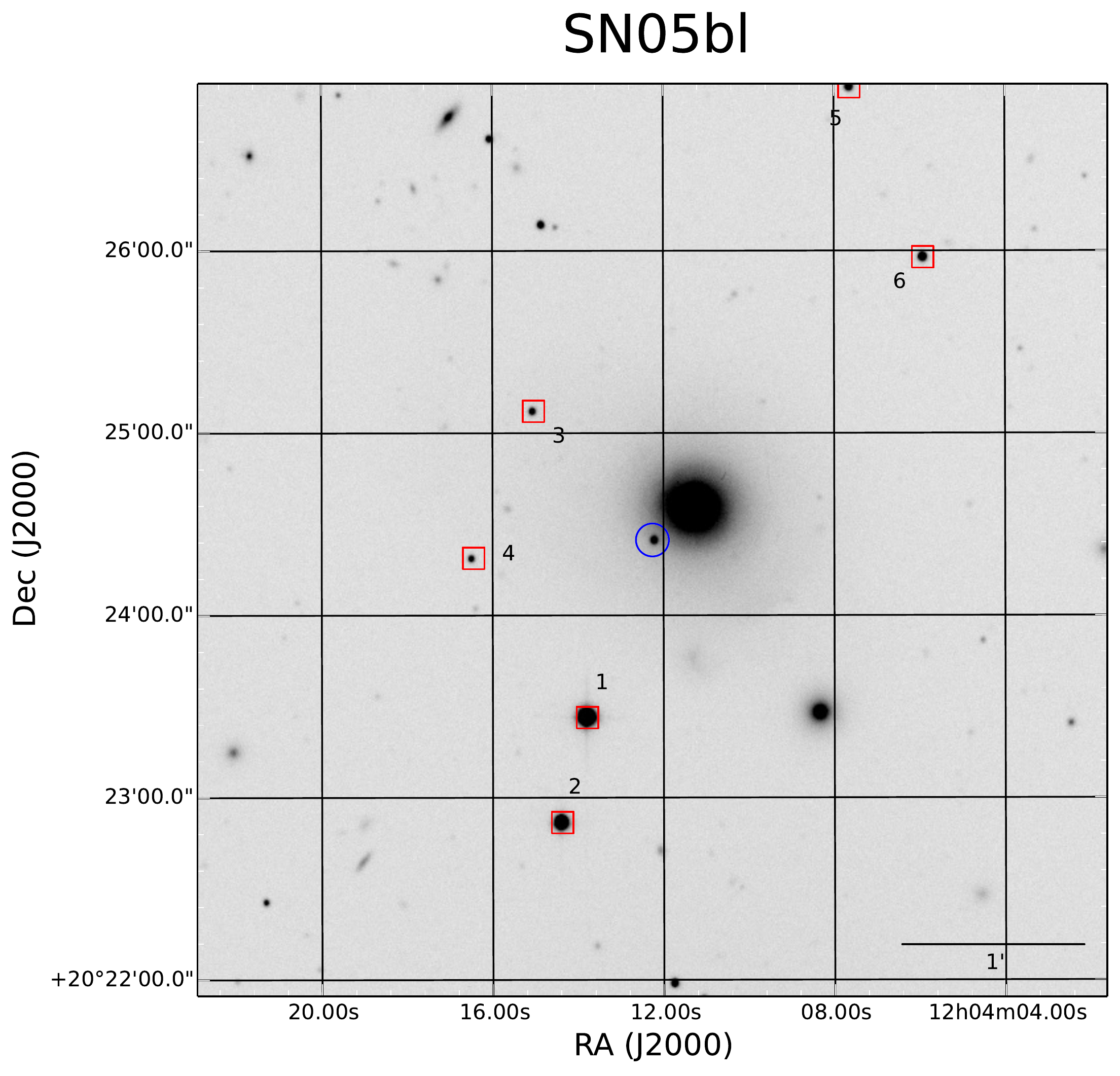}
\newline
\plottwo{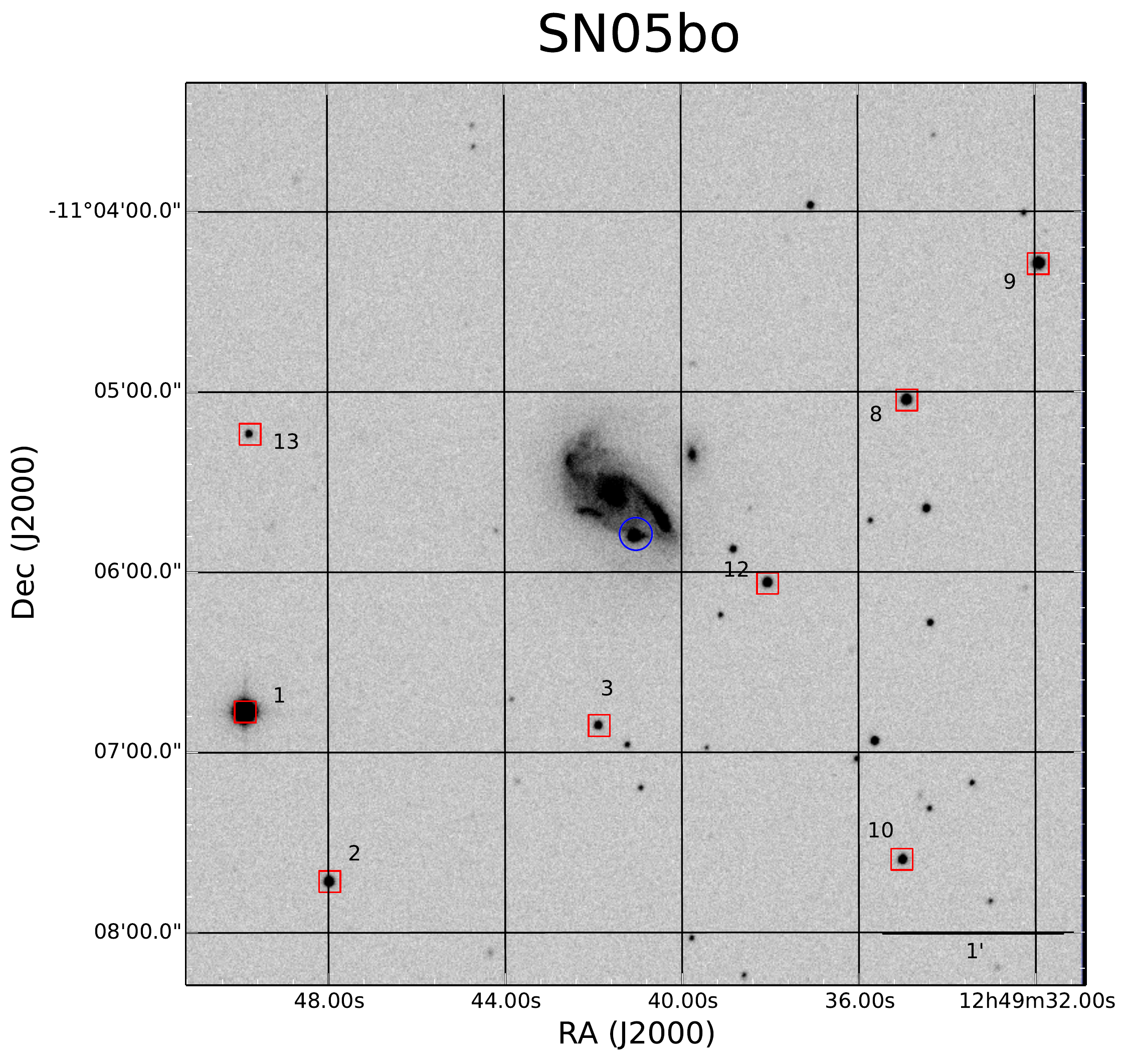}{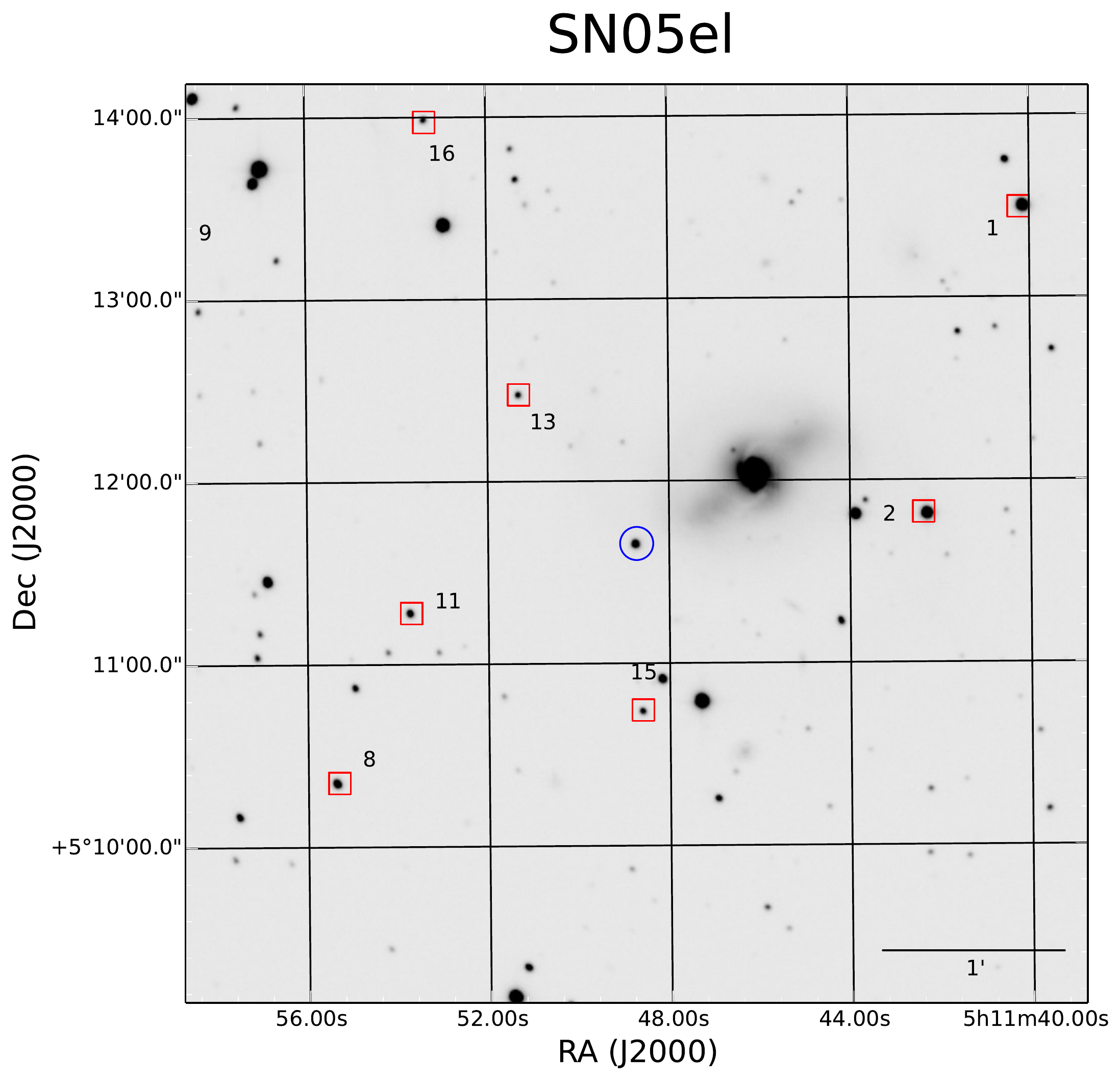}
\plottwo{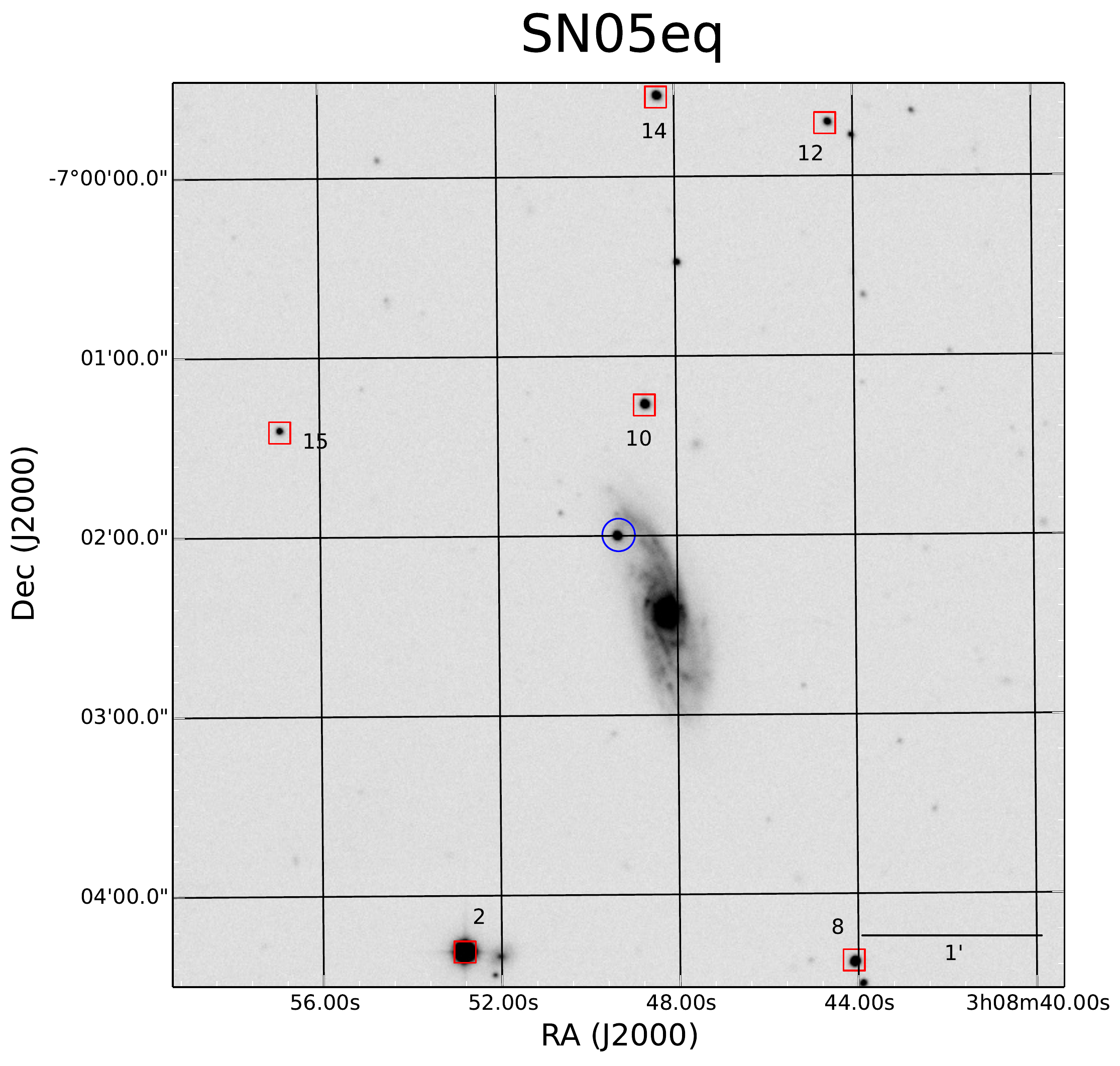}{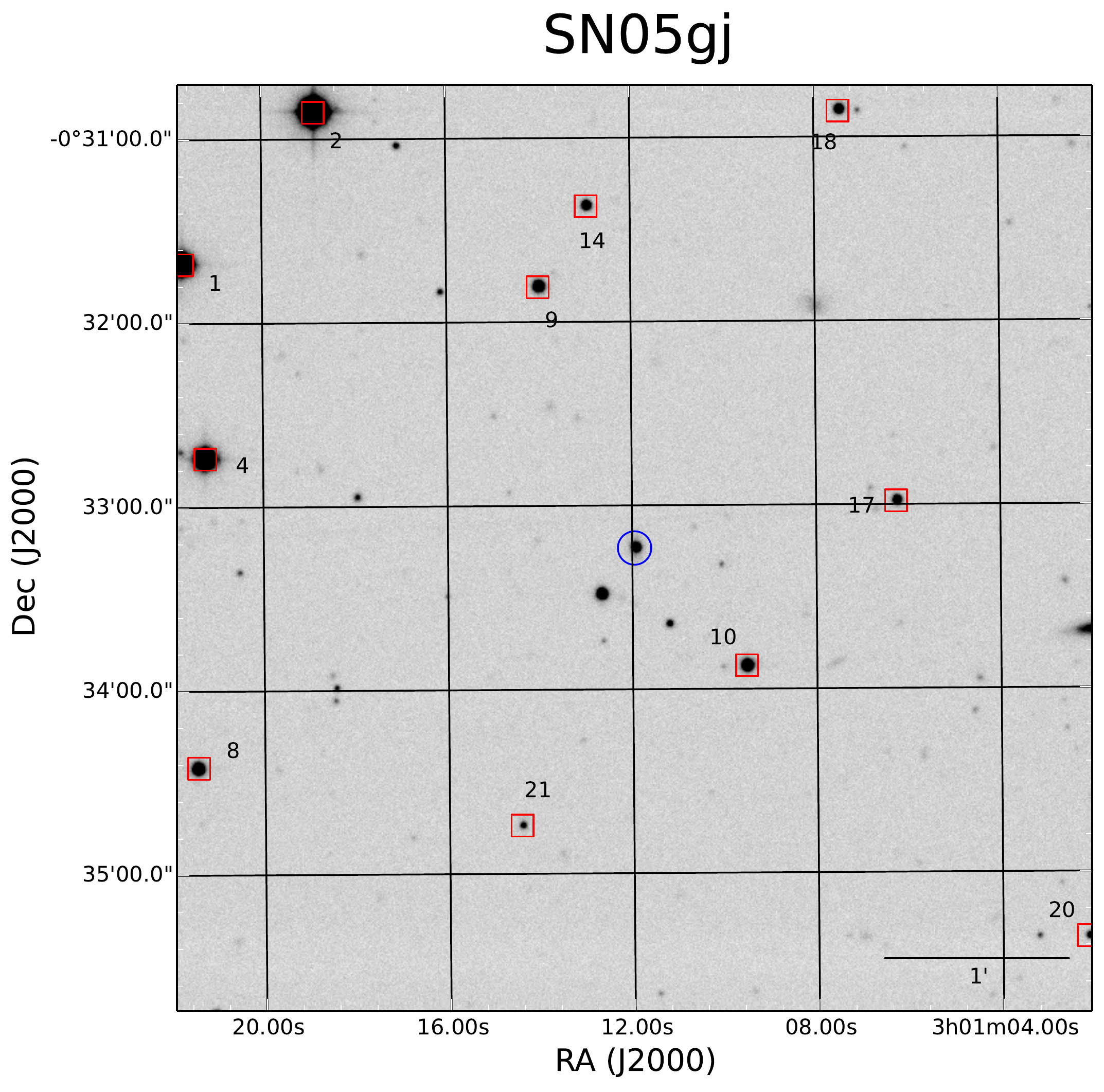}
{\center Krisciunas {\it et al.} Fig.~\ref{fig:fcharts}}
\end{figure}
\clearpage
\newpage

\begin{figure}[t]
\epsscale{1.0}
\plotone{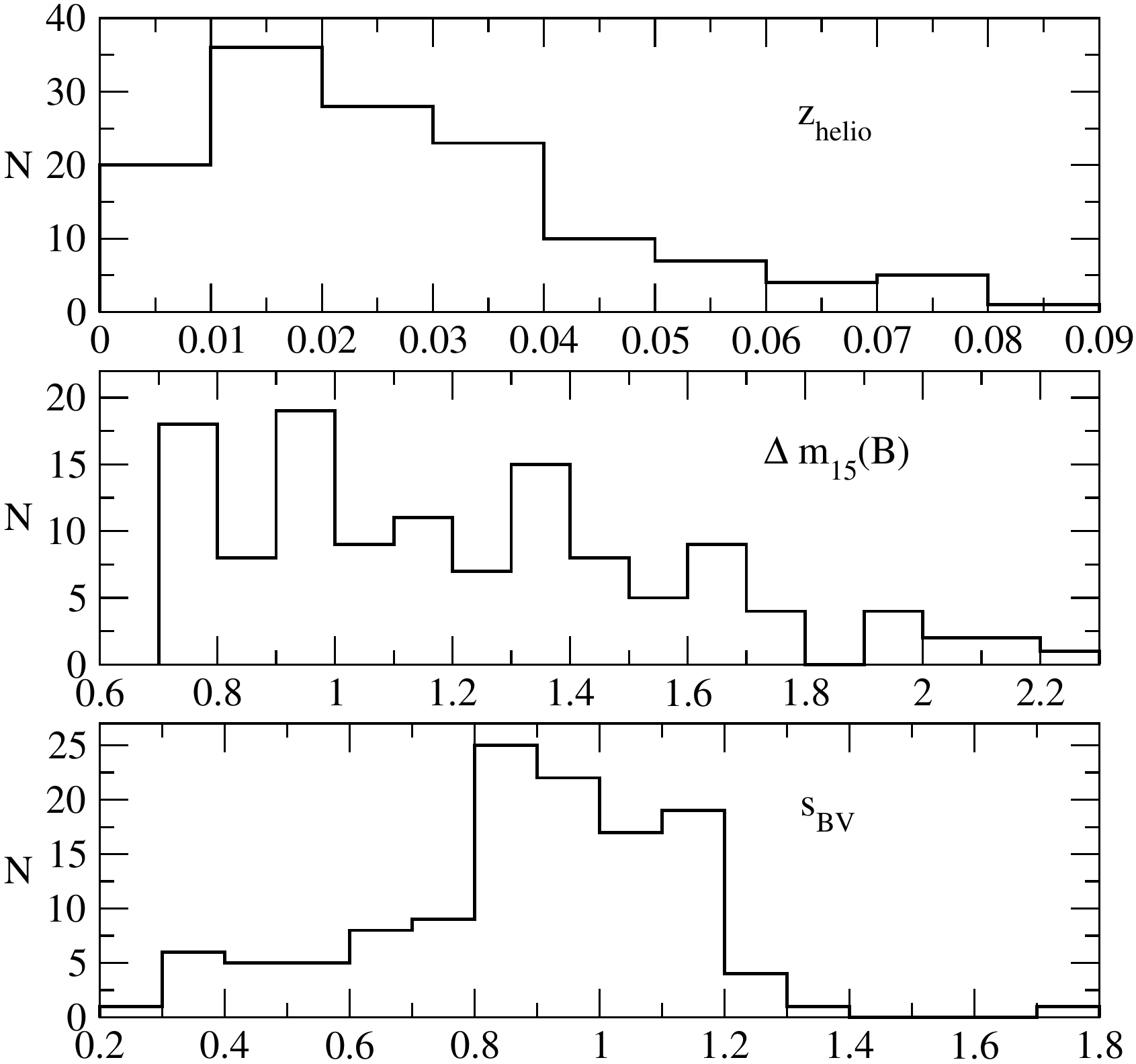}{\center Krisciunas {\it et al.}
Fig.~\ref{fig:histograms}}
\end{figure}


\begin{figure}[t]
\epsscale{1.0}
\plotone{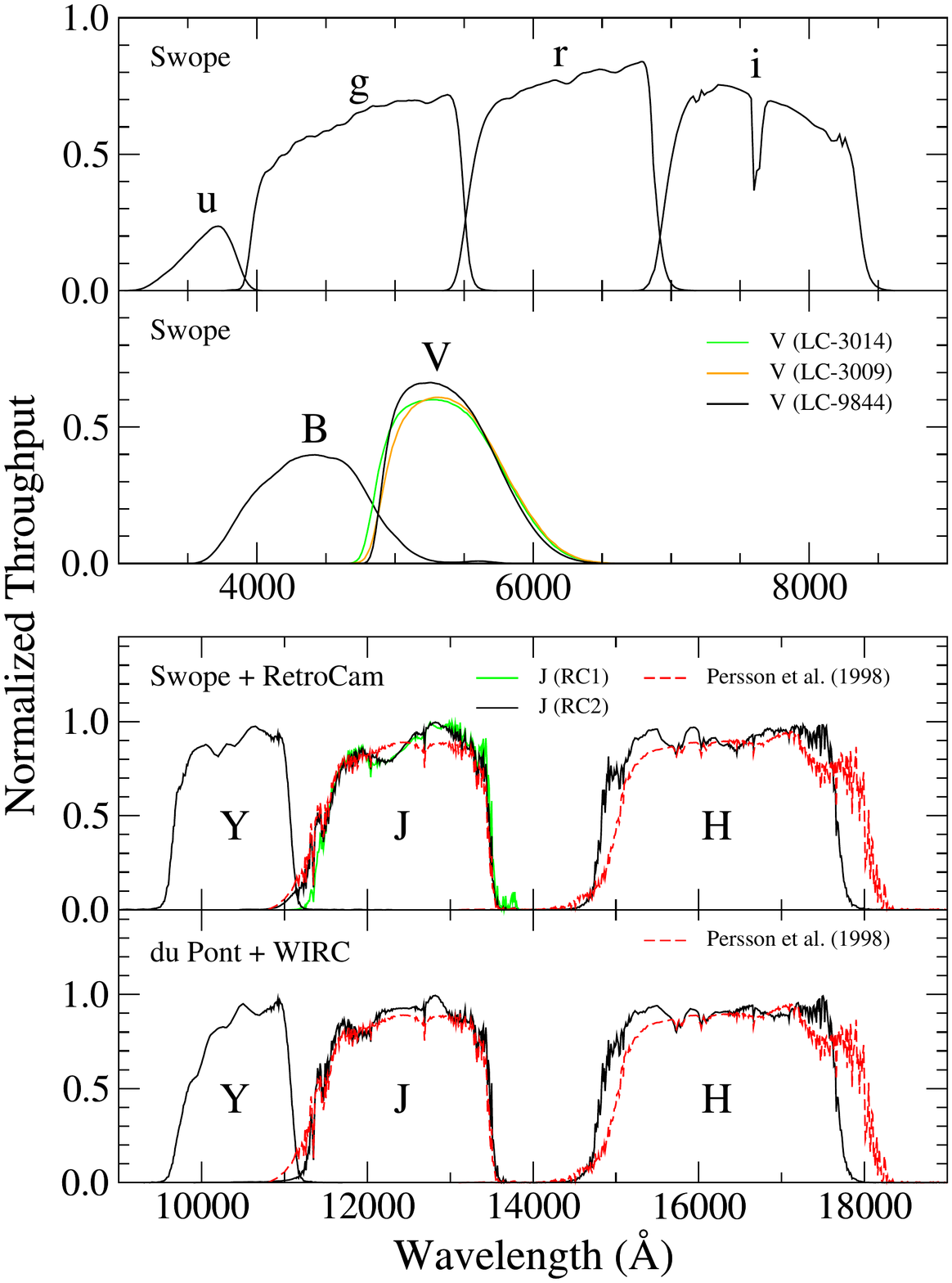}{\center Krisciunas {\it et al.} 
Fig.~\ref{fig:filters}}
\end{figure}


\begin{figure}[t]
\plotone{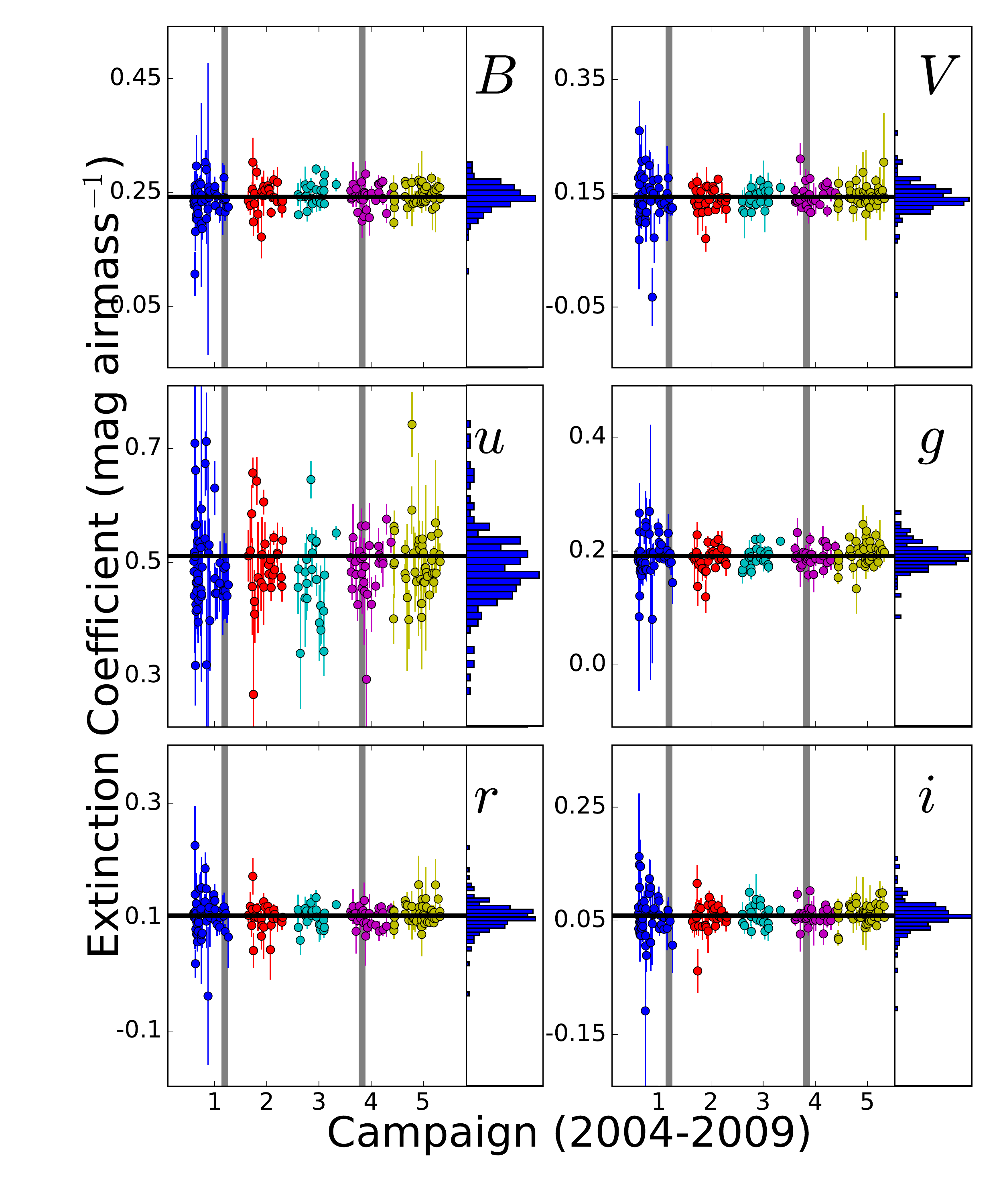}{\center Krisciunas {\it et al.} 
Fig.~\ref{fig:extinction}}
\end{figure}


\begin{figure}[t]
\plotone{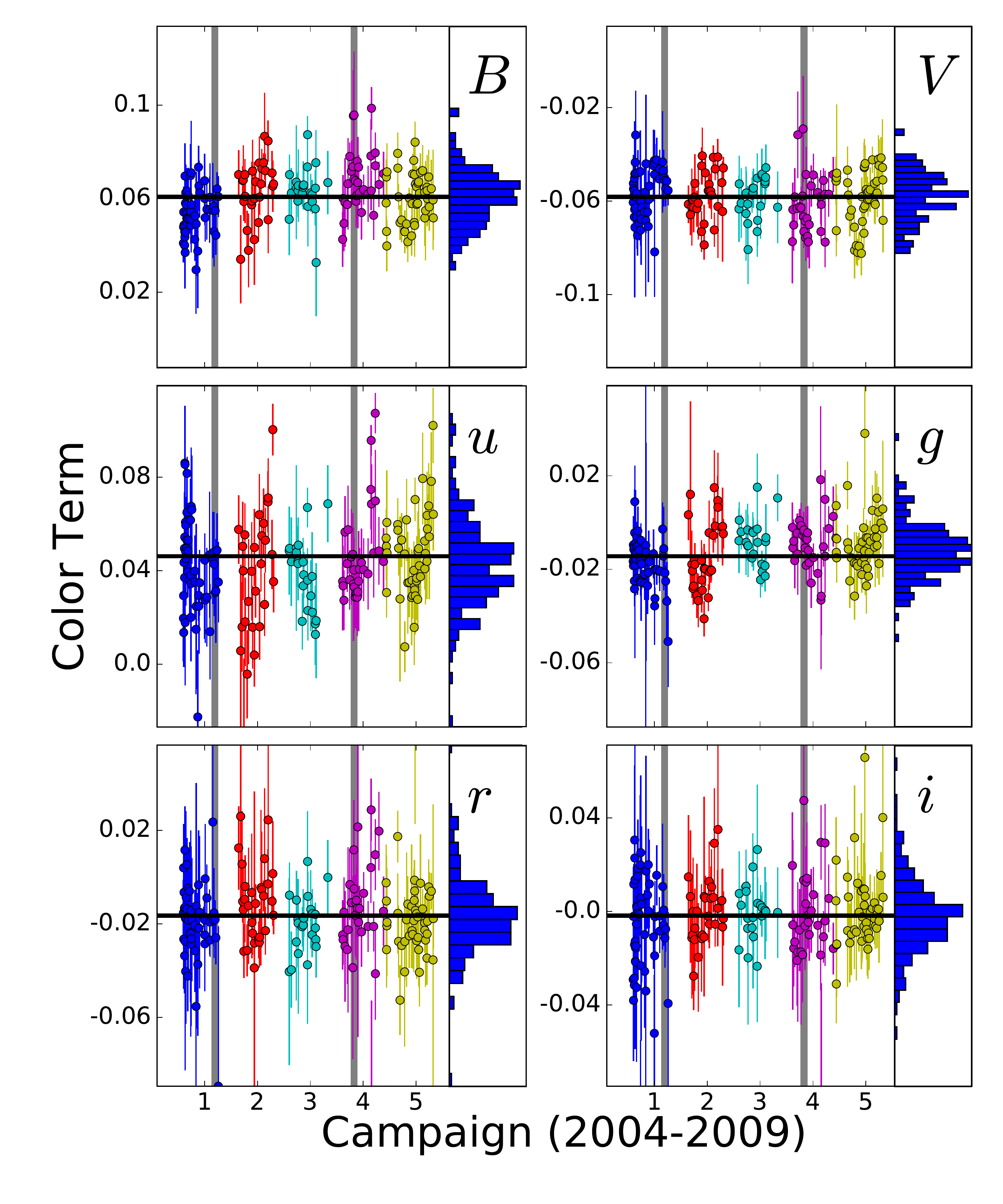}{\center Krisciunas {\it et al.} 
Fig.~\ref{fig:CT}}
\end{figure}


\begin{figure}[t]
\epsscale{1.0}
\plotone{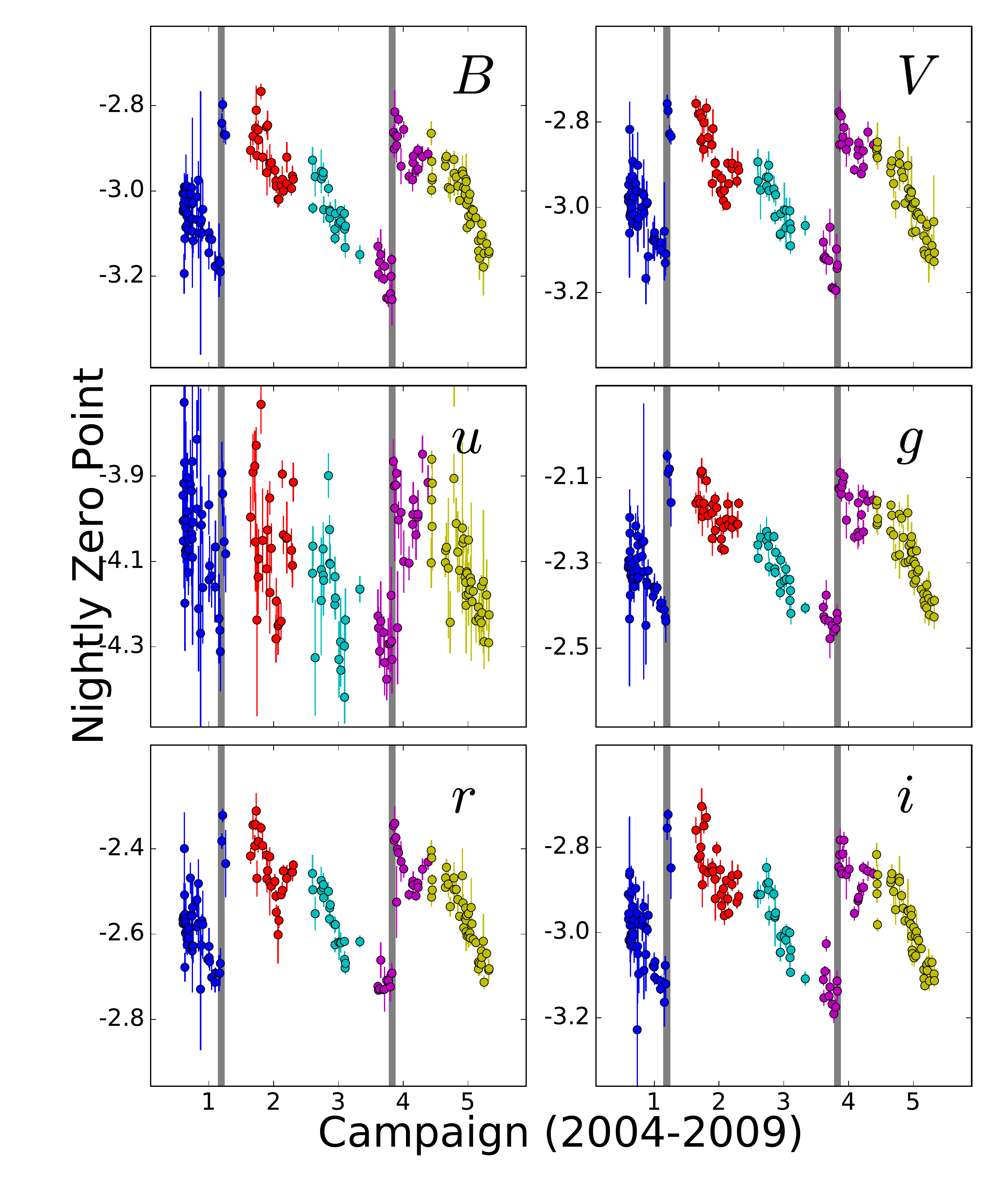}{\center Krisciunas {\it et al.} 
Fig.~\ref{fig:ZP}}
\end{figure}



\begin{figure}[t]
\epsscale{1.0}
\plotone{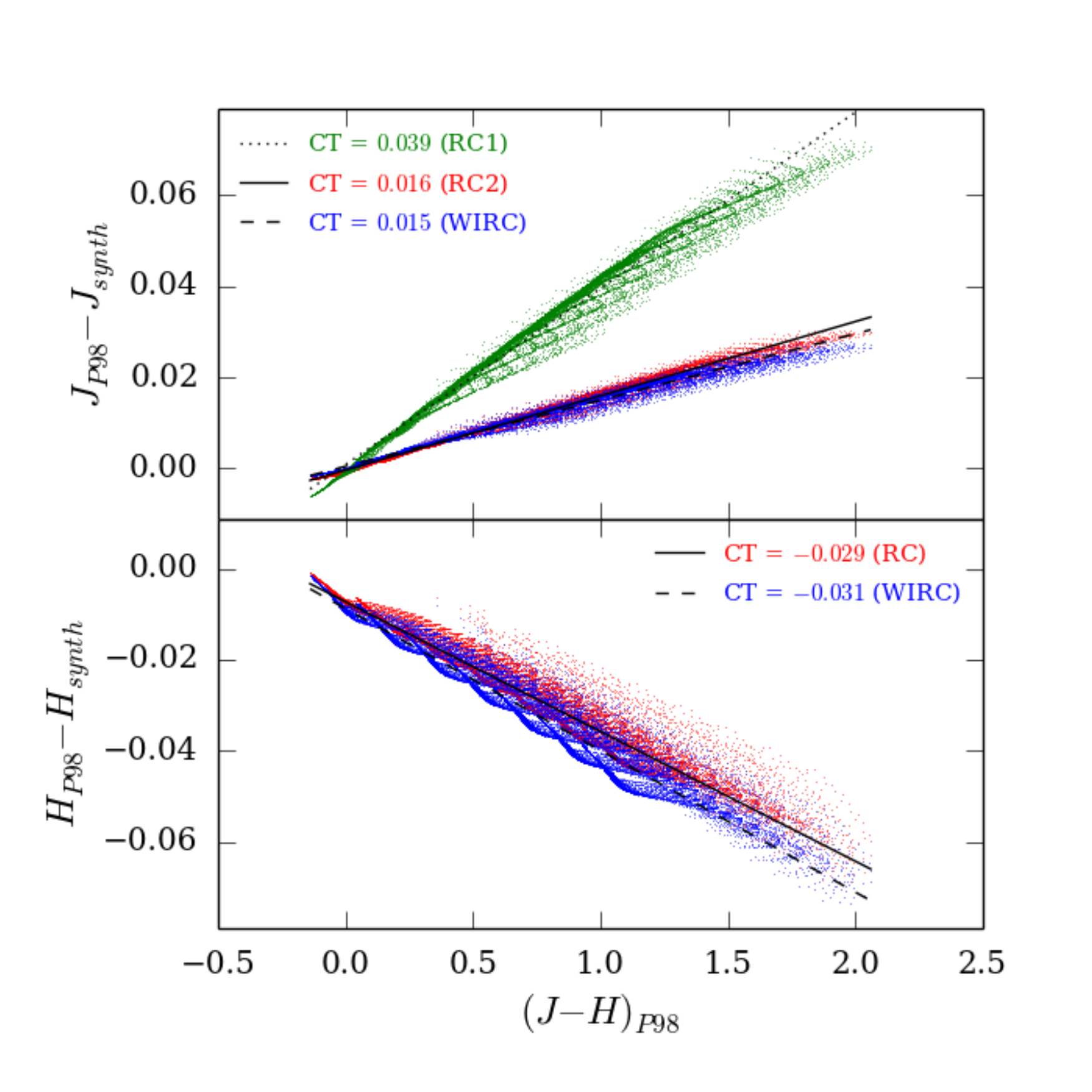}{\center Krisciunas {\it et al.} 
Fig.~\ref{fig:castelli_cts}}
\end{figure}


\begin{figure}[t]
\epsscale{1.0}
\plotone{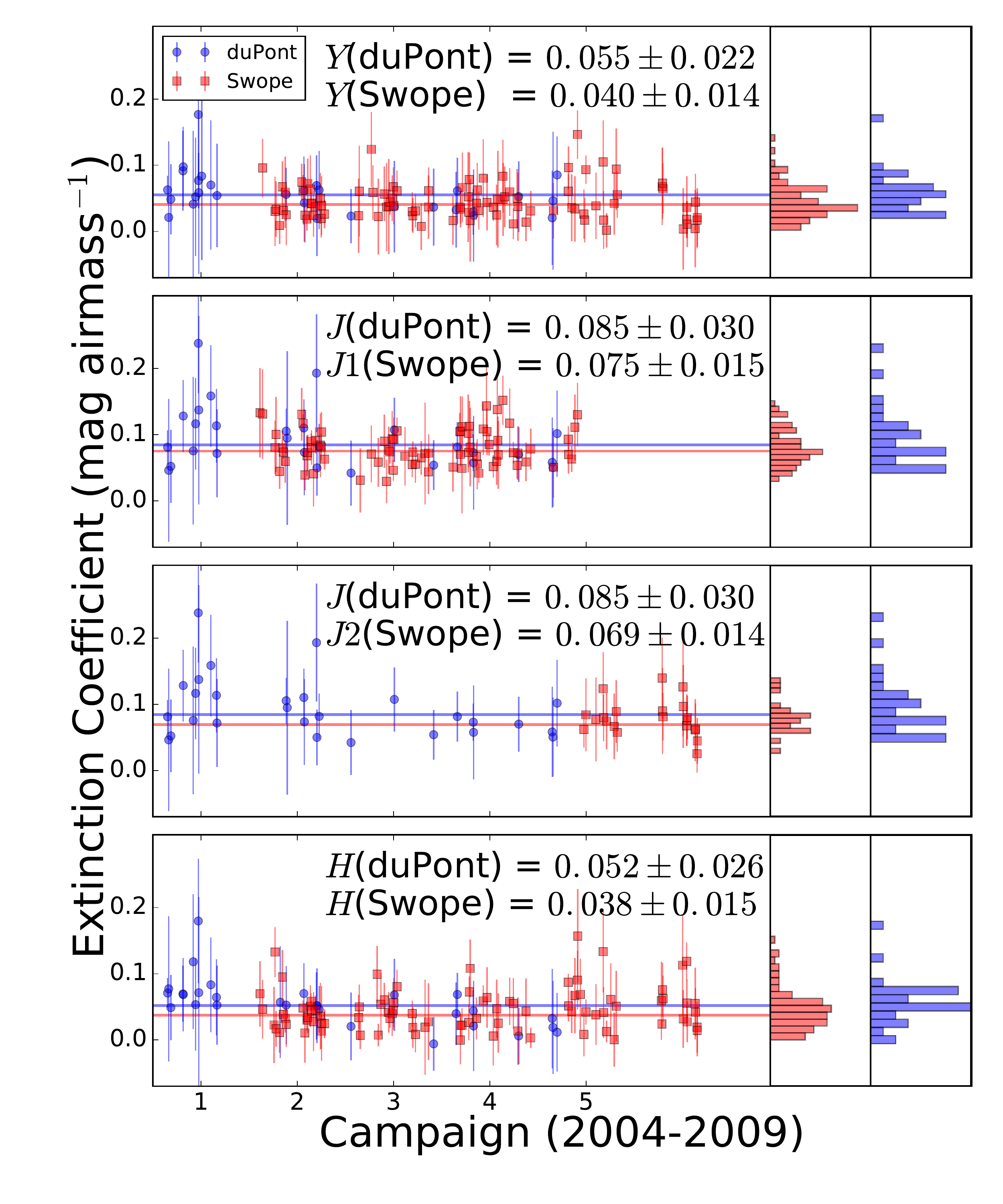}{\center Krisciunas {\it et al.} 
Fig.~\ref{fig:nir_extinction}}
\end{figure}



\begin{figure}[t]
\epsscale{1.0}
\plotone{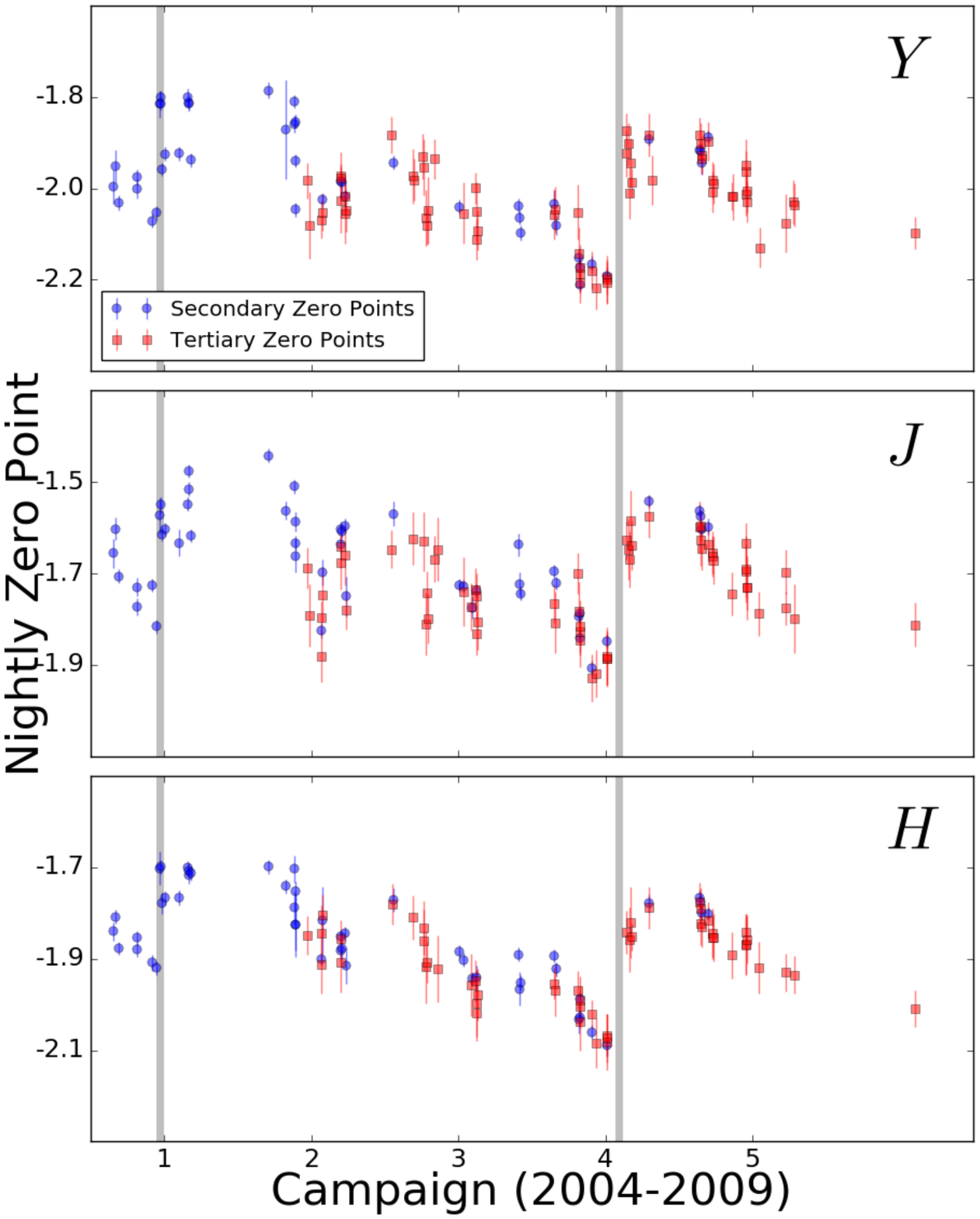}{\center Krisciunas {\it et al.} 
Fig.~\ref{fig:wirc_zero_pts}}
\end{figure}


\begin{figure}[t]
\epsscale{1.0}
\plotone{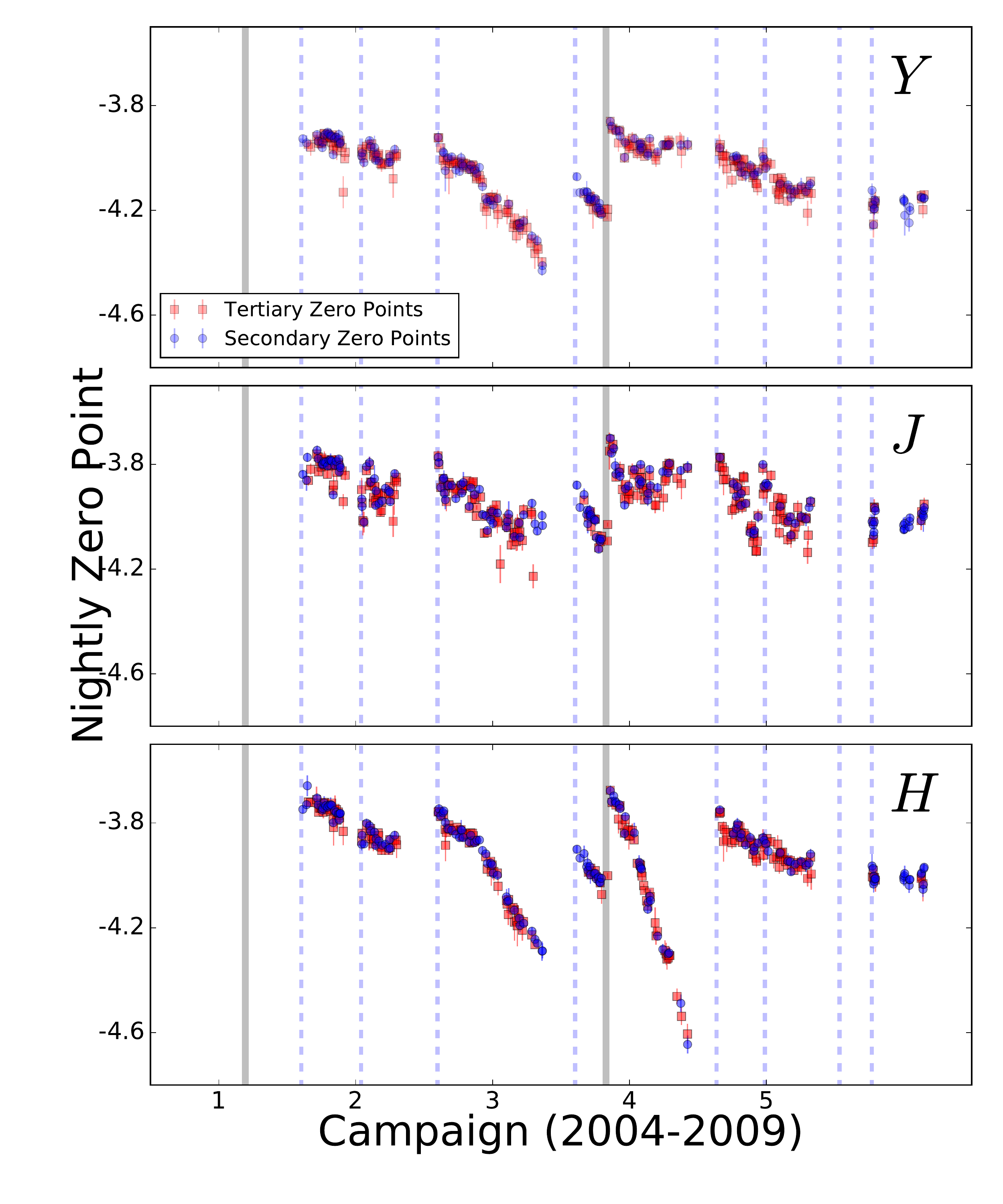}{\center Krisciunas {\it et al.} 
Fig.~\ref{fig:rc_zero_pts}}
\end{figure}


\begin{figure}[t]
\epsscale{1.0}
\plotone{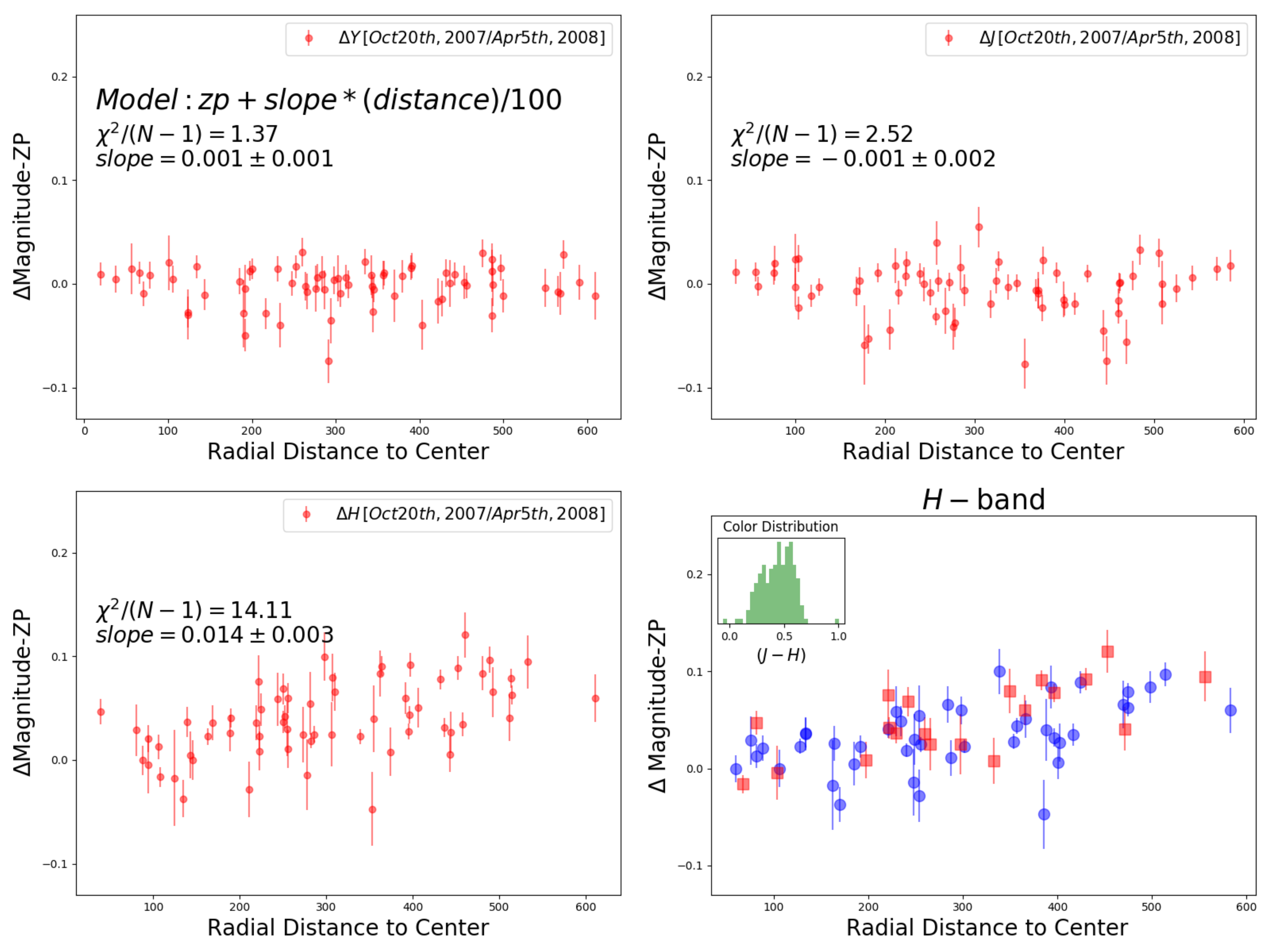}{\center Krisciunas {\it et al.} 
Fig.~\ref{fig:phot_diff_campaign4}}
\end{figure}


\begin{figure}[t]
\epsscale{1.0}
\plotone{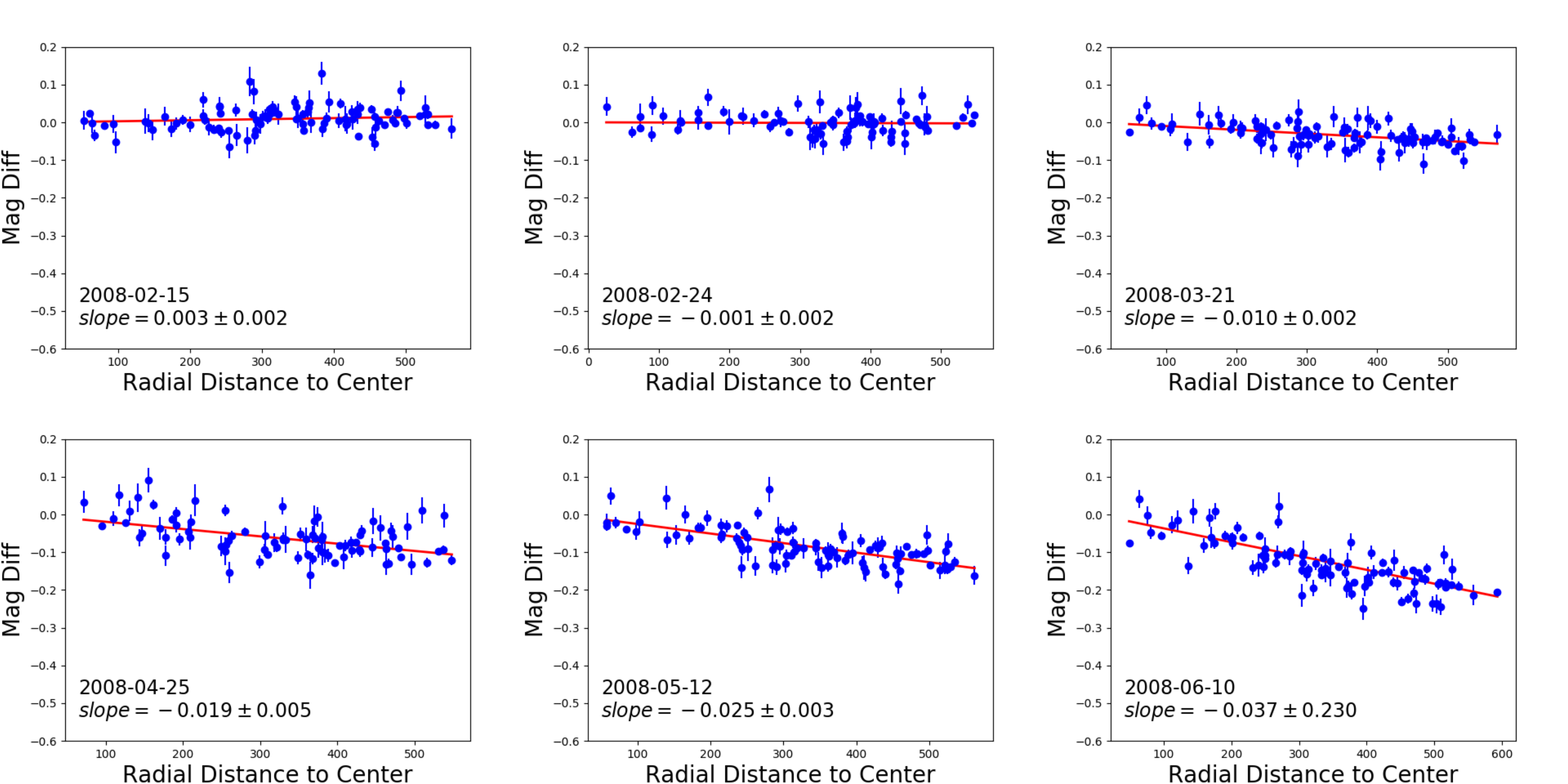}{\center Krisciunas {\it et al.} 
Fig.~\ref{fig:mag_diff_slopes}}
\end{figure}


\begin{figure}[t]
\epsscale{1.1}
\plottwo{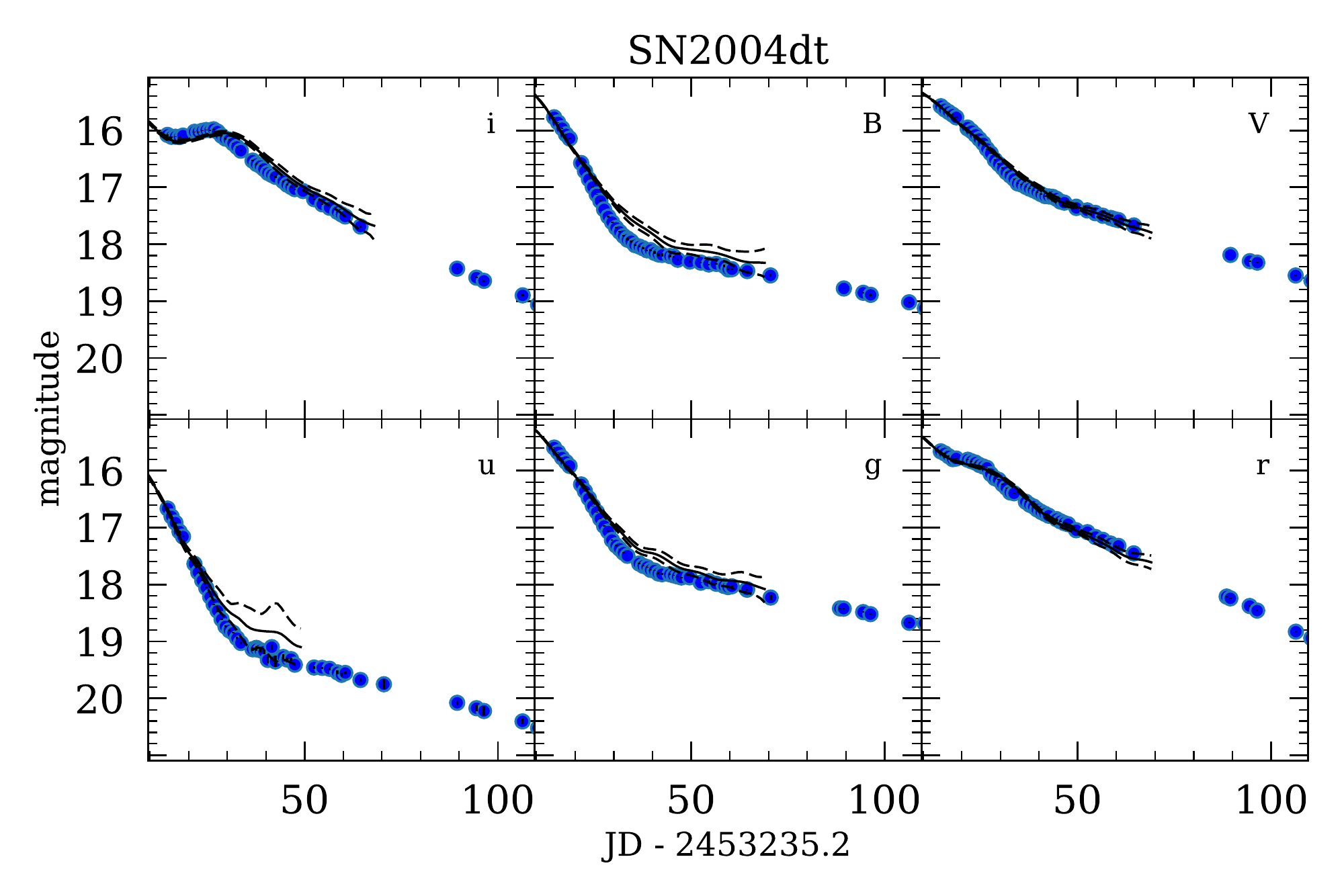}{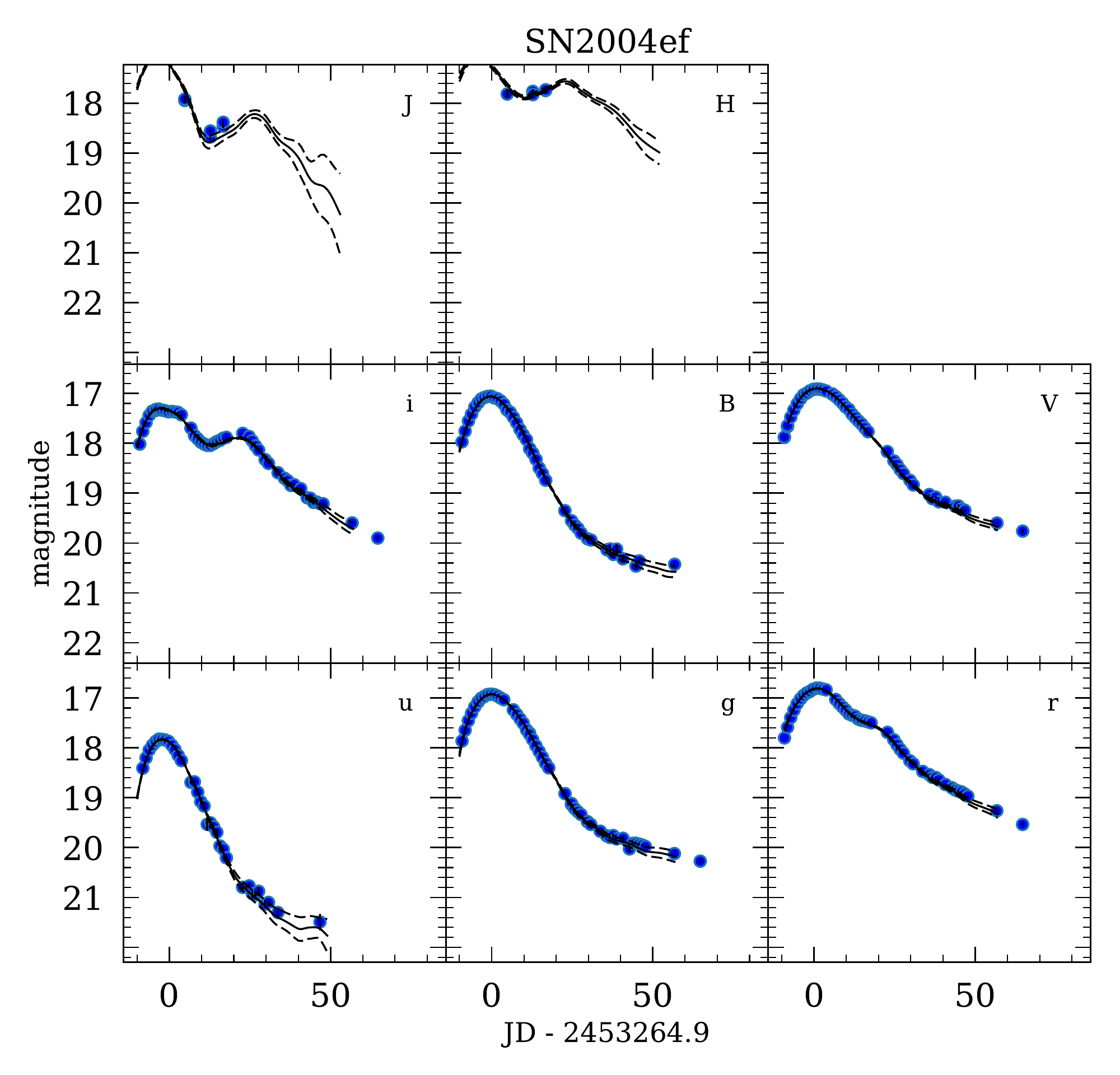}
\newline
\plottwo{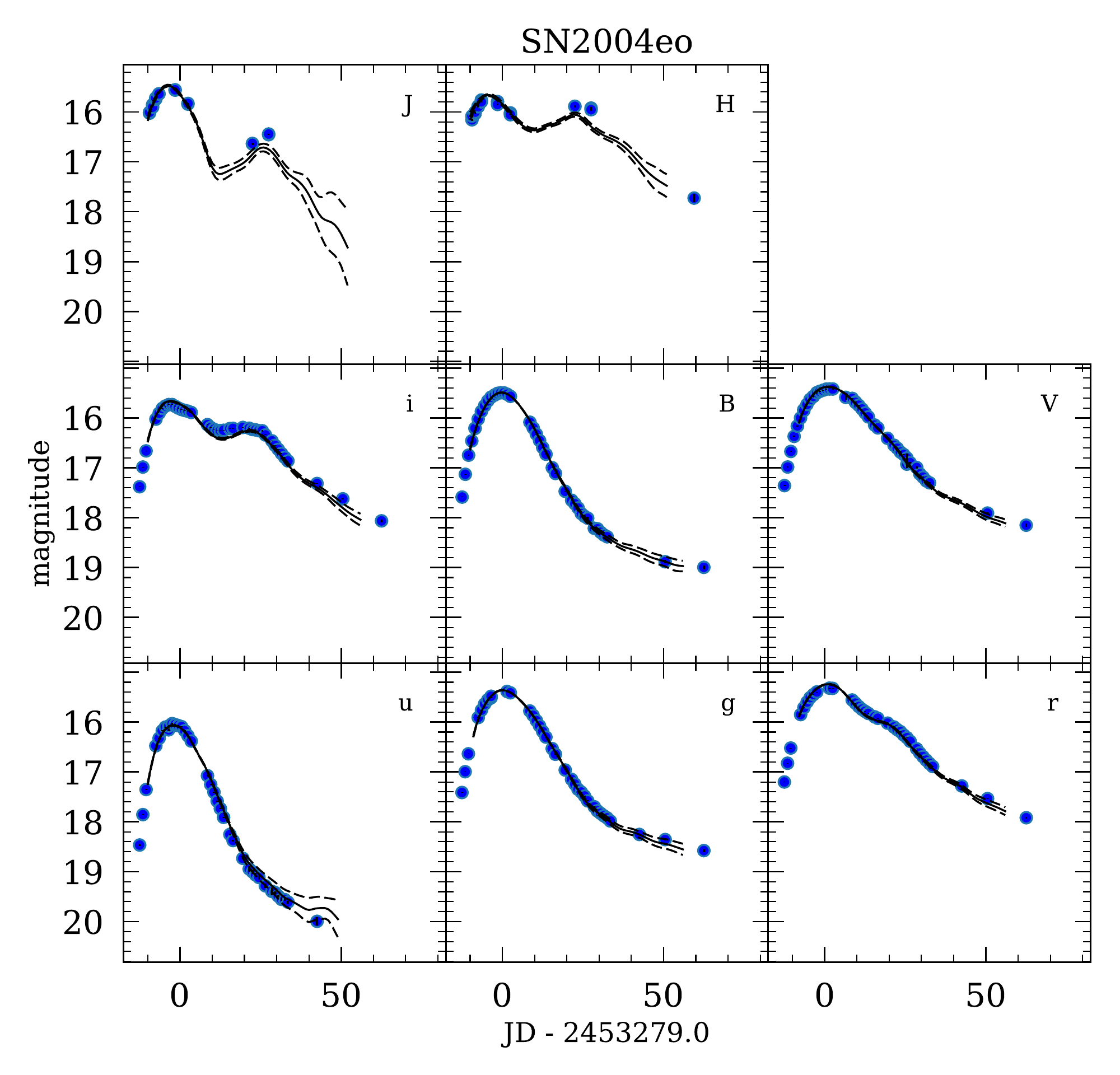}{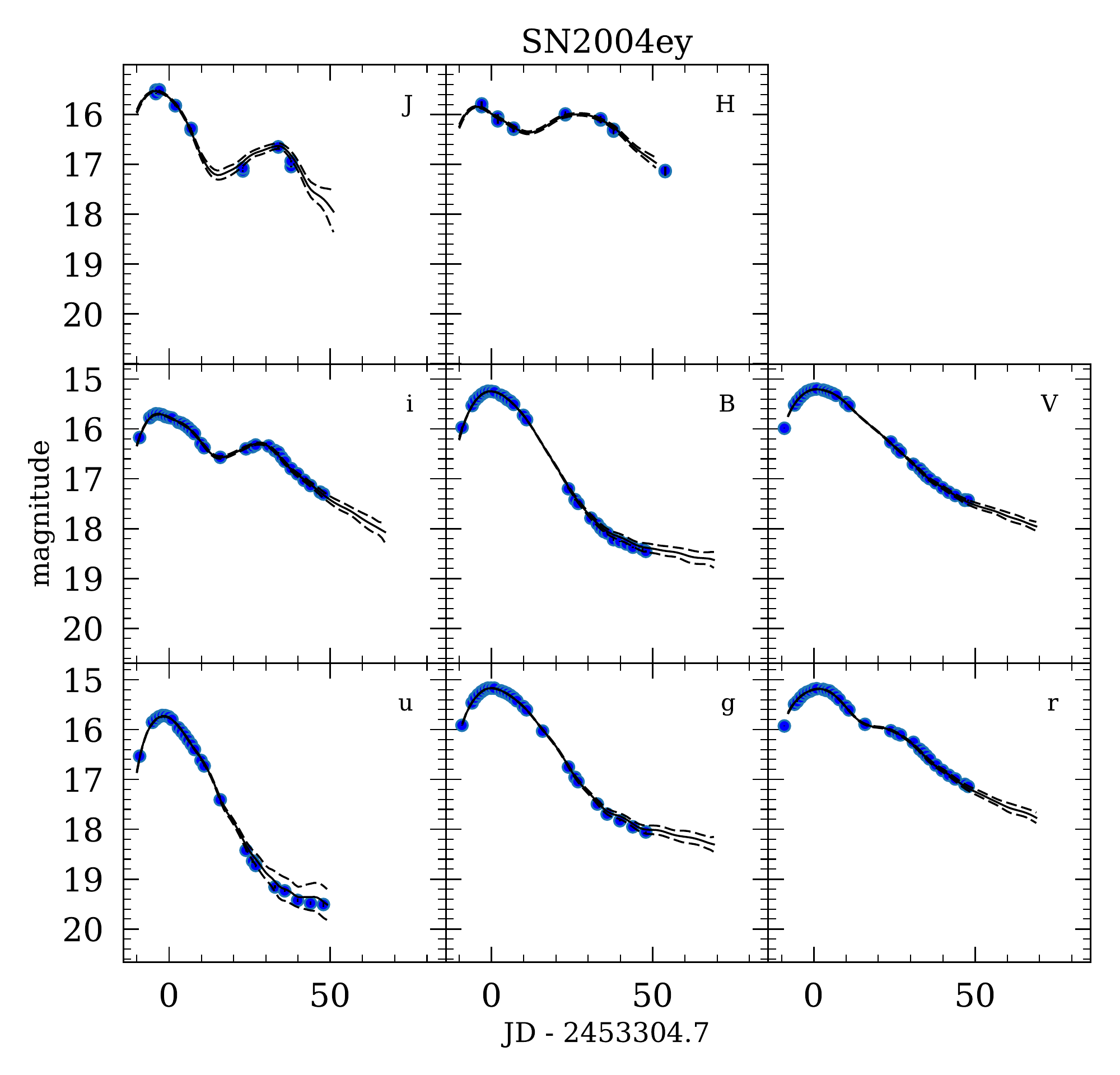}
{\center Krisciunas {\it et al.} Fig.~\ref{fig:sample_lcs}}
\end{figure}
\clearpage
\newpage

\begin{figure}[t]
\epsscale{1.1}
\plottwo{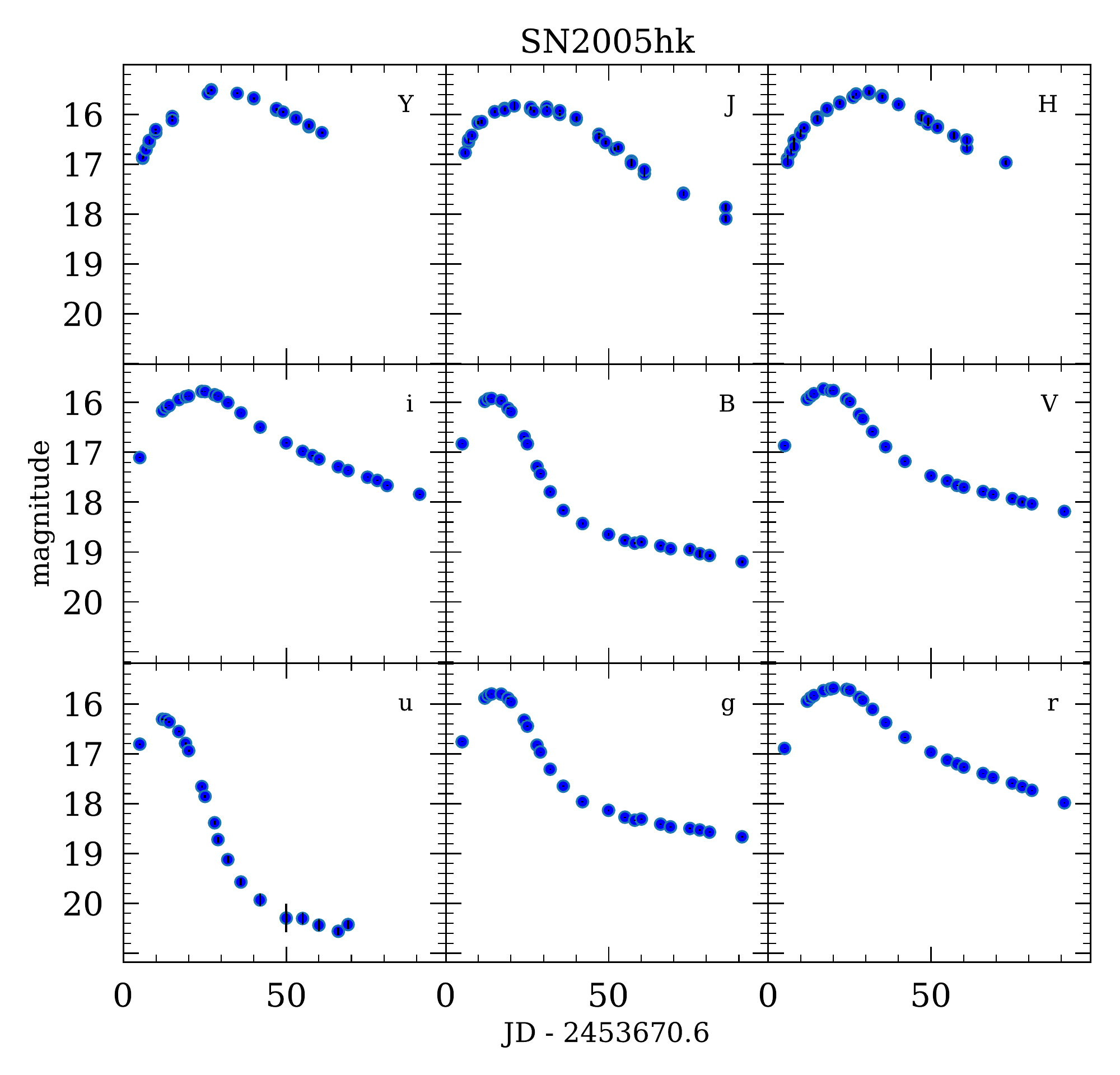}{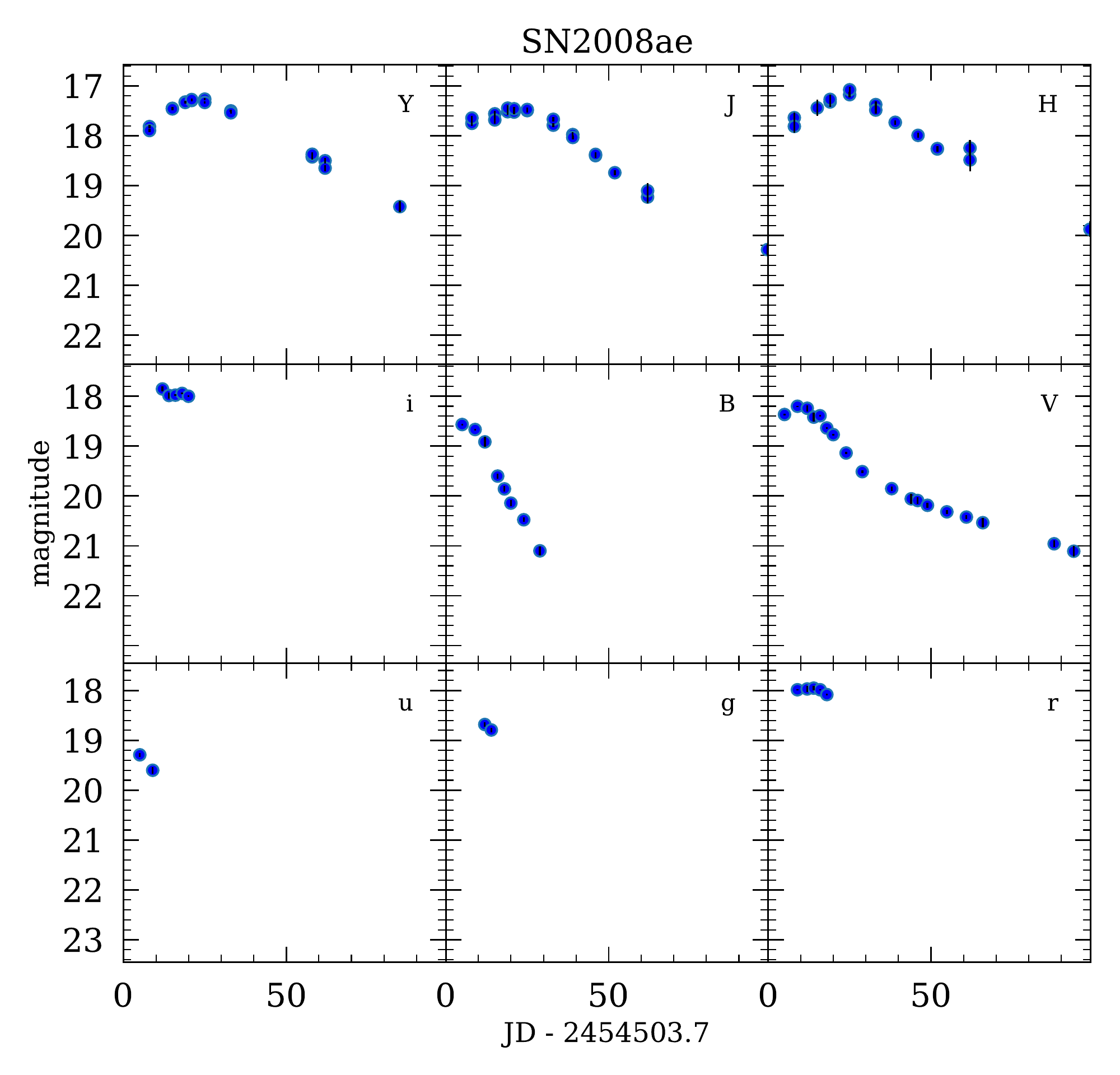}
\newline
\plottwo{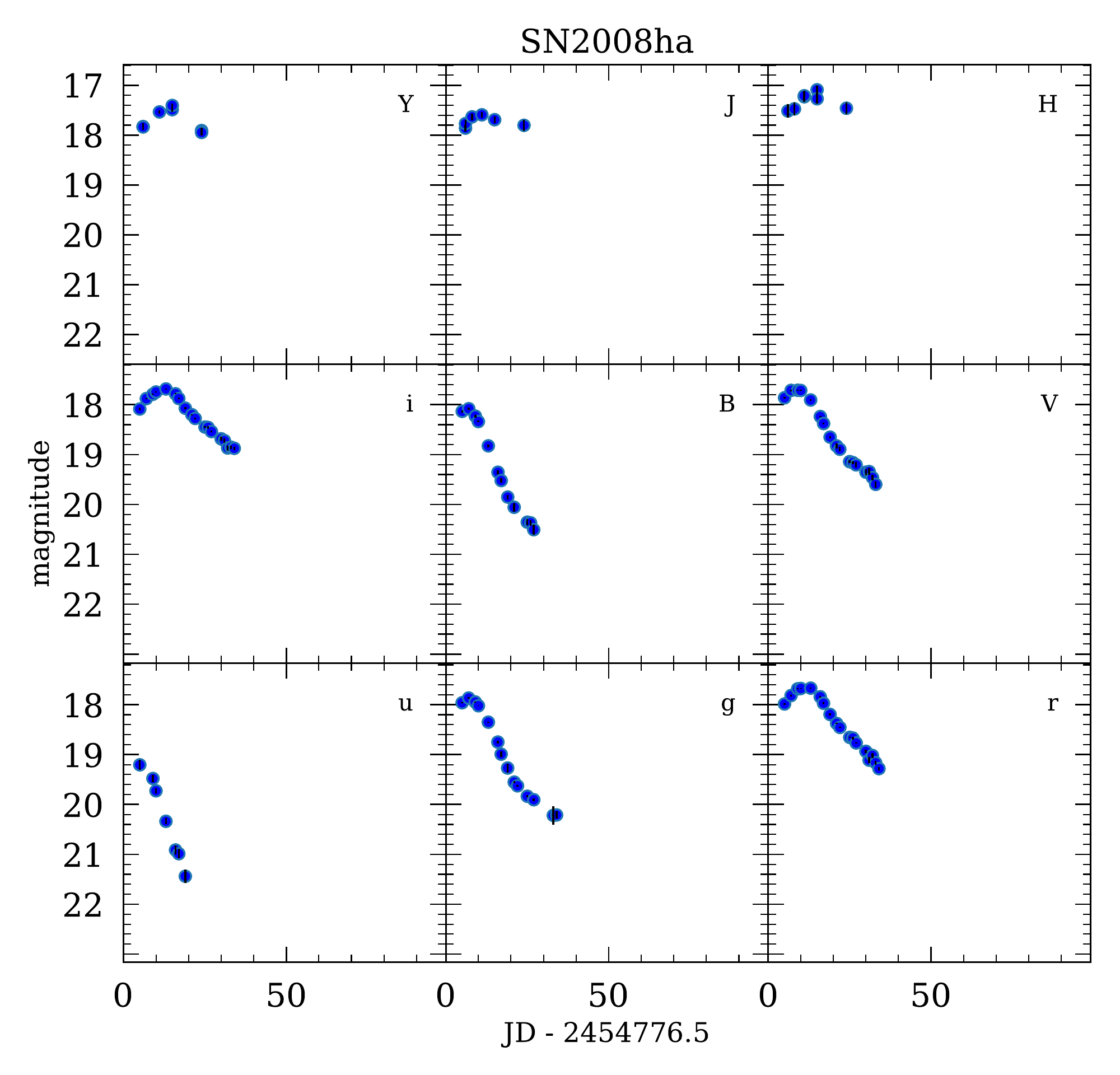}{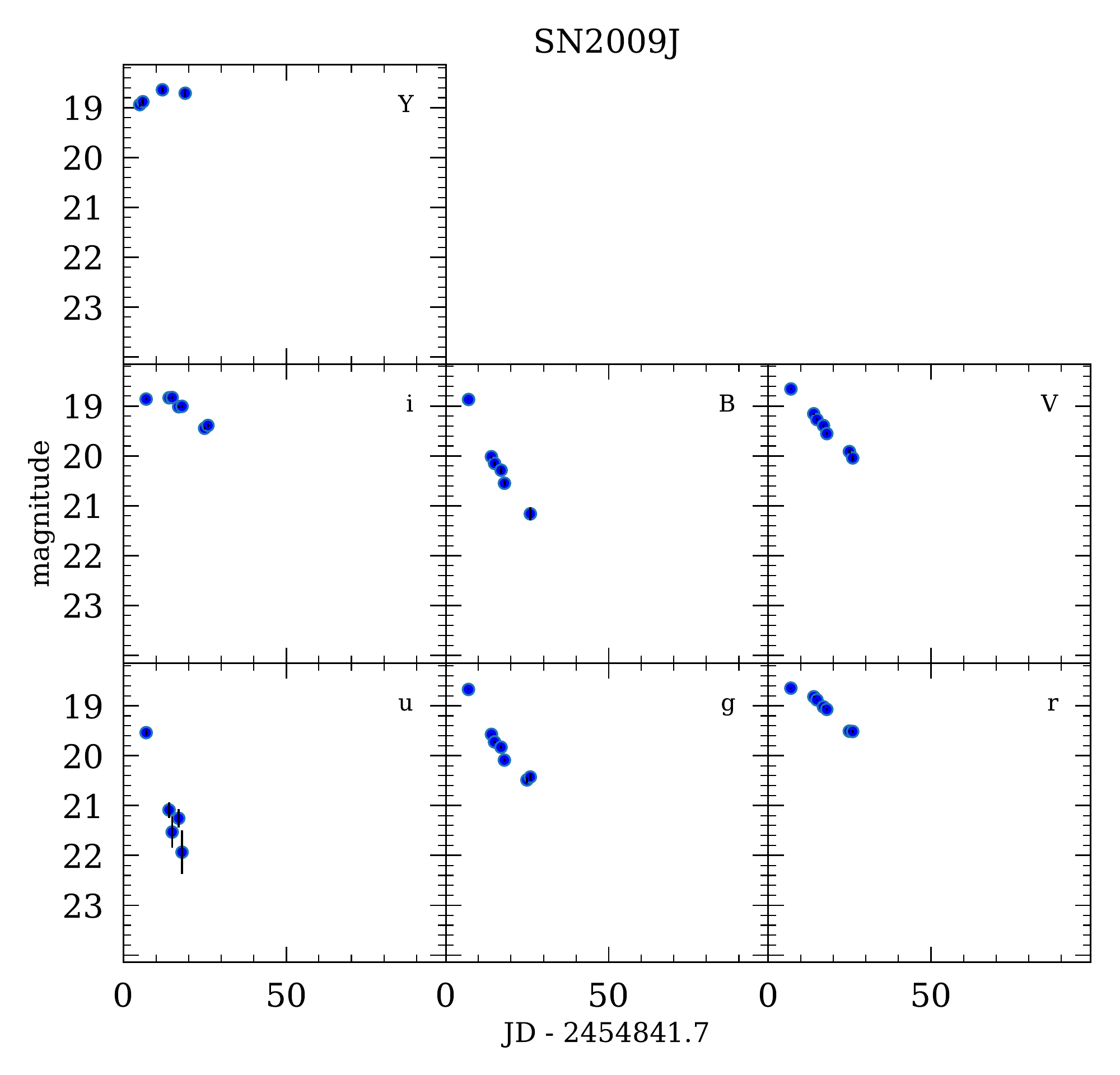}
{\center Krisciunas {\it et al.} Fig.~\ref{fig:Iax_lcs}}
\end{figure}
\clearpage
\newpage

\begin{figure}[t]
\epsscale{0.52}
\plotone{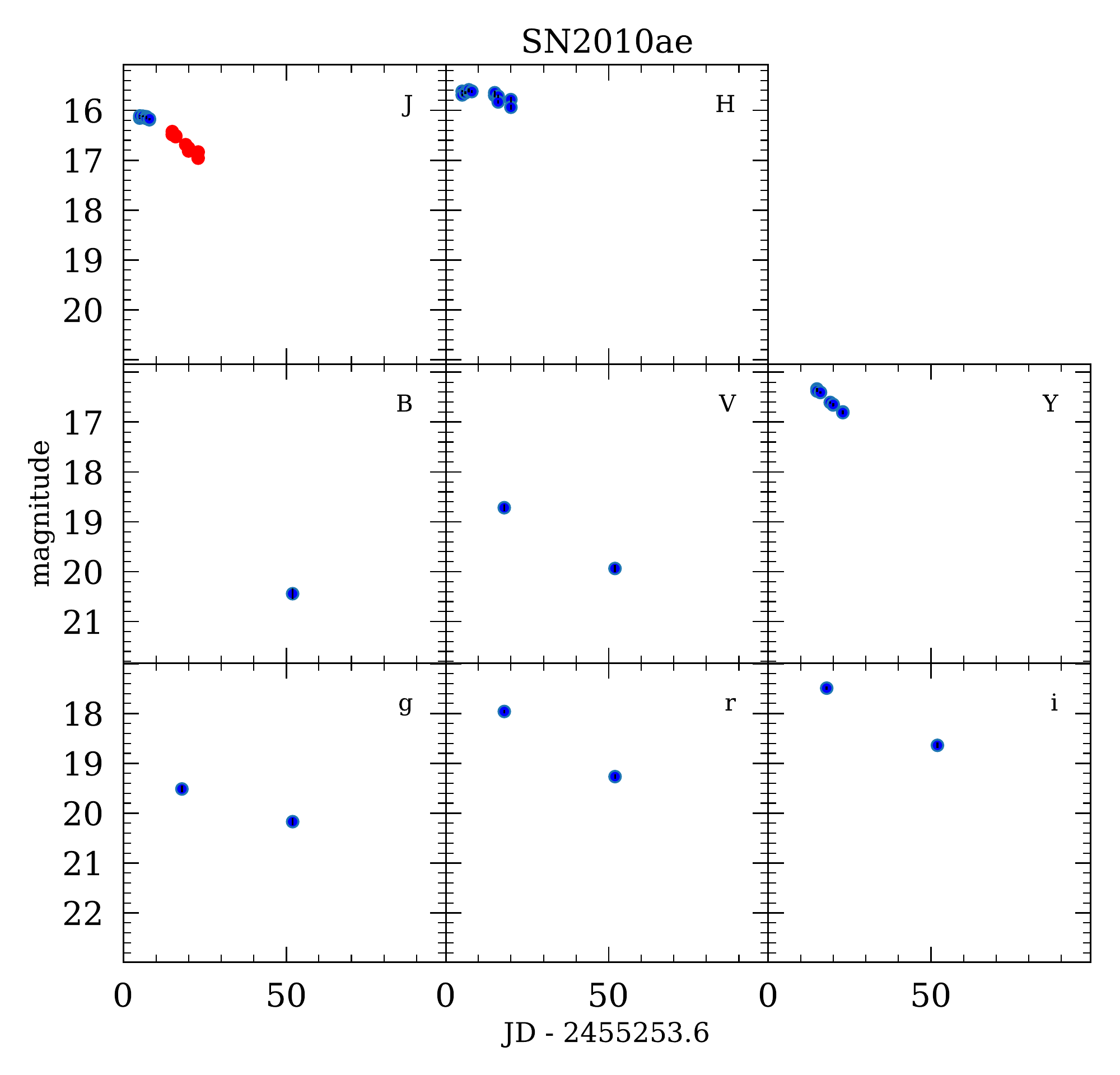}
{\center Krisciunas {\it et al.} Fig.~\ref{fig:Iax_lcs} (Continued)}
\end{figure}


\begin{figure}[t]
\epsscale{1.1}
\plottwo{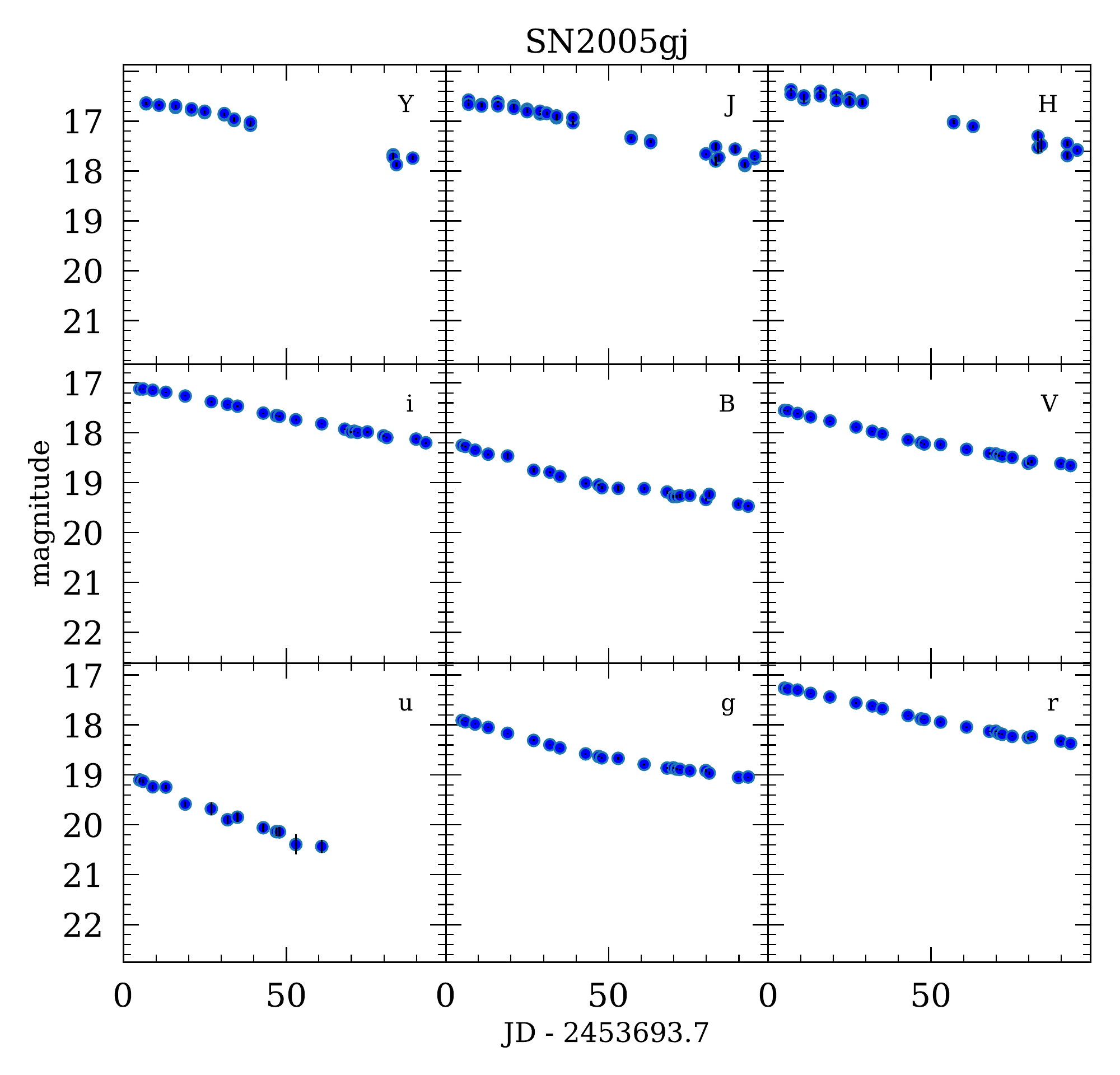}{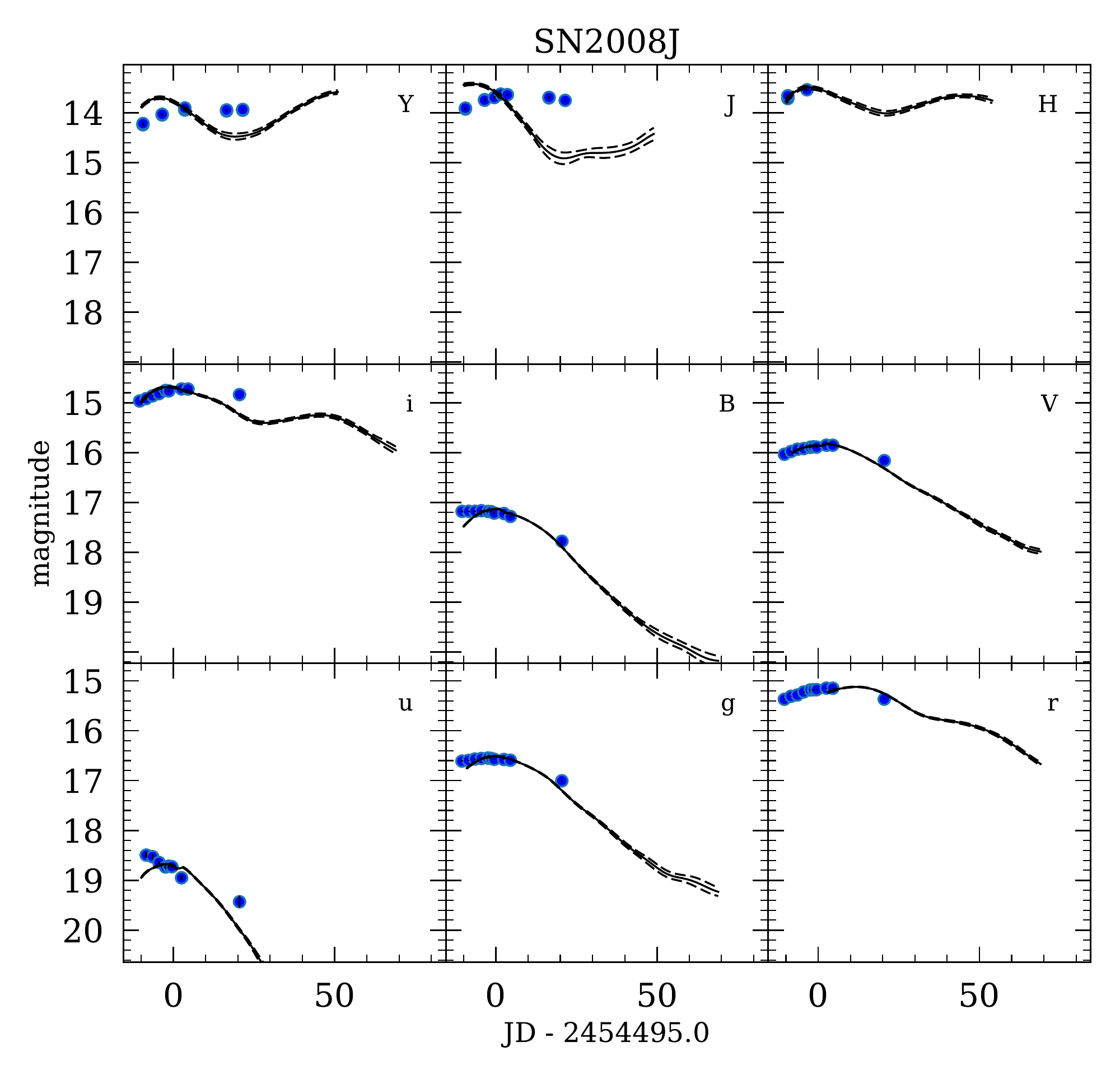}
\newline
\plottwo{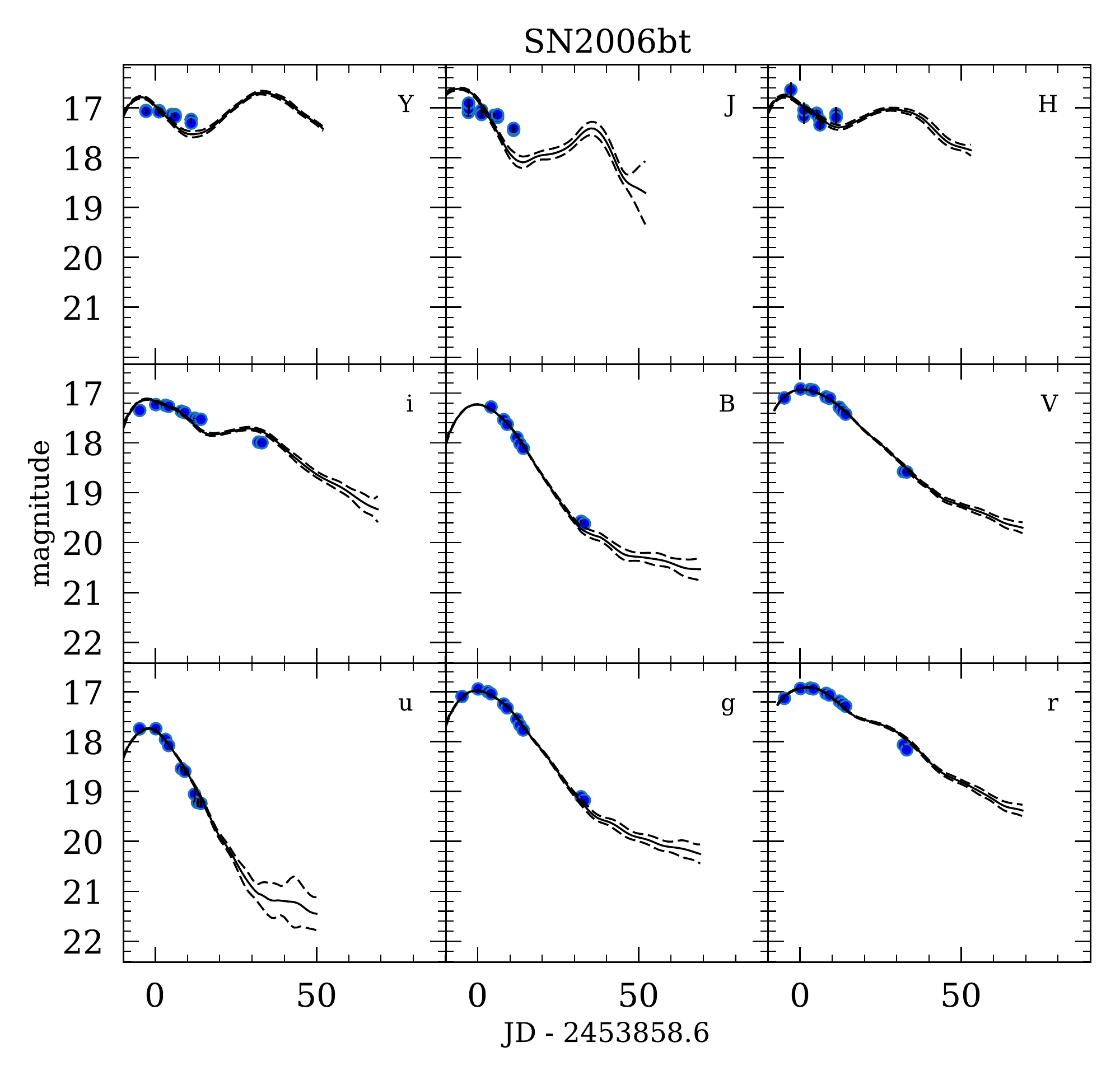}{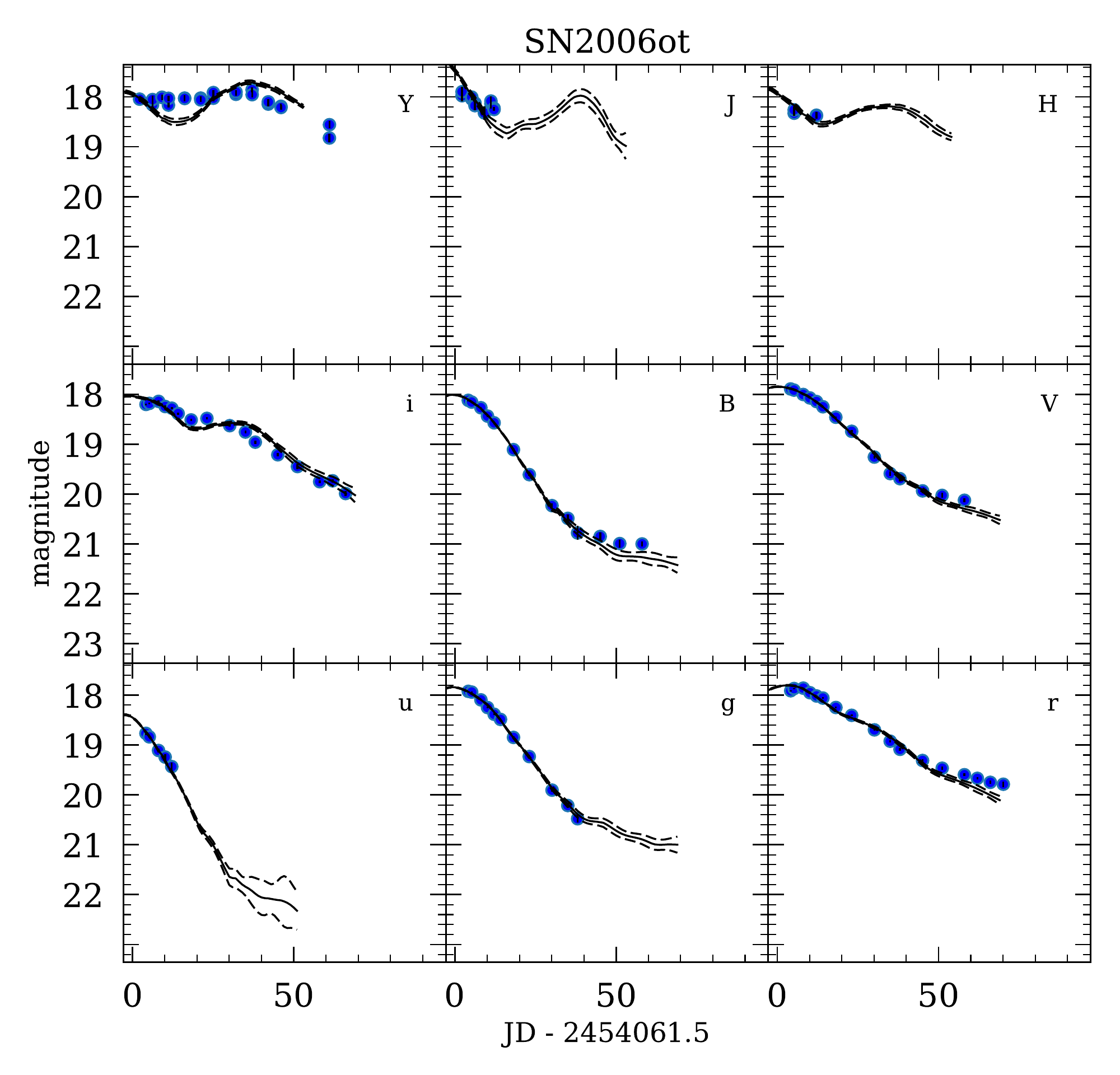}
{\center Krisciunas {\it et al.} Fig.~\ref{fig:other_lcs}}
\end{figure}
\clearpage
\newpage

\begin{figure}[t]
\epsscale{1.1}
\plottwo{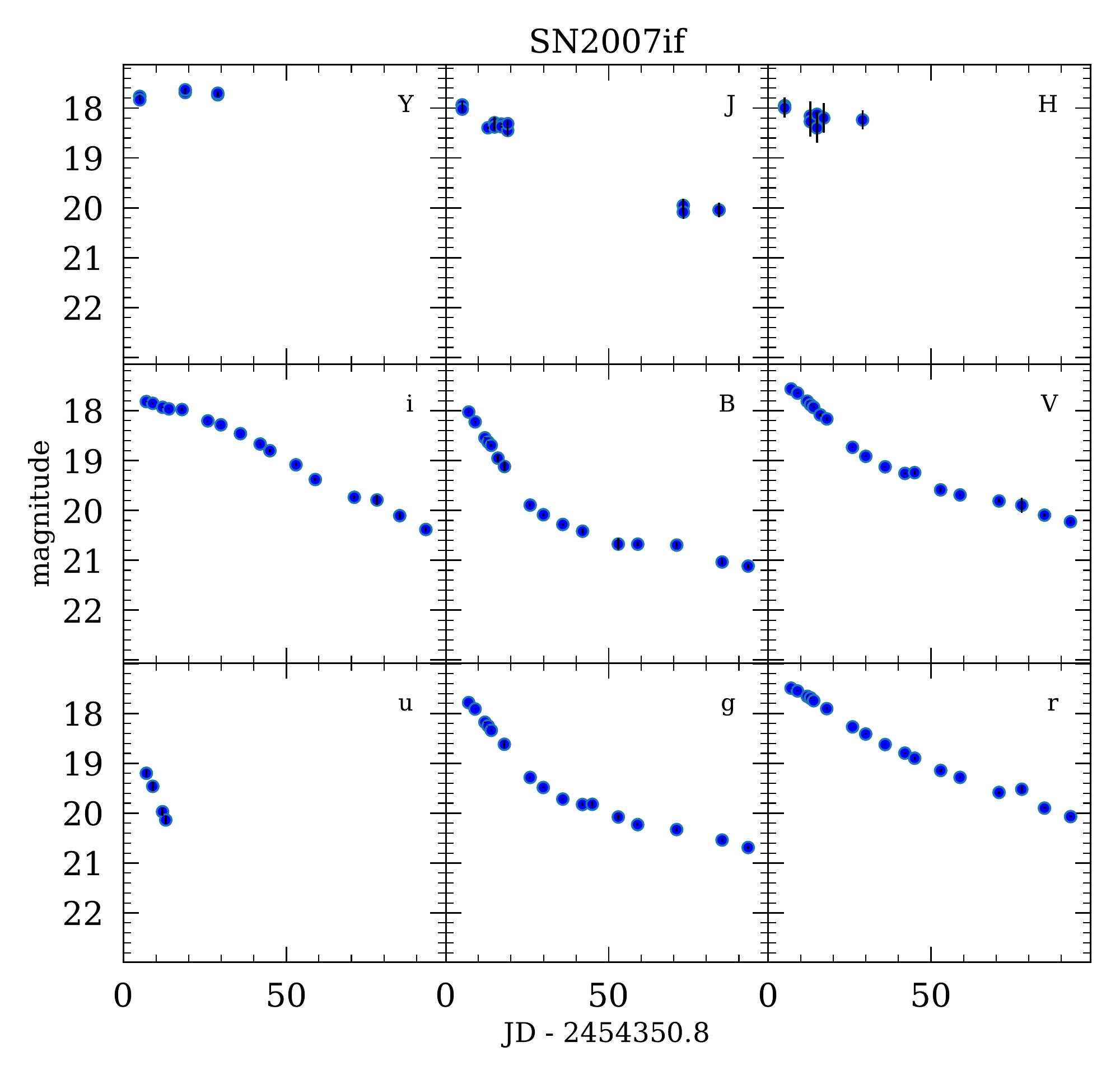}{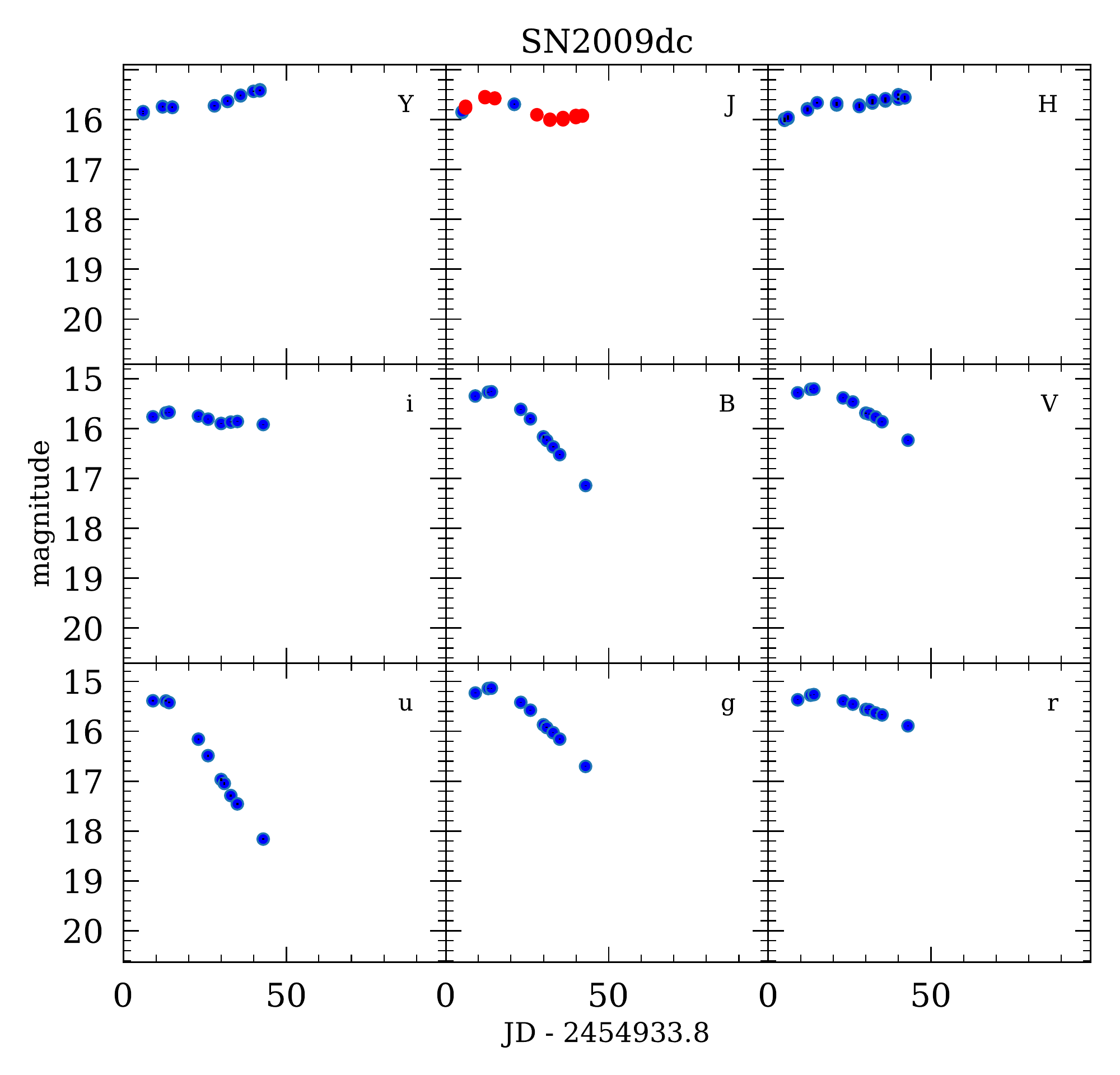}
{\center Krisciunas {\it et al.} Fig.~\ref{fig:other_lcs} (Continued)}
\end{figure}

\clearpage
\newpage


\begin{figure}[t]
\epsscale{1.0}
\plotone{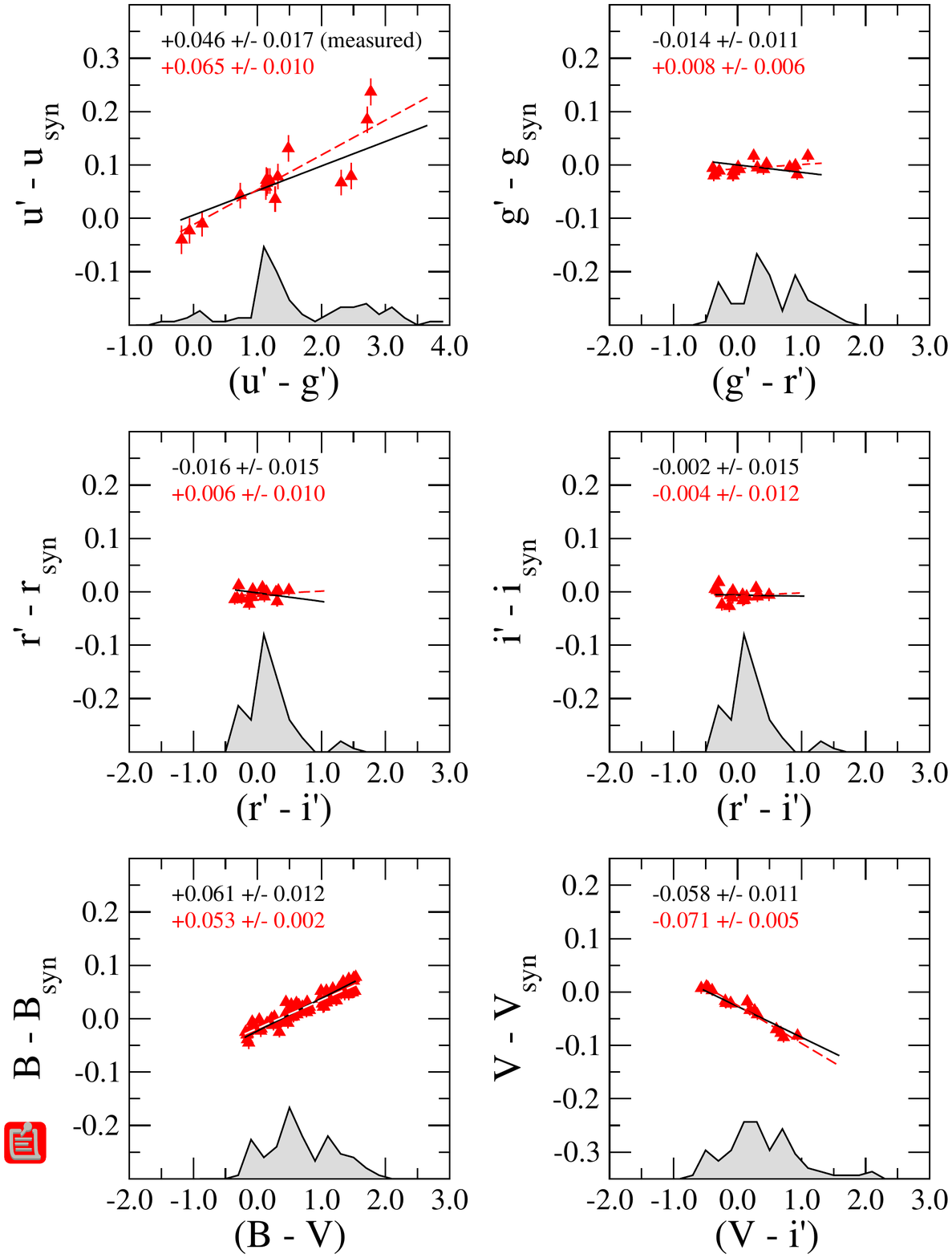}{\center Krisciunas {\it et al.} 
Fig.~\ref{fig:opt_syn_cts}}
\end{figure}


\begin{figure}[t]
\epsscale{1.0}
\plotone{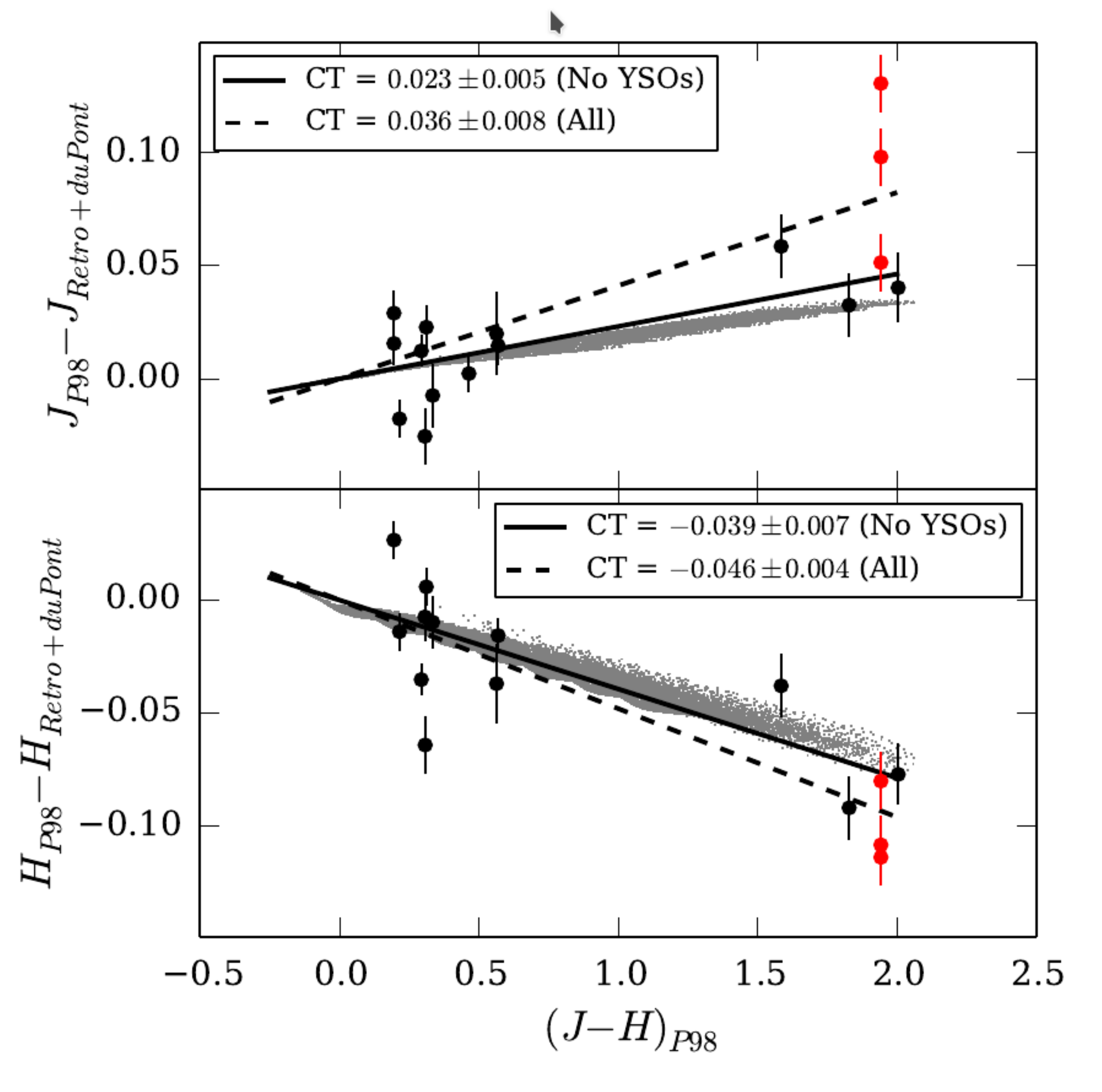}{\center Krisciunas {\it et al.} 
Fig.~\ref{fig:nir_ct_dupont_rc}}
\end{figure}



\begin{figure}[t]
\epsscale{0.7}
\plotone{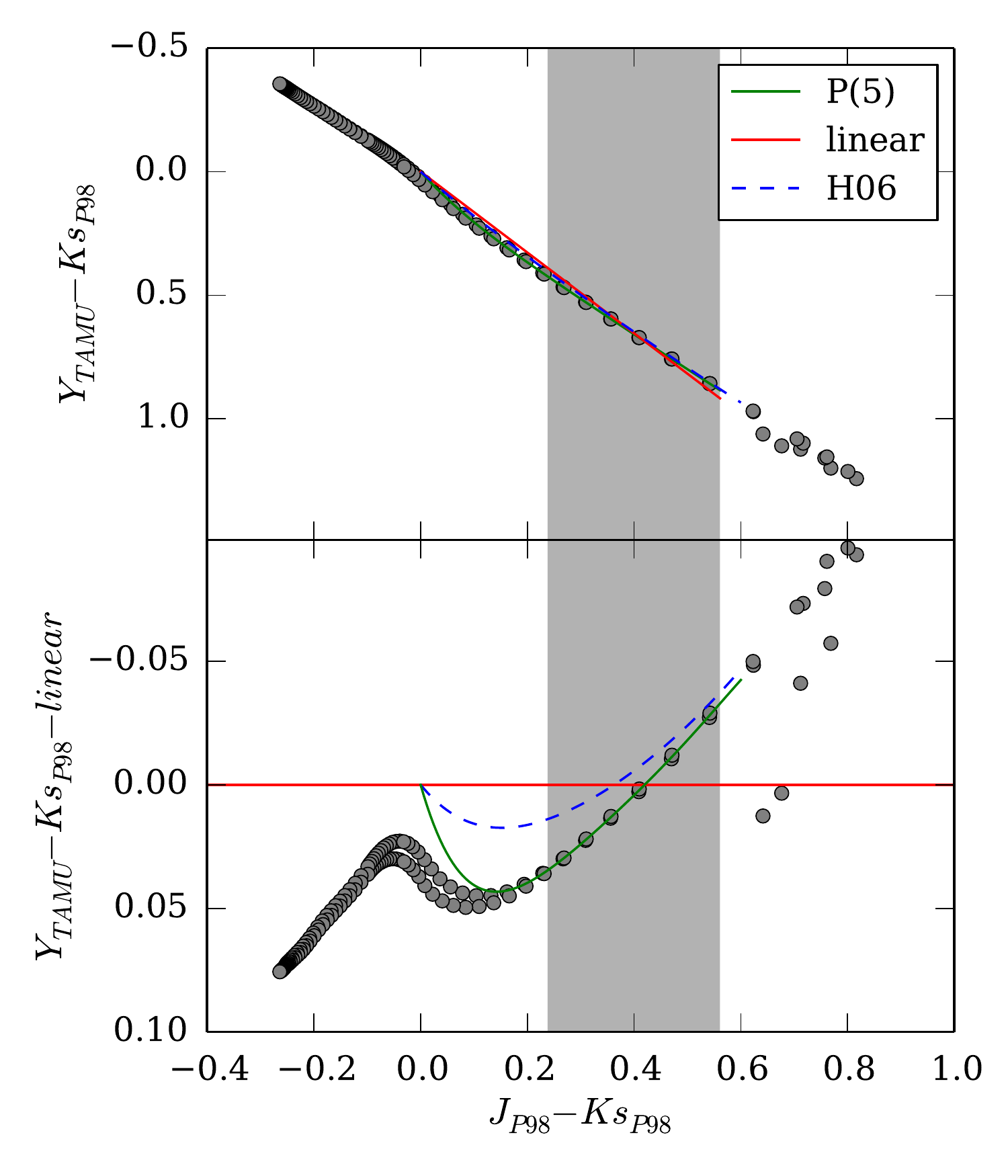}{\center Krisciunas {\it et al.} 
Fig.~\ref{fig:Y_calibration}}
\end{figure}


\begin{figure}[t]
\epsscale{1.0}
\plotone{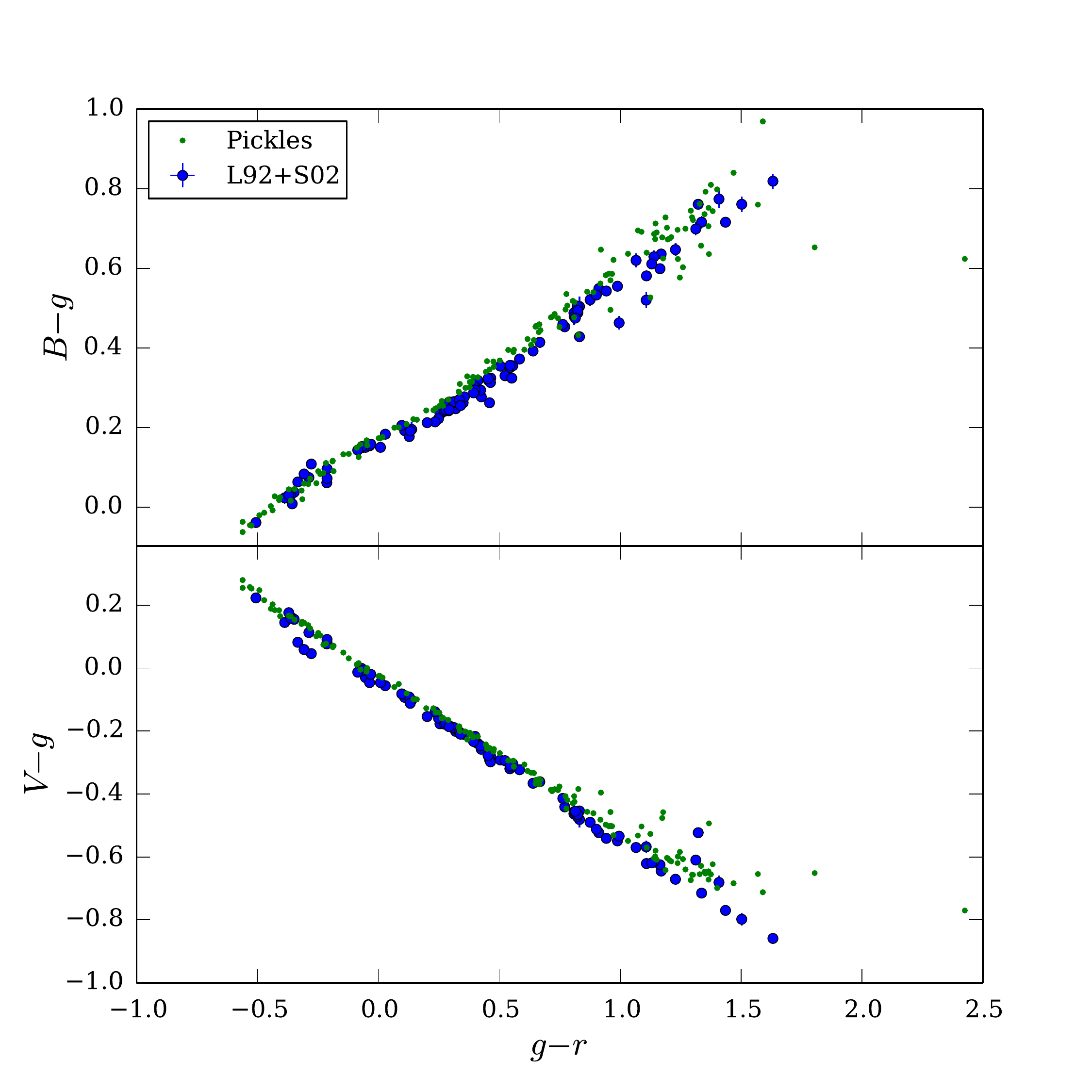}{\center Krisciunas {\it et al.} 
Fig.~\ref{fig:pickles}}
\end{figure}

\end{document}